\documentclass[11pt,titlepage,a4paper]{article}
\usepackage{bozhomac}	
\usepackage{bozhlogo}
\usepackage{cite}

\usepackage{xr1}
\title{\bfseries	\vspace*{-1.678902345in}
\vspace*{-7ex}
{
\begin{flushright}
	  \textbf{\large LANL xxx E-print archive No. quant-ph/0004041}\\[5ex]
\end{flushright}
}
{\huge Fibre bundle formulation of
 \\[6pt] nonrelativistic quantum mechanics\\  (full version)
}
}

\vspace{3ex}

\author{
Bozhidar Z. Iliev
\thanks{Department Mathematical Modeling,
Institute for Nuclear Research and \mbox{Nuclear} Energy,
Bulgarian Academy of Sciences,
Boul. Tzarigradsko chauss\'ee~72, 1784 Sofia, Bulgaria}
\thanks{E-mail address: bozho@inrne.bas.bg}
\thanks{URL: http://www.inrne.bas.bg/mathmod/bozhome/}
}

\date{
\vspace{2.27ex}\ShortTitle{Bundle quantum mechanics}	\\[0.27ex]
\vspace{3.27ex}
	\begin{tabular}{r@{$\colon\to~$}l}
\vspace{0.09ex} Basic ideas	& March 1996		\\[0.09ex]
\vspace{0.09ex} Began		& May 19, 1996		\\[0.09ex]
\vspace{0.09ex} Ended		& July 12, 1996		\\[0.09ex]
\vspace{0.09ex} Revised		& December 1996 -- January 1997,\\[0.09ex]
\vspace{0.09ex} Revised		& April 1997, September 1998, June 1999
								\\[0.09ex]
\vspace{0.09ex} Last updated	& April 4, 2000  	\\[0.09ex]
\vspace{0.27ex} Produced	& \fbox{\today}		\\[0.27ex]
	\end{tabular} \\[1.27ex]
	\begin{tabular}{r@{$\colon~$}l}
\vspace{0.27ex} LANL xxx archive server E-print No.& quant-ph/0004041
						\\[0.27ex]
	\end{tabular} \\[-0.27ex]
\vspace{4.27ex}{\Huge\BOZHO}	\\[4.27ex]
\vspace{0.27ex}\Subject{Quantum mechanics; Differential geometry} \\[2.27ex]
	\begin{tabular}{r@{\hspace{0.512em}}|@{\hspace{0.512em}}l}
\vspace{0.27ex}\MSC[1991]{81P05, 81P99, 81Q99, 81S99} 		  
&
\vspace{0.27ex}\PACS[1996]{02.40.Ma, 04.60.-m, 03.65.Ca, 03.65.Bz}
	\end{tabular} \\[1.27ex]
\vspace{0.27ex}\KeyWords{Quantum mechanics; Geometrization of quantum
		mechanics;\\ Fibre bundles}	\\[0.27ex]
}
\newcommand{\Hil}{\mathcal{F}}	
\newcommand{\HilB}{(\bHil,\proj,\base)}	
	\newcommand{\bHil}{\mathit{F}}	
	\newcommand{\proj}{\pi}		
	\newcommand{\base}{\mathit{M}}	

\newcommand{\Ham}{\mathcal{H}}	
\newcommand{\bHam}{\mathit{H}}	

\newcommand{\HamM}{\boldsymbol{\Ham}} 
\newcommand{\bHamM}{\boldsymbol{\bHam}} 

\newcommand{\mHam}{\boldsymbol{\Ham}^\mathbf{m}}   
\newcommand{\mbHam}{\boldsymbol{\bHam}\!^\mathbf{m}}

\newcommand{\dyn}[1]{\pmb{\mathbb{#1}}}	
	\newcommand{\ope}[1]{\mathcal{#1}}		 
	\newcommand{\mope}[1]{\boldsymbol{\mathcal{#1}}} 
	\newcommand{\mor}[1]{\mathit{#1}}		 
	\newcommand{\mmor}[1]{\boldsymbol{\mathit{#1}}}	 

\providecommand{\iu}{\mathrm{i}} 
\newcommand{\ih}{\mathrm{i}\hbar}
\newcommand{\iih}{\frac{1}{\ih}} 

\newcommand{\Rho}{\mathrm{P}} 

\begin{document}		

\renewcommand{\thefootnote}{\fnsymbol{footnote}}
\maketitle			
\renewcommand{\thefootnote}{\arabic{footnote}}

\tableofcontents		


\pagestyle{myheadings}
\markright{\itshape\bfseries Bozhidar Z. Iliev:
	\upshape\sffamily\bfseries Bundle nonrelativistic quantum mechanics}

\begin{abstract}

	We propose a new systematic fibre bundle formulation of
nonrelativistic quantum mechanics. The new form of the theory is equivalent
to the usual one and is in harmony with the modern trends in theoretical
physics and potentially admits new generalizations in different directions.
In it the Hilbert space of a quantum system (from conventional quantum
mechanics) is replaced with an appropriate Hilbert bundle of states and a
pure state of the system is described by a lifting of paths or section along
paths in this bundle. The evolution of a pure state is determined through the
bundle (analogue of the) Schr\"odinger equation.  Now the dynamical variables
and the density operator are described via liftings of paths or morphisms
along paths in suitable bundles. The mentioned quantities are connected by a
number of relations derived in this work.

\end{abstract}

\section {Introduction}
\label{I}
\setcounter{equation} {0}

	Usually the standard nonrelativistic quantum mechanics of pure states
is formulated in terms of vectors and operators in a Hilbert
space~\cite{Dirac-PQM, Fock-FQM, Messiah-QM, Prugovecki-QMinHS, L&L-3}.
This is in discrepancy and not in harmony with the new trends in
(mathematical) physics~\cite{Schutz, Coquereaux, Konopleva&Popov}
in which the theory of fibre
bundles~\cite{Husemoller, R_Hermann/Geom-phys-systems},
in particular vector
bundles~\cite{R_Hermann-1, R_Hermann-2},
is essentially used. This paper (and its further continuation(s)) is intended
to incorporate the quantum theory in the family of fundamental physical
theories based on the background of fibre bundles.

	The idea of geometrization of quantum mechanics is an old one (see,
e.g.,~\cite{Kibble-79} and the references therein). A good motivation for
such approach is given in~\cite{Anandan-90a,Kibble-79}.
Different geometrical structures in quantum mechanics were
introduced~\cite{Ashtekar&Schilling,Brody&Hughston},
for instance such as inner
products(s)~\cite{Fock-FQM,Messiah-QM,Kibble-79,Anandan-90b},
(linear) connection~\cite{Anandan-90a,Anandan-90b,Uhlmann-91a},
symplectic structure~\cite{Anandan-90a},
complex structure~\cite{Kibble-79}, etc.
The introduction of such structures admits a geometrical treatment of some
problems, for instance, the dynamics in the (quantum) phase
space~\cite{Kibble-79}
and the geometrical phase~\cite{Anandan-90a}.
In a very special case, a gauge structure, i.e.\ a parallel transport
corresponding to a linear connection, in quantum mechanics is
pointed out in~\cite{Wilczek-Zee}.
For us this work is remarkable with the fact that the equation~(10)
in it is a very `ancient' special version of our
equation~(\ref{5.12}) (see also below Sect.~\ref{V}). It, together with
equation~(\ref{5.13}), shows that (up to a constant) with respect to the
quantum evolution the Hamiltonian plays the r\^{o}le of a gauge field
(linear connection). In~\cite{Uhlmann-91a,Uhlmann-91b} one finds
different (vector) bundles defined on the base of the (usual) Hilbert space
of quantum mechanics or its modifications. In these works different parallel
transports in the corresponding bundles are introduced too. Some interesting
ideas concerning the theory of fibre bundles and quantum mechanics can be
found in~\cite{Bernstein&Phillips}.  A recent review of the foundations of
geometric quantization can be found in~\cite{Enriquez&et_al.}.

	A general feature of all of the references above-cited is that in
them all geometric concepts are introduced by using in one or the other way
the accepted mathematical foundation of quantum mechanics, viz.\  a
suitable Hilbert or projective Hilbert space and operators acting in it.
The Hilbert space may be
extended in a certain sense or replaced by a more general space, but this
does not change the main ideas. One of the aims of this work is namely to
change this mathematical background of quantum mechanics.

	Separately we have to mention the approach of Prugove\v{c}ki to the
quantum theory, a selective summary of which can be found
in~\cite{Drechsler/Tuckey-96} (see also the references therein) and
in~\cite{Coleman-96}. It can be characterized as `stochastic' and
`bundle'. The former feature will not be discussed in the present
investigation; thus we lose some advantages of the stochastic quantum theory
to which we shall return elsewhere. The latter `part' of the Prugove\v{c}ki's
approach has some common aspects with our present work but, generally, it is
essentially different. For instance, in both cases the quantum evolution from
point to point (in space-time) is described via a kind of (parallel or
generic linear) transport (along paths) in a suitable Hilbert fibre bundle.
But the notion of a `Hilbert bundle' in our and Prugove\v{c}ki's approach is
different regardless that in both cases the typical (standard) fibre is
practically the same (when one and the same theory is concerned). Besides, we
need not even to introduce the Poincar\'e (principal) fibre bundle over the
space-time or the phase space which play an important r\^{o}le in
Prugove\v{c}ki's theory. Also we have to notice that the used in it concepts
of quantum and parallel transport are special cases of the notion of a
`linear transport along paths'
introduced in~\cite{bp-normalF-LTP,bp-LTP-general}.
The application of the last concept, which is accepted in the present
investigation, has a lot of advantages, significantly simplifies some proofs
and makes certain results `evident' or trivial (e.g.\ the last part of
section~2 and the whole section~4 of~\cite{Drechsler/Tuckey-96}). At last, at
the present level (nonrelativistic quantum mechanics) our bundle formulation
of the quantum theory is insensitive with respect to the space-time
curvature. A detail comparison of Prugove\v{c}ki's and our approaches to the
quantum theory will be done elsewhere.

	Another geometric approach to quantum mechanics is proposed
in~\cite{Graudenz-94} and partially in~\cite{Graudenz-96}, the letter of
which is, with a few exceptions, almost a review of the former.
These works suggest two ideas which are quite important for us.
First, the quantum evolution could be described as a (kind of) parallel
transport in an infinitely dimensional (Hilbert) fibre bundle over the
space-time. And second, the concrete description of a quantum system should
explicitly depend on (the state of) the observer with respect to which it is
depicted (or who `investigates' it). These ideas are incorporated and
developed in our work.

	From the known to the author literature, the work
is~\cite{Asorey&Carinena&Paramio} closest to the approach developed here.
It contains an excellent motivation for applying the fibre bundle
technique to nonrelativistic quantum mechanics.%
\footnote{%
The author thanks J.F.~Cori\~nena (University of Zaragoza, Zaragoza, Spain)
for drawing his attention to reference~\cite{Asorey&Carinena&Paramio}
in May 1998.%
}
Generally said, in this paper the evolution of a quantum system is described
as a `generalized parallel transport' of appropriate objects in a Hilbert
fibre bundle over the 1\ndash dimensional manifold
$\mathbb{R}_+:=\{t :  t\in\mathbb{R}, t\ge0\}$,
interpreted as a `time' manifold (space). We shall comment on
reference~\cite{Asorey&Carinena&Paramio} later, after developing the
formalism required for its analysis (see below Sect.~\ref{V}).
Besides, we emphasizes once again, the paper~\cite{Asorey&Carinena&Paramio}
contains an excellent motivation why the apparatus of fibre bundle theory is
a natural scene for quantum mechanics.

	An attempt to formulate quantum mechanics in terms of a fibre bundle
over the phase space is made in~\cite{Reuter}. Regardless of some common
features, this paper is quite different from the present investigation on
which we shall comment later (see Sect.~\ref{discussion}). In particular,
in~\cite{Reuter} the gauge structure of the arising theory is governed by a
non\ndash dynamical connection related to the symplectic structure of the
system's phase space, while in this work analogous structure (linear
transport along paths) is uniquely connected with system's Hamiltonian,
playing here the r\^{o}le of a gauge field itself.

	The present work is a direct continuation of the considerations
in~\cite{bp-BQM-preliminary} which paper, in fact, may be regarded as its
preliminary version. Here we suggest a \emph{purely fibre bundle formulation
of the nonrelativistic quantum mechanics}.
The proposed geometric formulation of quantum mechanics is \emph{dynamical}
in a sense that all geometrical structures employed for the description of
a quantum system depend on and are determined form the dynamical
characteristics of the system. This new form of the theory is
\emph{entirely equivalent} to the usual one, which is a consequence or our
step by step equivalent reformulation of the quantum theory. The bundle
description is obtained on the base of the developed by the author theory of
transports along paths in fibre
bundles~\cite{bp-normalF-LTP,bp-TP-general,bp-LTP-general}, generalizing the
theory of parallel transport. It is partially generalized here to the
infinitely dimensional case.

        The main object in quantum mechanics is the Hamiltonian (operator)
which, through the Schr\"odinger equation, governs the evolution of a quantum
system~\cite{Fock-FQM,Messiah-QM,Prugovecki-QMinHS,L&L-3}. In our novel
approach its r\^{o}le is played by a suitable linear transport along paths
in an appropriate Hilbert bundle. It turns out that up to a
constant the matrix-bundle Hamiltonian, which is uniquely
determined by the Hamiltonian in a given field of bases, coincides with
the matrix of the coefficients of this transport
(cf.\ an analogous result in~\cite[sect.~5]{bp-BQM-preliminary}).
This fact, together with the replacement of the usual Hilbert space with a
Hilbert bundle, is the corner-stone for the possibility for the new
formulation of the nonrelativistic quantum mechanics.

\vspace{2ex}
	The present paper is organized as follows.

        In Sect.~\ref{II} are reviewed some well known facts from the quantum
standard mechanics and partially is fixed our notation. Here, as well as
throughout this work, we follow the established in the physical literature
degree of rigor.  But, if required, the present work can be reformulated to
meet the present\ndash day mathematical standards. For this purpose one can
use, for instance, the quantum-mechanical formalism described
in~\cite{Prugovecki-QMinHS} or in~\cite{Neumann-MFQM}
(see also~\cite{Reed&Simon,Steeb-modernQM}).

	Sect.~\ref{III} contains a preliminary mathematical material required
for the goals of this work. In Subsect.~\ref{III.1} are collected some basic
definitions concerning vector and Hilbert bundles. Next, in
Subsect.~\ref{III.2}, the notion of a linear transport along paths in vector
bundles is recalled and some its peculiarities in the Hilbert bundle case are
pointed. In Subsect.~\ref{III.3} the concepts of liftings of paths, sections
along paths and derivations along paths are introduced.

	Sect.~\ref{new-I} begins the building of the new bundle approach to
quantum mechanics. After a brief review of some references
(Subsect.~\ref{new-I.1}), a motivation for the application of fibre bundle
formalism to quantum mechanics is presented (Subsect.~\ref{new-I.2}). Also
some heuristic considerations of elements of the new theory are given.
Subsect.~\ref{new-I.3} introduces the basic initial assertions of the bundle
version of quantum mechanics. They are formulated in a form of two postulates
which are enough for the bundle description of the evolution of a quantum
system. In the bundle approach the analogues of the Hilbert space of states
and state vector of a system are the system's Hilbert bundle (of states) and
the (state) lifting of paths (or section along paths) in it. In
Subsect.~\ref{new-I.4} a preliminary summary of some results of this
investigation is presented.

	In Sect.~\ref{IV} is proved that in the bundle description the
evolution operator of a quantum system is (equivalently) replaced by a
suitable linear transport along paths, called \emph{evolution transport}.

	Sect.~\ref{V} is devoted to the bundle analogues of the Schr\"odinger
equation which are fully equivalent to it. In particular, in it is introduced
the
\emph{matrix-bundle Hamiltonian}
which governs the quantum evolution through the
\emph{matrix-bundle Schr\"odinger equation}.
The corresponding matrix of the bundle-evolution transport is found. It is
proved that
\emph{up to a constant the matrix of the coefficients of the bundle
evolution transport coincides with the matrix-bundle Hamiltonian}.
On this basis is derived the
\emph{(invariant) bundle-Schr\"odinger equation}.
Geometrically it simply means that the corresponding state sections are
(parallelly, or, more precisely, linearly) transported by means of the
bundle evolution transport (along paths).

	In Sect.~\ref{VI} is considered the question for the bundle
description of observables. It turns out that
\emph{to any observable there corresponds a unique Hermitian bundle
morphism (along paths) and vice versa}.

	The bundle description of the different pictures of motion is
presented in Sect.~\ref{VII}. The Schr\"odinger picture, which, in
fact, is investigated until Sect.~\ref{VII}, is reviewed in
Subsection~\ref{VII.1}. To the bundle Heisenberg picture is devoted
Subsection~\ref{VII.2}. The corresponding equations of motion for the
observables are derived and discussed. In Subsection~\ref{VII.3} is
investigated the `general' picture of motion obtained by means of an
arbitrary linear unitary bundle map (along paths). In it are derived
and discussed different equations for the state sections and
observables.

	In Sect.~\ref{VIII} are investigated problems concerning the
integrals of motion from fibre bundle point of view. An interesting result
here is that a dynamical variable is an integral of motion iff the
corresponding to it bundle morphism is transported along the observer's world
line with the help of the (bundle) evolution transport.

	The bundle approach to the quantum mechanics of mixed states is
investigated in Sect.~\ref{XI}. Subsection~\ref{XI.1} is a brief review of
the conventional concepts of mixed state(s) and density operator (matrix).
Their bundle description is presented in Subsection~\ref{XI.2}. It turns out
that to the density operator there corresponds a suitable
\emph{density morphism along paths}. The equations for its time evolution
are derived. In Subsection~\ref{XI.3} are studied problems connected with the
representations and description of mixed quantum states in the different
pictures of motion. The equations of motion for density morphisms and
operators are derived.

	Sect.~\ref{IX} is devoted to the curvature of the (bundle) evolution
transport.

	In Sect.~\ref{XII} is paid attention on the observer's r\^{o}le in the
theory and are considered and interpreted some modifications of the proposed
approach to quantum mechanics. Possible fields for further research are
sketched too.

	In Sect.~\ref{summary} are briefly summarized our
results and  it is presented a comparison table between the conventional,
Hilbert space, and the new, Hilbert bundle, formulations of nonrelativistic
quantum mechanics.

	In Sect.~\ref{discussion} are discussed certain aspects of the bundle
formulation of nonrelativistic quantum mechanics and are pointed some its
possible generalizations and applications.

%
%

\section {Evolution of pure quantum states (review)}
\label{II}
\setcounter{equation} {0}

	In quantum mechanics~\cite{Messiah-QM,Fock-FQM,L&L-3,Steeb-modernQM}
a pure state of a quantum system is described by a state vector $\psi(t)$ (in
Dirac's~\cite{Dirac-PQM} notation $|t\rangle$). Generally, depends on the
time $t\in\mathbb R$ and belongs to a Hilbert space $\Hil$ (specific to any
concrete system) generically endowed with a nondegenerate Hermitian scalar
product
\(
\langle\cdot | \cdot \rangle\colon  \Hil\times\Hil \to \mathbb{C}.
\)%
\footnote{%
For some (\eg unbounded) states the system's state vectors form a more
general space than a Hilbert one. This is insignificant for the following
presentation. A sufficient for our purposes summary of Hilbert space theory
is given in~\cite[Appendix]{Lang/Manifolds}.%
}
For every two instants of time $t_2$ and $t_1$ the corresponding state vectors
are connected by the equality
\begin{equation}	\label{2.1}
\psi(t_2) = \ope{U}(t_2,t_1)\psi(t_1)
\end{equation}
where $\ope{U}$ is the \emph{evolution operator} of the
system~\cite[chapter~IV, Sect.~3.2]{Prugovecki-QMinHS}.  It is supposed to be
linear and unitary, i.e.\
	\begin{align}   \label{2.20}
\ope{U}(t_2,t_1)( \lambda\psi(t_1) + \mu\xi(t_1) )	&=
\lambda \ope{U}(t_2,t_1)(\psi(t_1)) + \mu \ope{U}(t_2,t_1)( \xi(t_1) ),
\ \ \ \ \ \ \ \         \\  \label{2.2}
\ope{U}^\dag(t_1,t_2) &= \ope{U}^{-1}(t_2,t_1)
	\end{align}
for every $\lambda,\mu\in\mathbb{C}$
and state vectors $\psi(t),\xi(t)\in\Hil$, and such that for any $t$
	\begin{equation}	\label{2.3}
\ope{U}(t,t) = \id_\Hil.
	\end{equation}
Here $\id_X$ means the identity map of a set $X$ and the dagger ($\dag$)
 denotes  Hermitian conjugation, i.e.\ if $\varphi,\psi\in\Hil$ and
$\ope{A}\colon \Hil\to\Hil$, then $\ope{A}^\dag$ is defined by
	\begin{equation}	\label{2.4}
\langle \ope{A}^\dag\varphi | \psi \rangle =
			\langle \varphi | \ope{A}\psi \rangle.
	\end{equation}
In particular $\ope{U}^\dag$ is defined by
\(
\langle \ope{U}^\dag(t_1,t_2)\varphi(t_2) | \psi(t_1) \rangle =
\langle \varphi(t_2) | \ope{U}(t_2,t_1)\psi(t_1) \rangle.
\)
So, knowing $\psi(t_0)=\psi_0$ for some moment~$t_0$, one knows the state
vector for every moment $t$ as
\(
\psi(t)=\ope{U}(t,t_0)\psi(t_0)=\ope{U}(t,t_0)\psi_0.
\)

	Let $\Ham(t)$ be the Hamiltonian (function) of the
system. It, generally, depends on the time~$t$ explicitly%
\footnote{%
Of course, the Hamiltonian depends also on the observer with respect to which
the evolution of the quantum system is described. This dependence is
usually implicitly assumed and not written
explicitly~\cite{Dirac-PQM,Messiah-QM}. This deficiency will be eliminated in
a natural way further in the present work. The Hamiltonian can also depend on
other quantities, such as the (operators of the) system's generalized
coordinates. This possible dependence is insignificant for our investigation
and will not be written explicitly.%
}
and it is supposed to be a Hermitian operator, i.e.\ $\Ham^\dag(t)=\Ham(t)$.
The Schr\"odinger equation
(see~\cite[\S~27]{Dirac-PQM} or~\cite[chapter~V, Sec.~3.1]{Prugovecki-QMinHS})
	\begin{equation}	\label{2.5}
\ih\frac{d \psi(t)}{d t} = \Ham(t) \psi(t),
	\end{equation}
with
$\iu\in\mathbb{C}$ and $\hbar$ being respectively the imaginary unit
and the Plank's constant (divided by $2\pi$),
together with some initial condition
	\begin{equation}	\label{2.5a}
\psi(t_0)=\psi_0\in\Hil
	\end{equation}
is postulated in the quantum mechanics. They determine the time-evolution of
the state vector $\psi(t)$.

		The substitution of~(\ref{2.1}) into~(\ref{2.5}) shows that
there is a bijective correspondence between~$\ope{U}$ and~$\Ham$ described by
	\begin{equation}	\label{2.6}
\ih\frac{\partial \ope{U}(t,t_0)}{\partial t} = \Ham(t)\circ\ope{U}(t,t_0),
\qquad
\ope{U}(t_0,t_0) = \id_\Hil
	\end{equation}
where $\circ$ denotes composition of maps.
If~$\ope{U}$ is given, then
	\begin{equation}	\label{2.7}
\Ham(t) =
\ih\frac{\partial \ope{U}(t,t_0)}{\partial t} \circ \ope{U}^{-1}(t,t_0) =
\ih\frac{\partial \ope{U}(t,t_0)}{\partial t} \circ \ope{U}(t_0,t),
	\end{equation}
where we have used the equality
\[
\ope{U}^{-1}(t_2,t_1) = \ope{U}(t_1,t_2)
\]
which follows from~(\ref{2.1})
(see also below~(\ref{2.8}) or Sect.~\ref{IV}). Conversely, if
$\Ham$ is given, then~\cite[chapter~VIII, \S\ 8]{Messiah-QM} $\ope{U}$
is the unique solution of the integral equation
\(
\ope{U}(t,t_0) = \id_\Hil +
{\iih} \int\limits_{t_0}^{t} \Ham(\tau)\ope{U}(\tau,t_0) d\tau,
\)
i.e.\ we have
	\begin{equation}	\label{2.8}
 \ope{U}(t,t_0) = \Texp\int\limits_{t_0}^{t}
{\iih} \Ham(\tau) d\tau,
	\end{equation}
where $\Texp\int\limits_{t_0}^{t}\cdots d\tau$ is the chronological
(called also T-ordered, P-ordered or path-ordered) exponent (defined, e.g., as
the unique solution of the initial-value problem~(\ref{2.6}); see
also~\cite[equation~(1.3)]{Asorey&Carinena&Paramio}).%
\footnote{%
The physical meaning of $\ope{U}$ as a propagation function, as well as its
explicit calculation (in component form) via $\Ham$ can be found, e.g.,
in~\cite[\S~21 and \S~22]{Bjorken&Drell-1}%
}
From here
follows that the Hermiticity of $\Ham$, $\Ham^\dag=\Ham$, is equivalent to
the unitarity of $\ope{U}$ (see~(\ref{2.2})).

	Let us note that for the rigorous mathematical understanding of the
derivations in~(\ref{2.5}), (\ref{2.6}), and~(\ref{2.7}), as well as of the
chronological (path-ordered) exponent in~(\ref{2.8}), one has to apply the
developed in~\cite{Prugovecki-QMinHS} mathematical apparatus, but this is out
of the subject of the present work.

	If $\ope{A}(t)\colon \Hil\to\Hil$ is the (linear Hermitian) operator
corresponding to a dynamical variable
$\dyn{A}$ at the moment $t$, then the mean value
(= the mathematical expectation) which it assumes at a state described by a
state vector $\psi(t)$ with a finite norm is
	\begin{equation}	\label{2.9}
\langle\ope{A}\rangle_\psi^t :=
\langle\ope{A}(t)\rangle_\psi^t :=
\langle\ope{A}(t)\rangle_{\psi(t)} :=
\frac{\langle\psi(t) | \ope{A}(t)\psi(t)\rangle}
{\langle\psi(t) | \psi(t)\rangle}.
	\end{equation}
It is interpreted as the observed value of $\dyn{A}$ that can be measured
experimentally.

	Often the operator $\ope{A}$ can be chosen independent
of the time $t$. (This is possible, e.g., if $\ope{A}$ does not depend on $t$
explicitly~\cite[chapter~VII, \S~9]{Messiah-QM} or if the spectrum of
$\ope{A}$ does not change in time~\cite[chapter~III, sect.~13]{Fock-FQM}.) If
this is the case, it is said that the system's evolution is depicted in the
Schr\"odinger picture of motion~\cite[\S~28]{Dirac-PQM}, \cite[chapter~VII,
\S~9]{Messiah-QM}.

\section{Mathematical preliminaries}
	\label{III}

	Before starting with the formulation of quantum mechanics is terms of
fibre bundles, several geometrical tools have to be known. In this section
is collected most of the pure mathematical material required for this  goal.
At first, we present some facts from the theory of Hilbert bundles.
These bundles will replace the Hilbert spaces in quantum mechanics. Next, we
give a brief introduction to the theory of linear transports along paths in
vector bundles and specify peculiarities of the Hilbert bundle case. The
linear transports are needed for the bundle description of the quantum
evolution. At the end, we pay attention to the liftings of paths and
section along paths. These objects will replace the state vectors of
ordinary quantum mechanics.

\subsection{Hilbert bundles}
	\label{III.1}

	At the beginning, to fix the terminology, we recall the definitions
of bundle, section, fibre map, morphism, and vector bundle.%
\footnote{%
For details, examples, etc.,
see~\cite{Husemoller,Steenrod,Greub&et_al.-1,Hartman,R_Hermann-1,R_Hermann-2,
Sze-Tsen}.%
}

	A \emph{bundle} is a triple $(E,\pi,B)$ of sets $E$ and $B$, called
(total) bundle space and base (space) respectively, and (generally)
surjective mapping $\pi\colon E\to B$, called projection. For every $b\in B$
the set $\pi^{-1}(b)$ is called fibre over $b$.
	If $X\subseteq B$, the bundle
$(E,\pi,B)|_X := (\pi^{-1}(X),\pi|_{\pi^{-1}(X)},X)$
is called the restriction on $X$ of a bundle $(E,\pi,B)$.
	A \emph{section} of the bundle $(E,\pi,B)$ is a mapping
$\sigma\colon B\to E$ such that $\pi\circ\sigma=\id_B$, \ie
$\sigma\colon b\mapsto \sigma(b)\in\pi^{-1}(b)$. The set of sections of
$(E,\pi,B)$ is denoted by $\Sec(E,\pi,B)$.

	A mapping $\varphi\colon E\to E$ is said to be a \emph{fibre map} if
it carries fibres into fibres. Precisely,  $\varphi$ is a fibre map iff, for
every $b\in B$, there exists a point $b'\in B$ such that $\varphi$ maps
$\pi^{-1}(b)$ into $\pi^{-1}(b')$,
\ie  $\varphi|_{\pi^{-1}(b)}\colon \pi^{-1}(b)\to\pi^{-1}(b')$.
A \emph{(fibre) morphism} of the bundle $(E,\pi,B)$ is a pair $(\varphi,f)$
of maps $\varphi\colon E\to E$ and $f\colon B\to B$ such that
$\pi\circ\varphi=f\circ\pi$. The set of morphisms of $(E,\pi,B)$ is denoted
by $\Morf(E,\pi,B)$, \ie
\[
\Morf(E,\pi,B)
 := \{
	(\varphi,f)| \varphi\colon E\to E,\ f\colon B\to B,\
	\pi\circ\varphi = f\circ\pi
    \} .
\]
A map $\varphi\colon E\to E$ is called \emph{morphism over $B$} or
\emph{$B$-morphism} if $(\varphi,\id_B)\in\Morf(E,\pi,B)$. The set of all
$B$-morphisms of $(E,\pi,B)$ will be denoted by $\Morf_B(E,\pi,B)$, \ie
\[
\Morf_B(E,\pi,B)
 := \{\varphi|\varphi\colon E\to E,\ \pi\circ\varphi=\pi \} .
\]

	For every morphism $(\varphi,f)\in\Morf(E,\pi,B)$, the map $\varphi$
is a fibre map since from $\pi\circ\varphi = f\circ\pi$ follows
 $\varphi|_{\pi^{-1}(b)}\colon\pi^{-1}(b)\to\pi^{-1}(f(b))$ for every $b\in
B$.  In particular, any $B$\ndash morphism $\varphi\in\Morf_B(E,\pi,B)$  is a
fibre\ndash preserving map as
 $\varphi|_{\pi^{-1}(b)}\colon\pi^{-1}(b)\to\pi^{-1}(b)$.
Conversely, every fibre map $\varphi\colon E\to E$ defines a morphism
$(\varphi,f)$ with $f:=\pi\circ\varphi\circ\pi^{-1}\colon B\to B$;  $f$ is
called \emph{induced map}  of the fibre map $\varphi$, resp.\ $(\varphi,f)$
is \emph{induced morphism}.%
\footnote{%
The transport along a path $\gamma\colon J\to B$ is an example of a fibre map
along $\gamma$ -- \emph{vide infra} Subsect.~\ref{III.2}.%
}

	Consider the set of point-restricted morphisms
	\begin{align*}
E_0 : &= \{
 (\varphi_b,f)\,|\,
	\varphi_b = \varphi|_{\pi^{-1}(b)},
	\ b\in B,\ (\varphi,f)\in\Morf(E,\pi,B)
	\}
\\
 & = \{
	(\varphi_b,f) \,|\, \varphi_b\colon\pi^{-1}(b)\to \pi^{-1}(f(b)),
	\ b\in B,\ f\colon B\to B
	\} ,
	\end{align*}
\ie $(\psi,f)\in E_0$ iff $f\colon B\to B$ and there exists a unique $b\in B$
such that $\psi\colon\pi^{-1}(b)\to\pi^{-1}(f(b))$ and we write $\psi_b$ for
$\psi$. Defining  $\pi_0\colon E_0\to B$ by $\pi_0(\varphi_b,f):=b$ for
$(\varphi_b,f)\in E_0$, we see that $(E_0,\pi_0,B)$ is a bundle over the same
base $B$ as $(E,\pi,B)$. This is the \emph{bundle of point\ndash restricted
morphisms} of $(E,\pi,B)$. It will be denoted by $\morf(E,\pi,B)$, \ie
$\morf(E,\pi,B):=(E_0,\pi_0,B)$. There exists a bijective correspondence
$\tau$ such that
\[
\Morf(E,\pi,B)  \xrightarrow{\ \tau\ } \Sec\bigl(\morf(E,\pi,B)\bigr) .
\]
In fact, for $(\varphi,f)\in\Morf(E,\pi,B)$, we put
$\tau\colon(\varphi,f)\mapsto\tau_{(\varphi,f)}$ with
\(
\tau_{(\varphi,f)}\colon b \mapsto \tau_{(\varphi,f)}(b)
 := (\varphi|_{\pi^{-1}(b)},f)\in \pi_0^{-1}(b)
\)
for every $b\in B$. Conversely, for $\sigma\in\Sec(\morf(E,\pi,B))$, we set
\(
\tau^{-1}\colon\sigma\mapsto\tau^{-1}(\sigma):=(\varphi,f)\in\Morf(E,\pi,B)
\)
where, if $b \in B$ and $\sigma(b)=(\varphi_b,f)$, the map
$\varphi\colon E\to E$ is defined by $\varphi|_{\pi^{-1}(b)}:=\varphi_b$.

	The above constructions can be modified for morphisms over the
bundle's base as follows. The \emph{bundle $\morf_B(E,\pi,B)$ of
point\ndash restricted morphisms over} $B$ of $(E,\pi,B)$ has a base $B$,
bundle space
	\begin{align*}
E_{0}^{B} : &=
 \{
   \varphi_b \,|\, \varphi_b=\varphi|_{\pi^{-1}(b)},\
	b\in B,\ \varphi\in\Morf_B(E,\pi,B)
 \}
\\
 & = \{ \varphi_b \,|\, \varphi_b\colon \pi^{-1}(b)\to\pi^{-1}(b),\ b\in B \}
	\end{align*}
and projection $\pi_{0}^{B}\colon E_{0}^{B}\to B$ such that
\[
\pi_{0}^{B}(\varphi_b) := b,\quad \varphi_b\in E_{0}^{B} .
\]
For brevity, the bundle $\morf_B(E,\pi,B)$ will be refereed as the
\emph{bundle of restricted morphisms} of $(E,\pi,B)$.
Evidently, the set $E_{0}^{B}$ coincides with the set of
point\ndash restricted fibre\ndash preserving fibre maps of $(E,\pi,B)$.
There is a bijection
\[
\Morf_B(E,\pi,B)  \xrightarrow{\ \chi\ } \Sec\bigl(\morf_B(E,\pi,B)\bigr)
\]
given by
$\chi\colon\varphi\mapsto\chi_\varphi$, $\varphi\in\Morf_B(E,\pi,B)$,
with $\chi_\varphi\colon b\mapsto\chi_\varphi(b):=\varphi|_{\pi^{-1}(b)}$,
$b\in B$. Its inverse is
$\chi^{-1}\colon\sigma\mapsto\chi^{-1}(\sigma)=\varphi$,
$\sigma\in\Sec(\morf_B(E,\pi,B))$, with $\varphi\colon E\to E$ given via
$\varphi|_{\pi^{-1}(b)}=\sigma(b)$ for every $b\in B$.

	If $E$ and $B$ are topological spaces, which is the most widely
considered case, the bundle $(E,\pi,B)$ is called \emph{topological}.  In this
case in the definition of a bundle is included the \emph{bundle property}:
there exists a (topological) space $\mathcal{E}$ such that for each $b\in B$,
there is an open neighborhood (`directory space') $W$ of $b$ in  $B$ and
homeomorphism (`decomposition function')
$\phi_W\colon W\times\mathcal{E}\to\pi^{-1}(W)$ of $W\times\mathcal{E}$ onto
$\pi^{-1}(W)$ satisfying the condition
 $(\pi\circ\phi_W)(w,e)=w$ for $\psi\in W$ and $e\in\mathcal{E}$. Besides, if
the restriction
$\phi_W|_b\colon\{b\}\times\mathcal{E}\to\pi^{-1}(b)$, $b\in B$,
is homeomorphism, the bundle property is called \emph{local triviality},
$\mathcal{E}$ is called \emph{(typical, standard) fibre of the bundle}, and
every fibre $\pi^{-1}(b)$ is homeomorphic to $\mathcal{E}$
for every $b\in B$.

	A \emph{vector bundle} is locally trivial bundle $(E,\pi,B)$ such
that: (i) the fibres $\pi^{-1}(b)$, $b\in B$ and the standard fibre
$\mathcal{E}$ are (linearly) isomorphic vector spaces and (ii) the
decomposition mappings $\phi_W$ and their restrictions $\phi_W|_b$ are
(linear) isomorphisms between vector spaces. The dimension of $\mathcal{E}$,
$\dim\mathcal{E}=\dim\pi^{-1}(b)$ for every $b\in B$, is called dimension of
the vector bundle, resp.\ it is called $\dim\mathcal{E}$\ndash dimensional.
Here the vector spaces are considered over some field, usually the real or
complex numbers; in the context of the present work, the complex case will be
employed.

	When vector bundles are considered, in the definition of a morphism
or $B$-morphism is included the condition the corresponding fibre maps to be
linear. For example, $\varphi\colon E\to E$ is morphism over $B$ of a vector
bundle $(E,\pi,B)$ if $\pi\circ\varphi=\pi$ and the restricted mapping
$\varphi|_{\pi^{-1}(b)}\colon\pi^{-1}(b)\to\pi^{-1}(b)$ is linear for every
$b\in B$.

	\begin{Defn}	\label{Defn3.2}
	A Hilbert (fibre) bundle is a vector bundle whose fibres over the
base are isomorphic Hilbert spaces or, equivalently, whose (standard) fibre
is a Hilbert space.
	\end{Defn}

	In the present investigation we shall show that the Hilbert bundles
can be taken as a natural mathematical framework for a geometrical
formulation of quantum mechanics.

	Some quite general aspects of the Hilbert bundles can be found
in~\cite[chapter~VII]{Lang/Manifolds}. Below we are going to consider only
certain specific properties and structures of the Hilbert bundle theory
required for the present investigation.

	Let $\HilB$ be a Hilbert bundle with bundle space $\bHil$, base
$\base$, projection $\proj$, and (typical) fibre $\Hil$. The fibre over
$x\in\base$ will be often denoted by $\bHil_x$, $\bHil_x:=\pi^{-1}(x)$. Let
$l_x\colon\bHil_x\to\Hil$, $x\in\base$, be the isomorphisms defined by the
restricted decomposition functions, viz., as
$\phi_W|_x\colon\{x\}\times\Hil\to\bHil_x$, we define $l_x$ via
 $\phi_W|_x(x,\psi)=:l_x^{-1}(\psi)\in\pi^{-1}(x)$ for every $\psi\in\Hil$.
We call the maps $l_x$ point\ndash trivializing maps (isomorphisms).

	Let $\langle\cdot | \cdot\rangle\colon\Hil\times\Hil\to\mathbb{R}$ be
the (non\ndash degenerate Hermitian) scalar product in the Hilbert space
$\Hil$ and, respectively, for every $x\in\base$ the map
\(
\langle\cdot | \cdot\rangle_x
	\colon\bHil_x\times\bHil_x\to\mathbb{R}
\)
be the scalar product in the fibre $\bHil_x$ considered as a Hilbert space.%
\footnote{%
The map $x\mapsto\langle\cdot | \cdot\rangle_x$, $x\in\base$, or the
collection of maps
$\{ \langle\cdot | \cdot\rangle_x,\ x\in\base \} $
is called a \emph{fibre metric} on $\HilB$.%
}
For a general Hilbert bundle $\HilB$, the scalar products
 $\langle\cdot | \cdot\rangle_x$, $x\in\base$,
and $\langle\cdot | \cdot\rangle$
are completely independent. Such a situation is unsatisfactory from the
view\nobreakdash point of many applications for which the Hilbert spaces
$\bHil_x$, $x\in\base$ and $\Hil$ is required to be \emph{isometric}. We say
that the vector structure of the Hilbert bundle $\HilB$ is \emph{compatible}
with its metric structure if the (linear) isomorphisms
$l_x\colon\bHil\to\Hil$ preserve the scalar products (are metric\ndash
preserving), \viz iff
\(
\langle\varphi_x | \psi_x\rangle_x
		= \langle l_x(\varphi_x) | l_x(\psi_x)\rangle
\)
for every $\varphi_x,\psi_x\in\bHil_x$. A Hilbert bundle with compatible
vector and metric structure will be called \emph{compatible Hilbert bundle}.
In such a bundle the linear isomorphisms $l_x|x\in\base$ not only
(isomorphicly) connect the vector structures of the fibres $\bHil_x$,
$x\in\base$ and $\Hil$, but they also transform the (Hermitian) metric structure
$\langle\cdot | \cdot\rangle$ from $\Hil$ to $\bHil$ for every $x\in\base$
according to
	\begin{equation}	\label{4.8}
\langle \cdot|\cdot \rangle _x=
\langle l_{x}\cdot|l_{x}\cdot \rangle, \qquad x\in \base
	\end{equation}
and, consequently, from $\bHil_x$ to $\Hil$ through
	\begin{equation}
	\tag{\ref{4.8}$^prime$}	\label{4.8a}
\langle \cdot|\cdot \rangle =
\langle l_{x}^{-1}\cdot|l_{x}^{-1}\cdot \rangle _x, \qquad x\in \base.
	\end{equation}
It is easy to see that the maps
$l_{x\to y}:=l_y^{-1}\circ l_x\colon\proj^{-1}(x)\to\proj^{-1}(y)$
are
(i) fibre maps for fixed $y$,
(ii) linear isomorphisms, and
(iii) isometric, \ie metric preserving in a sense that
	\begin{equation}	\label{4.8new}
\langle l_{x\to y} \cdot | l_{x\to y} \cdot \rangle_y
 = \langle\cdot  | \cdot \rangle_x.
	\end{equation}
Rewording, we can say that  $l_{x\to y}$ are fibre-isometric isomorphisms.
Consequently, all of the fibres over the base and the standard fibre of a
compatible Hilbert bundle are (linearly) isometric and isomorphic Hilbert
spaces.

	Beginning from now on in the present investigation, \emph{only
compatible Hilbert bundles will be employed}. To save writing and for the sake
of brevity, we shall call them simply Hilbert bundles.

	Now some definitions in compatible Hilbert bundles are in order.
Notice, below we present the minimum of material concerning Hilbert bundles
which is absolutely required for formulation of quantum mechanics in terms of
bundles.

	Defining the \emph{Hermitian conjugate} map (operator)
\( \mor{A}_{x}^{\ddag}\colon \Hil\to \bHil_x \) of a map
\( \mor{A}_x\colon  \bHil_x \to \Hil \) by
	\begin{equation}	\label{4.9}
\langle \mor{A}_{x}^{\ddag}\varphi|\chi_x \rangle_x :=
\langle \varphi | \mor{A}_x\chi_x\rangle,
\qquad \varphi\in\Hil,\quad \chi_x\in \bHil_x,
	\end{equation}
we find (see~(\ref{4.8}))
	\begin{equation}	\label{4.10}
\mor{A}_{x}^{\ddag} =
l_{x}^{-1}\circ \left( \mor{A}_x\circ l_{x}^{-1} \right)^\dag
	\end{equation}
where the dagger denotes Hermitian conjugation in $\Hil$ (see~(\ref{2.4})).

	We call a map $\mor{A}_x\colon\bHil_x\to\Hil$ \emph{unitary} if
	\begin{equation}	\label{4.10a}
\mor{A}_{x}^{\ddag} = \mor{A}_{x}^{-1} .
	\end{equation}
Evidently, the isometric isomorphisms $l_x\colon\bHil_x\to\Hil$ are unitary
in this sense:

	\begin{equation}	\label{4.10b}
l_{x}^{\ddag} = l_{x}^{-1}.
	\end{equation}

	Similarly, the \emph{Hermitian conjugate}
map to a map
\(
\mor{A}_{x\to y}\in\{\mor{C}_{x\to y}\colon\bHil_x\to\bHil_y,\ x,y\in\base\}
\)
is a map
\( \mor{A}_{x\to y}^{\ddag}\colon \bHil_x\to \bHil_y \)
defined via
	\begin{equation}	\label{4.11}
\langle \mor{A}_{x\to y}^{\ddag} \Phi_x| \Psi_y \rangle_y :=
\langle \Phi_x| \mor{A}_{y\to x}\Psi_y \rangle_x,
\qquad \Phi_x\in \bHil_x,\quad \Psi_y\in \bHil_y.
	\end{equation}
Its explicit form is
	\begin{equation}	\label{4.12}
\mor{A}_{x\to y}^{\ddag} =
l_{y}^{-1}\circ
\left( l_x\circ \mor{A}_{y\to x}\circ l_{y}^{-1} \right)^\dag
\circ l_x.
	\end{equation}
As $(\ope{A}^\dag)^\dag\equiv \ope{A}$ for any
$\ope{A}\colon \Hil\to\Hil$, we have
	\begin{equation}	\label{4.12a}
\bigl( \mor{A}_{x\to y}^{\ddag} \bigr)^\ddag = \mor{A}_{x\to y} .
	\end{equation}

	 If
\(
\mor{B}_{x\to y}\in\{\mor{C}_{x\to y}\colon\bHil_x\to\bHil_y,\ x,y\in \base\}
\),
then a simple verification shows
	\begin{equation}	\label{4.12b}
\left(\mor{B}_{y\to z}\circ \mor{A}_{x\to y} \right)^\ddag =
\mor{A}_{y\to z}^\ddag\circ\mor{B}_{x\to y}^\ddag, \qquad x,y,z\in \base.
	\end{equation}

	A map $\mor{A}_{x\to y}$ is called \emph{Hermitian} if
	\begin{equation}	\label{4.12c}
\mor{A}_{x\to y}^\ddag = \mor{A}_{x\to y}.
	\end{equation}
A simple calculation proves that the maps
\( l_{x\to y}:=l_y^{-1}\circ l_x \)
are Hermitian.

	A map
\(
\mor{A}_{x\to y}\colon \bHil_x\to\bHil_y
\)
is called \emph{unitary}
if it has a left inverse map and
	\begin{equation}	\label{4.12d}
\mor{A}_{x\to y}^\ddag = \mor{A}_{y\to x}^{-1},
	\end{equation}
where
\(
\mor{A}_{x\to y}^{-1}\colon \bHil_y\to\bHil_x
\)
is the \emph{left} inverse of
\(\mor{A}_{x\to y}\),
i.e.\
\(
\mor{A}_{x\to y}^{-1}\circ \mor{A}_{x\to y} := \id_{\bHil_x}
\).

	A simple verification by means of~\eref{4.11} shows the equivalence
of~\eref{4.12d} with
	\begin{equation}	\tag{\ref{4.12d}$^\prime$}	\label{4.12d'}
\langle\mor{A}_{y\to x}\cdot | \mor{A}_{y\to x}\cdot\rangle _x
   = \langle\cdot | \cdot\rangle _y
   \colon\bHil_y\times\bHil_y\to\mathbb{C},
	\end{equation}
\ie the unitary maps are fibre-metric compatible in a sense that they
preserve the fibre scalar (inner) product. Such maps will be called
\emph{fibre\ndash isometric} or simply \emph{isometric}.

	It is almost evident that the maps $l_{x\to y}=l_{y}^{-1}\circ l_x$
are unitary, that is we have:%
\footnote{%
The Hermiticity and at the same time unitarity of $l_{x\to y}$ is not
incidental as they define a (flat) linear transport (along paths or along the
identity map of $M$) in $\HilB$ (see~(\ref{3.4}), the paragraph
after~(\ref{4.12f}), and footnote~\ref{FlatTransport}).%

}
	\begin{equation}	\label{4.12h}
l_{x\to y}^{\ddag}=l_{x\to y}=l_{y\to x}^{-1},
\qquad
l_{x\to y}:=l_{y}^{-1}\circ l_x  \colon\proj^{-1}(x)\to\proj^{-1}(y) .
	\end{equation}

	Let $\mor{A}$ be a morphism over $\base$ of $\HilB$, i.e.\
$\mor{A}\colon \bHil\to \bHil$ and $\proj\circ \mor{A} = \pi$,
and $\mor{A}_x:=\left.\mor{A}\right|_{\bHil_x}$.
The \emph{Hermitian conjugate}
bundle morphism $\mor{A}^\ddag$ to $\mor{A}$ is defined by (cf.~(\ref{4.11}))
	\begin{equation}	\label{4.morphism}
\langle \mor{A}^\ddag\Phi_x | \Psi_x \rangle_x :=
\langle \Phi_x | \mor{A}\Psi_x \rangle_x,
\qquad \Phi_x,\Psi_x\in \bHil_x.
	\end{equation}
Thus (cf.~(\ref{4.12}))
	\begin{equation}	\label{4.15}
\mor{A}_{x}^{\ddag} := \mor{A}^\ddag\big|_{\bHil_x} =
l_{x}^{-1}\circ\left(l_x\circ \mor{A}_x\circ l_{x}^{-1}\right)^\dag\circ l_x.
	\end{equation}

	A bundle morphism $\mor{A}$ is called
\emph{Hermitian}
if \(\mor{A}_{x}^{\ddag}=\mor{A}_x\) for every $\ x\in \base$, i.e.\  if
	\begin{equation}	\label{4.16}
\mor{A}^\ddag=\mor{A},
	\end{equation}
and it is called
\emph{unitary}
if \(\mor{A}_{x}^{\ddag}=\mor{A}_x^{-1}\) for every $\ x\in \base$, i.e.\  if
	\begin{equation}	\label{4.17}
\mor{A}^\ddag=\mor{A}^{-1}.
	\end{equation}
Using~\eref{4.morphism}, we can establish the equivalence of~\eref{4.17} and
	\begin{equation}	\tag{\ref{4.17}$^\prime$}	\label{4.17'}
\langle\mor{A}\cdot | \mor{A}\cdot\rangle_x = \langle\cdot | \cdot\rangle_x
   \colon\bHil_x\times\bHil_x\to\mathbb{C}.
	\end{equation}
Consequently the unitary morphisms are fibre-metric compatible, \ie they are
\emph{isometric} in a sense that they preserve the fibre Hermitian scalar
(inner) product.

	Starting with Sect.~\ref{V} of the present work,
we will need to deal with the differentiable properties of the employed
Hilbert bundle $\HilB$. To make this possible, we will require the (total)
bundle space $\bHil$ to be (at least) $C^1$ manifold.%
\footnote{%
As the fibres $\bHil_x\subset\bHil$ are, generally, infinitely dimensional,
the dimension of $\bHil$ is generically infinity. On the theory of such
manifolds, see, for instance,~\cite{Lang/Manifolds}.%
}
Besides, we shall need the paths in $\base$ to have a continuous tangent
vectors; in our interpretation of $\base$ as a space\nobreakdash time
model, this corresponds to the existence of velocities of the (point\ndash
like) particles and observers. To ensure this natural requirement, we assume
$\base$ to be a $C^1$ differentiable manifold.%
\footnote{%
In the most applications $\base$ is supposed to be of class $C^2$ or $C^3$
(\eg the Riemannian manifold of general relativity) or even  $C^\infty$ (\eg
the Minkowski space\ndash time of special relativity or the Euclidean space
of classical/quantum mechanics).%
}
Moreover, we shall need the point\ndash trivializing isomorphisms $l_x$ to
have a $C^1$ dependence of $x\in\base$, \ie the mapping
$l\colon\bHil\to\Hil$ given by $l\colon u\mapsto l_{\proj(u)}u$ for
$u\in\bHil$, to be of class $C^1$ as a map between manifolds. A Hilbert
bundle with the last property will be called  $C^1$ bundle (or bundle of
class $C^1$).

	Let us summarize the basic requirements for the bundle $\HilB$ that
will be employed in this work:
(i) it is a compatible Hilbert bundle;
(ii) the bundle space $\bHil$ and the base $\base$ are  $C^1$ differentiable
manifolds;
and (iii) it is of class $C^1$ and the set  $\{l_x,\ x\in \base\}$ of point
trivializing ($C^1$ isometric) isomorphisms is fixed.%
\footnote{\label{FlatTransport}%
the last condition is equivalent on $\HilB$ to be fixed a
path\ndash independent, of class $C^1$, and isometric linear transport along
paths (\emph{vide infra} Subsect.~\ref{III.2}). From such position,
the formalism will be studied elsewhere. Here we notice that the maps
$l_{x\to y}$ (see~\eref{4.12h}) define such a transport: if
$\gamma\colon J\to\base$ and $s,t\in J$, by proposition~\ref{Prop3.1} the map
$l\colon\gamma\mapsto l^\gamma$ with
\(
l^\gamma\colon(s,t)\mapsto l_{\gamma(s)\to\gamma(t)}
	= l_{\gamma(t)}^{-1}\circ l_{\gamma(s)}
\)
is a linear transport in $\HilB$; it is obviously path\ndash independent, of
class $C^1$, and isometric as $l_x$ are such.%
}
The isomorphisms $l_x$ will frequently and explicitly be used throughout the
present work. The formalism of the theory  is not invariant under their choice
but the corresponding transformation formulae are easily derivable and the
physical predictions are independent of them. For instance, if $\{m_x\}$ is
another set of point\ndash trivializing isomorphisms, the scalar products
$\langle \cdot|\cdot \rangle_{x}^{l}$ and
$\langle \cdot|\cdot \rangle_{x}^{m}$
defined respectively by $\{l_x\}$ and $\{m_x\}$ are connected via the equality
\(
\langle \cdot|\cdot \rangle_{x}^{l}
 = \langle \varphi_{l,m}\cdot|\varphi_{l,m}\cdot \rangle_{x}^{m}
\)
where the (fibre\ndash preserving) bundle morphism
 $\varphi_{l,m}\colon\bHil\to\bHil$
is given by
\(
\varphi_{l,m}|_{\bHil_x}
 := \varphi_{l,m;x}
 ;= m_{x}^{-1}\circ l_x
\colon \bHil_x\to\bHil_x.
\)
By means of the morphisms $\varphi_{l,m}=\varphi_{m,l}^{-1}$ the formalism
can be transformed from a particular choice of $\{l_x\}$ to any other one
$\{m_x\}$. Running some steps ahead, we have to say  that the set $\{l_x\}$
can not be fixed on the base of conventional quantum mechanics, its
particular choice is external to it. In this sense $\{l_x\}$ is a free
parameter in the bundle formulation of quantum mechanics. Regardless of this,
as we shall see, the predictions of the resulting theory are independent of
the concrete choice of $\{l_x\}$ and coincide with the ones of conventional
quantum mechanics.

\subsection{Linear transports along paths}
\label{III.2}

	The general theory of linear transports along paths in vector bundles
is developed at length in~\cite{bp-normalF-LTP,bp-LTP-general}. In the
present investigation we shall need only a few definitions and results from
these papers when the bundle considered is a Hilbert one (\emph{vide supra}
definition~\ref{Defn3.2}). To their partial introduction and motivation is
devoted the current section.

	Let $(E,\pi,B)$ be a complex%
\footnote{%
All of our definitions and results hold also for real vector bundles. Most of
them are valid for vector bundles over more general fields too but this is
inessential for the following.%
}
vector bundle (see Subsect.~\ref{III.1} or,
e.g.,~\cite{Greub&et_al.-1,Husemoller}) with bundle
(total) space $E$, base $B$, projection $\pi\colon E\to B$, and isomorphic
fibres $\pi^{-1}(x)\subset E$, $x\in B$.  Let $\mathcal{E}$ be the (standard,
typical) fibre of the bundle, \ie a vector space to which all $\pi^{-1}(x)$,
$x\in B$ are isomorphic. By $J$ and $\gamma\colon J\to B$ we
denote, respectively, a real interval and path in $B$.

	\begin{Defn}	\label{Defn3.1}
	A linear transport along paths in the  bundle $(E,\pi,B)$ is a map $L$
assigning to any path $\gamma\colon J\to B$ a map $L^\gamma$, transport along
$\gamma$, such that $L^\gamma\colon (s,t)\mapsto L^\gamma_{s\to t}$ where the
map
	\begin{equation}				   \label{3.fibre}
L^\gamma_{s\to t} \colon  \pi^{-1}(\gamma(s)) \to \pi^{-1}(\gamma(t))
	\qquad s,t\in J,
	\end{equation}
called transport along $\gamma$ from $s$ to $t$, has the properties:
	\begin{alignat}{2}				    \label{3.1}
L^\gamma_{s\to t}\circ L^\gamma_{r\to s} &=
			L^\gamma_{r\to t},&\qquad  r,s,t&\in J, \\
L^\gamma_{s\to s} &= \id_{\pi^{-1}(\gamma(s))}, & s&\in J,   \label{3.2}
\\
L^\gamma_{s\to t}(\lambda u + \mu v) 			     \label{3.linear}
  &= \lambda L^\gamma_{s\to t}u + \mu L^\gamma_{s\to t}v,
	& \lambda,\mu &\in \mathbb{C},\quad u,v\in{\pi^{-1}(\gamma(s))},
	\end{alignat}
where  $\circ$ denotes composition of maps and
$\id_X$ is the identity map of a set $X$.
	\end{Defn}

	\begin{Rem}
	Equations~\eref{3.1} and~\eref{3.2} mean that $L$ is a
\emph{transport along paths} in the bundle
$(E,\pi,B)$~\cite[definition~2.1]{bp-TP-general}, while~\eref{3.linear}
specifies that it is \emph{linear}~\cite[equation~(2.8)]{bp-TP-general}. In
the present paper only linear transports will be used.
	\end{Rem}

	This definition generalizes the concept of a parallel transport in
the theory of (linear) connections (see~\cite{bp-TP-general,bp-TP-parallelT}
and the references therein for details and comparison).

	A few comments on definition~\ref{Defn3.1} are now in order.
According to equation~\eref{3.fibre}, a linear transport along paths may be
considered as a path-depending connection: it establishes a fibre (isomorphic
- see below) correspondence between the fibres over the path along which it
acts. By virtue of equation~\eref{3.linear} this correspondence is linear.
Such a condition is a natural one when vector bundles are involved, it simply
represents a compatibility condition with the vectorial structure of the
bundle (see~\cite[sect.~2.3]{bp-TP-general} for details). Equation~\eref{3.2}
is a formal realization of our intuitive and na\"{\i}ve understanding that if
we `stand' at some point of a path without `moving' along it, then `nothing'
should happen with the fibre over that point. This property fixes a
0\ndash{ary} operation in the set of (linear) transports along paths,
defining in it the `unit' transport. At last, the equality~\eref{3.1}, which
may be called a group property of the (linear) transports along paths, is a
rigorous expression of the intuitive representation that the `composition'
of two (linear) transports along one and the same path must be a (linear)
transport along the same path.

	In general, different forms of~\eref{3.fibre}--\eref{3.linear} are
well know properties of the parallel transports generated by (linear)
connections (see~\cite{bp-TP-parallelT}). By this reason these transports
turn to be special cases of the general (linear) transport along
paths~\cite[theorem~3.1]{bp-TP-parallelT}. In particular,
	comparing definition~\ref{Defn3.1}
with~\cite[definition~2.1]{bp-LT-Deriv-tensors} and taking into
account~\cite[proposition~4.1]{bp-LT-Deriv-tensors}, we conclude that special
types of linear transports along paths are: the parallel transport assigned to
a linear connection (covariant differentiation) of the tensor algebra of a
manifold~\cite{K&N-1,Schouten/Ricci},
Fermi-Walker transport~\cite{Hawking&Ellis,
Synge}, Fermi transport~\cite{Synge},
Truesdell transport~\cite{Walwadkar,Walwadkar&Virkar},
Jaumann transport~\cite{Radhakrishna&et_al.},
Lie transport~\cite{Hawking&Ellis,Schouten/Ricci},
the modified Fermi\ndash Walker and Frenet\ndash Serret
transports~\cite{Dandoloff&Zakrzewski},
etc.
Consequently definition~\ref{Defn3.1} is general enough to cover a list of
important transports used in theoretical physics and mathematics. Thus
studying  the properties of the linear transports along paths, we can make
corresponding conclusions for any one of the transports mentioned.%
\footnote{%
The concept of linear transport along paths in vector bundles can be
generalized to the transports along paths in arbitrary
bundles~\cite{bp-TP-general} and to transports along maps in
bundles~\cite{bp-TM-general}. An interesting considerations of the concept of
(parallel) `transport' (along closed paths) in connection with homotopy
theory and the classification problem of bundles can be found
in~\cite{Stasheff-PT}. These generalizations will not be used in the
present work.%
}

	From~\eref{3.1} and~\eref{3.2}, we get that
$L^\gamma_{s\to t}$ are invertible and
	\begin{equation}	\label{3.3}
\left(L^\gamma_{s\to t}\right)^{-1} = L^\gamma_{t\to s},
	\qquad s,t\in J.
	\end{equation}
Hence the linear transports along paths are in fact linear isomorphisms
between the fibres over the path along which they act.

	The following two propositions establish the general structure of
the linear transports along paths.

	\begin{Prop}	\label{Prop3.1}
	A map~\eref{3.fibre} is a linear transport along $\gamma$ from $s$ to
$t$ for every $s,t\in J$ if and only if there exist an isomorphic with
$\pi^{-1}(x),\ x\in B$ vector space $V$ and family of linear isomorphisms
$\{F(s;\gamma)\colon \pi^{-1}(\gamma(s))\to V,\ s\in J\}$ such that
	\begin{equation}	\label{3.4}
L_{s\to t}^{\gamma} =
	F^{-1}(t;\gamma) \circ  F(s;\gamma),\qquad s,t\in J.
	\end{equation}
	\end{Prop}

	\begin{Proof}
	If~\eref{3.fibre} is a linear transport along $\gamma$ from $s$ to
$t$, then fixing some $s_0\in J$ and using~\eref{3.2} and~\eref{3.3}, we get
\(
L_{s\to t}^{\gamma} = L_{s_0\to t}^{\gamma} \circ L_{s\to s_0}^{\gamma}
	= \bigl(L_{t\to s_0}^{\gamma}\bigr)^{-1} \circ L_{s\to s_0}^{\gamma}.
\)
So~\eref{3.4} holds for $V=\pi^{-1}(\gamma(s_0))$ and
$F(s;\gamma)=L_{s\to s_0}^{\gamma}$. Conversely, if~\eref{3.4} is valid for
some linear isomorphisms $F(s;\gamma)$, then a straightforward calculation
shows that it converts~\eref{3.1} and~\eref{3.2} into identities
and~\eref{3.linear} holds due to the linearity of $F(s;\gamma)$.
	\end{Proof}

	\begin{Prop}	\label{Prop3.2}
	Let in the vector bundle $(E,\pi,B)$ be given linear transport
along paths with a representation~\eref{3.4} for some vector space $V$
and linear isomorphisms $F(s;\gamma)\colon \pi^{-1}(\gamma(s))\to V,\ s\in J$.
Then for a vector space $\lindex[V]{}{\star}$ there exist linear
isomorphisms
\(
\lindex[\mspace{-2mu}F]{}{\star}(s;\gamma)\colon \pi^{-1}(\gamma(s))\to
	\lindex[V]{}{\star},
\ s\in J
\)
for which
	\begin{equation}	\label{3.6}
L_{s\to t}^{\gamma} =
	\lindex[\mspace{-2mu}F]{}{\star}^{-1}(t;\gamma) \circ
	\lindex[\mspace{-2mu}F]{}{\star}(s;\gamma),\qquad s,t\in J.
	\end{equation}
iff there exists a linear isomorphism
$D(\gamma)\colon V\to\lindex[V]{}{\star}$ such that
	\begin{equation}	\label{3.7}
\lindex[\mspace{-2mu}F]{}{\star}(s;\gamma) = D(\gamma)\circ F(s;\gamma),
				\qquad s\in J.
	\end{equation}
	\end{Prop}

	\begin{Proof}
	If~\eref{3.7} holds, then the substitution of
\(
F(s;\gamma) = D^{-1}(\gamma)\circ \lindex[\mspace{-2mu}F]{}{\star}(s;\gamma)
\)
into~\eref{3.4} results in~\eref{3.6}. Vice versa, if~\eref{3.6} is valid,
then from its comparison with~\eref{3.4} follows that
\(
D(\gamma) = \lindex[\mspace{-2mu}F]{}{\star}(t;\gamma)
				\circ \bigl(F(t;\gamma)\bigr)^{-1}
	  = \lindex[\mspace{-2mu}F]{}{\star}(s;\gamma)
				\circ \bigl(F(s;\gamma)\bigr)^{-1}
\)
is the required (independent of $s,t\in J$) isomorphism.
	\end{Proof}

	Let $(E,\pi,B)$ be a vector bundle whose bundle space  $E$ is a $C^1$
differentiable manifold. A linear transport $L^\gamma$ along
$\gamma\colon J\to B$ is called \emph{differentiable of class} $C^k$,
$k=0,1$, or simply $C^k$ transport, if for arbitrary $s\in J$ and
$u\in\pi^{-1}(\gamma(s))$, the path
$\overline{\gamma}_{s;u} \colon J \to E$ with
 $\overline{\gamma}_{s;u}(t) := L^\gamma_{s\to t} u \in \pi^{-1}(\gamma(t))$,
 $t\in J$, is a $C^k$ mapping in the bundle space $E$.%
\footnote{%
If $E$ is of class $C^r$ with $r=0,1,\ldots,\infty,\omega$, we can define in
an evident way a $C^k$ transport for every $k\le r$.%
}
If a $C^k$ linear transport has a representation~\eref{3.4}, the mapping
$s\mapsto F(s;\gamma)$ is of class $C^k$. So, the transport $L^\gamma$  is of
class $C^k$ iff $L^\gamma_{s\to t}$ has $C^k$ dependence on  $s$ and $t$
simultaneously.
If $\{e_i(\cdot;\gamma)|i=1,\ldots,\dim\pi^{-1}(\gamma(s))\}$
is a $C^k$ frame along $\gamma$,
\ie $\{e_i(s;\gamma)\}$ is a basis in $\pi^{-1}(\gamma(s))$ and the mapping
$s\mapsto e_i(s;\gamma)$ is of class $C^k$ for every $i$,
$L^\gamma$ is of class $C^k$ iff its matrix $\Mat{L}(t,s;\gamma)$
with respect to $\{e_i(s;\gamma)\}$, $s\in J$ has $C^k$ dependence on $s$
and $t$. Here  the elements of $\Mat{L}(t,s;\gamma)$ are defined via the
expansion
	\begin{equation}	\label{3-2.8}
L_{s\to t}^{\gamma} \bigl(e_i(s;\gamma)\bigr)
		=:\Sprindex[L]{i}{j}(t,s;\gamma) e_j(t;\gamma)
		\qquad s,t\in J.
	\end{equation}

	A transport $L$ along paths in $(E,\pi,B)$, $E$ being $C^1$ manifold,
is said to be of class $C^k$, $k=0,1$, if the corresponding transport
$L^\gamma$ along $\gamma$ is of class $C^k$ for every $C^1$ path
$\gamma\colon J\to B$.
Further we consider only $C^1$ linear transports along paths whose matrices
will be referred to smooth frames along paths.

	The above definition and results for linear transports along paths
deal with the general case concerning arbitrary vector bundles and are
therefore insensitive to the dimensionality of the bundle's base or fibres.
Below we point out some peculiarities of the case of a Hilbert bundle whose
fibres are generally infinitely dimensional.

	For linear transports in a Hilbert bundle are valid all results
of~\cite{bp-normalF-LTP,bp-LTP-general, bp-TP-general} with a possible
exception of the ones in which (local) bases in the fibres are involved. The
cause for this is that the dimension of a Hilbert space is (generally)
infinity. So, there arise problems connected with the convergence or
divergence of the corresponding sums or integrals.  Below we try to avoid
these problems and to formulate our assertions and results in an invariant
way.

	Of course, propositions~\ref{Prop3.1} and~\ref{Prop3.2} remain valid
on Hilbert bundles; the only addition is that the vector spaces $V$ and
$\lindex[V]{}{\star}$ are now Hilbert spaces.

	In~\cite[sect.~3]{bp-normalF-LTP} are introduced the so-called
\emph{normal} frames for a linear transport along paths as a (local) field of
bases in which (on some set) the matrix of the transport is unit. Further in
this
 work, in subsection~\ref{VII.2}, 
we shall see that the normal frames realize the Heisenberg picture of motion
in the Hilbert bundle formulation of quantum mechanics.

	Now (see below the paragraph after equation~\eref{4.12f}) we shall
establish a result specific for the Hilbert bundles that has no analogue in
the general theory: a transport along paths is Hermitian if and only if it is
unitary. This assertion is implicitly contained
in~\cite[sect.~3]{bp-BQM-preliminary} (see the paragraph after equation~(3.6)
in it).

	We call a (possibly linear) transport along paths in $\HilB$
\emph{Hermitian} or \emph{unitary} if it satisfies
respectively~(\ref{4.12c}) or~(\ref{4.12d}) in which $x$, and $y$ are
replaced with arbitrary values of the parameter of the transportation
path, i.e.\  if respectively
	\begin{align}	\label{4.12e}
\left( L_{s\to t}^{\gamma} \right)^\ddag &= L_{s\to t}^{\gamma} ,
\qquad s,t\in J,\quad \gamma\colon J\to \base,
\\	\label{4.12f}
\left( L_{s\to t}^{\gamma} \right)^\ddag &=
				\left( L_{t\to s}^{\gamma} \right)^{-1}.
	\end{align}

	A simple corollary from~(\ref{3.3}) is the equivalence
of~(\ref{4.12e}) and ~(\ref{4.12f}); therefore, a
\emph{%
transport along paths in a Hilbert bundle is Hermitian
if and only if it is unitary%
},
i.e.\  these concepts are equivalent. For such transports we say that they are
\emph{consistent} or \emph{compatible} with the Hermitian structure (metric
(inner product)) of the Hilbert bundle~\cite{bp-TP-morphisms}.
Evidently, they are \emph{isometric} fibre maps along the paths they act.
Therefore, a transport along paths in a Hilbert bundle is isometric iff it is
Hermitian or iff it is unitary.%
\footnote{%
The author thanks prof.\ James Stasheff (Math-UNC, Chapel Hill, NC, USA) for
suggesting in July 1998 the term `isometric transport' in the context
given.%
}

\subsection{Liftings of paths, sections, and derivations along paths}
	\label{III.3}

	A \emph{lifting}%
\footnote{%
For detail see, e.g.,~\cite{Sze-Tsen}.%
}
(in a vector bundle $(E,\pi,B)$) of a map $g\colon X\to B$, $X$ being a
set, is a map $\overline{g}\colon X\to E$ such that $\pi\circ\overline{g}=g$;
in particular, the liftings of the identity  map $\id_B$ of the base $B$ are
called \emph{sections} and their set is
$\Sec(E,\pi,B):=\{\sigma|\sigma\colon B\to E,\ \pi\circ\sigma=\id_B\}$.
Let $\Path(A):=\{\gamma|\gamma\colon J\to A\}$ be the
\emph{set of paths} in a set $A$ and
\(
\PLift(E,\pi,B) :=
        \{\lambda|\lambda \colon \Path(B)\to\Path(E),\
        (\pi\circ\lambda)(\gamma) = \gamma
   \text{ for }\gamma\in\Path(B) \}
\)
be the \emph{set of liftings of paths} from $B$ to $E$.%
\footnote{%
Every linear transport $L$ along paths provides a lifting of paths: for every
$\gamma\colon J\to B$ fix some $s\in J$ and $u\in\pi^{-1}(\gamma(s))$, the
mapping $\gamma\mapsto \overline{\gamma}_{s;u}$ with
$\overline{\gamma}_{s;u}(t):=L_{s\to t}^{\gamma}u$, $t\in J$ is a
lifting of paths from $B$ to $E$.%

}
The set $\PLift(E,\pi,B)$
is:
(i) A natural $\mathbb{C}$\ndash vector space if we put
$ (a\lambda+b\mu)\colon\gamma\mapsto a\lambda_\gamma + b\mu_\gamma $
for $a,b\in\mathbb{C}$, $\lambda,\mu\in\PLift(E,\pi,B)$, and
$\gamma\in\Path(B)$,
where, for brevity, we write $\lambda_\gamma$ for $\lambda(\gamma)$,
$\lambda\colon\gamma \mapsto \lambda_\gamma$;
(ii) A natural left module with respect to complex functions on $B$: if
$f,g\colon B\to \mathbb{C}$, we define
 $(f\lambda+g\mu)\colon\gamma\mapsto(f\lambda)_\gamma+(g\mu)_\gamma$
with $(f\lambda)_\gamma(s):=f(\gamma(s))\lambda_\gamma(s)$ for
$\gamma\colon J\to B$ and $s\in J$;
(iii) A left module with respect to the set
\(
\PF(B) := \{ \varphi|\varphi\colon\gamma\mapsto\varphi_\gamma, \
\gamma\colon J\to B,\ \varphi_\gamma\colon J\to \mathbb{C} \}
\)
of \emph{functions along paths} in the base $B$:
for $\varphi,\psi\in\PF(B)$, we set
\(
(\varphi\lambda+\psi\mu) \colon\gamma\mapsto
	(\varphi\lambda)_\gamma + (\psi\mu)_\gamma
\)
where
\(
(\varphi\lambda)_\gamma(s)
  := (\varphi_\gamma\lambda_\gamma)(s)
  := \varphi_\gamma(s)\lambda_\gamma(s).
\)

	The dimension of $\PLift(E,\pi,B)$ as a $\mathbb{C}$-vector space is
infinity but as a left $\PF(B)$\ndash module is equal to the one of
$(E,\pi,B)$ (\ie of its fibres). In the last case a basis in $\PLift(E,\pi,B)$
can be constructed as follows. For every $\gamma\colon J\to B$ and $s\in J$,
choose a basis
$\{e_i(s;\gamma) | i=1,\ldots,\dim\pi^{-1}(\gamma(s))\}$
in $\pi^{-1}(\gamma(s))$; if $E$ is a
$C^1$ manifold, we suppose $e_i(s;\gamma)$ to have a $C^1$ dependence on $s$.
Define $e_i\in\PLift(E,\pi,B)$ by
 $e_i\colon\gamma\mapsto e_i|_\gamma:=e_i(\cdot;\gamma)$, \ie
 $e_i|_\gamma\colon s\mapsto e_i|_\gamma(s):=e_i(s;\gamma)$.
The set $\{e_i\}$ is a basis in $\PLift(E,\pi,B)$, \ie for every
$\lambda\in\PLift(E,\pi,B)$ there are $\lambda^i\in\PF(B)$ such that
$\lambda=\sum_{i}\lambda^ie_i$ and $\{e_i\}$
are $\PF(B)$\ndash linearly independent.
Actually, for $\gamma\colon J\to B$ and $s\in J$, we have
$\lambda_\gamma(s)\in\pi^{-1}(\gamma(s))$, so there exists numbers
$\lambda_\gamma^i(s)\in\mathbb{C}$ such that
$\lambda_\gamma(s)=\sum_{i}\lambda_\gamma^i(s) e_i(s;\gamma)$.
Defining $\lambda^i\in\PF(B)$ by
$\lambda^i\colon\gamma\mapsto\lambda_\gamma^i$ with
$\lambda_\gamma^i\colon s\mapsto\lambda_\gamma^i(s)$,
we get $\lambda=\sum_{i}\lambda^ie_i$;
if $e_i(\cdot;\gamma)$ is of class $C^1$, such are $\lambda_\gamma^i$.
The $\PF(B)$\ndash linear independence of
$\{e_i\}$ is evident corollary of the $\mathbb{C}$\ndash linear independence
of $\{e_i(s;\gamma)\}$. As we notice above, if $E$ is  $C^1$ manifold, we
choose $e_i$, \ie $e_i|_\gamma$, to be $C^1$ and, consequently, the components
$\lambda^i$, \ie $\lambda_\gamma^i$, are of class $C^1$ too.

	Let $(E,\pi,B)$ be a vector bundle whose bundle space $E$ is  $C^1$
manifold.
	A lift $\lambda\in\PLift(E,\pi,B)$ is said to be of class $C^k$,
$k=0,1$, if in some (and hence in any) $C^k$ frame in $\PLift(E,\pi,B)$ its
components are of class $C^k$ along any $C^k$ path, \ie $\lambda$ is of class
$C^k$ if $\lambda_\gamma$ is of class $C^k$ for every $C^k$ path $\gamma$.
Analogously, $\varphi\in\PF(B)$ is of class $C^k$ if $\varphi_\gamma$ is of
class $C^k$ for a $C^k$ path $\gamma$.
	Denote by $\PLift^k(E,\pi,B)$, $k=0,1$, the set of $C^k$ liftings
of paths from $B$ to $E$ and by $\PF^k(B)$, $k=0,1$, the set of $C^k$
functions along paths in $B$. If also the base $B$ is $C^1$ manifold, we
denote by $\Sec^k(E,\pi,B)$ the set of $C^k$ sections of the bundle
$(E,\pi,B)$.

	\begin{Defn}	\label{Defn3.3}
			\index{derivation along paths|defined}
	A derivation along paths in $(E,\pi,B)$ or a derivation of
liftings of paths in $(E,\pi,B)$ is a map
	\begin{subequations}	\label{3-2.19}
	\begin{equation}	\label{3-2.19a}
	D\colon\PLift^1(E,\pi,B) \to \PLift^0(E,\pi,B)
	\end{equation}
	\end{subequations}
which is $\mathbb{C}$-linear,
	\begin{subequations}	\label{3-2.24}
	\begin{equation}	\label{3-2.24a}
D(a\lambda+b\mu) = aD(\lambda) + bD(\mu)
	\end{equation}
	\end{subequations}
for $a,b\in\mathbb{C}$ and $\lambda,\mu\in\PLift^1(E,\pi,B)$,
and the mapping
	\begin{equation}
	\tag{\ref{3-2.19}b}	\label{3-2.19b}
D_{s}^{\gamma}\colon \PLift^1(E,\pi,B) \to \pi^{-1}(\gamma(s)),
	\end{equation}
defined via
\(
D_{s}^{\gamma}(\lambda)
 := \bigl( (D(\lambda))(\gamma) \bigr) (s)
  = (D\lambda)_\gamma(s)
\)
and called derivation along $\gamma\colon J\to B$ at $s\in J$, satisfies the
`Leibnitz rule':
	\begin{equation}
	\tag{\ref{3-2.24}b}	\label{3-2.24b}
D_s^\gamma(f\lambda)
 = \frac{\od f_\gamma(s)}{\od s} \lambda_\gamma(s)
	+ f_\gamma(s) D_s^\gamma(\lambda)
	\end{equation}
for every $f\in \PF^1(B)$. The mapping
	\begin{equation}
	\tag{\ref{3-2.19}c}	\label{3-2.19c}
D^{\gamma}\colon \PLift^1(E,\pi,B) \to \Path\bigl(\pi^{-1}(\gamma(J))\bigr),
	\end{equation}
defined by $D^\gamma(\lambda):=(D(\lambda))|_\gamma=(D\lambda)_\gamma$, is
called derivation along $\gamma$.
	\end{Defn}

	Before continuing with the study of linear transports along
paths, we want to say a few words on the links between sections (along paths)
and liftings of paths.

	The set $\PSec(E,\pi,B)$ of \emph{sections along paths} of $(E,\pi,B)$
consists of mappings
$\bs{\sigma}\colon\gamma\mapsto\bs{\sigma}_\gamma$ assigning
to every path $\gamma\colon J\to B$ a section
$\bs{\sigma}_\gamma\in\Sec\bigl( (E,\pi,B)|_{\gamma(J)} \bigr)$ of the
restricted on $\gamma(J)$ bundle. Every (ordinary) section
$\sigma\in\Sec(E,\pi,B)$ generates a section $\bs{\sigma}$ along paths via
\(
\bs{\sigma}\colon\gamma\mapsto\bs{\sigma}_\gamma:=\sigma|_{\gamma(J)},
\)
 \ie
$\bs{\sigma}_\gamma$ is simply the restriction of $\sigma$ on $\gamma(J)$;
hence $\bs{\sigma}_\alpha=\bs{\sigma}_\gamma$ for every path
$\alpha\colon J_\alpha\to B$ with $\alpha(J_\alpha)=\gamma(J)$. Every
$\bs{\sigma}\in\PSec(E,\pi,B)$ generates a lifting
$\hat{\bs{\sigma}}\in\PLift(E,\pi,B)$ by
\(
\Hat{\bs{\sigma}}\colon \gamma\mapsto \Hat{\bs{\sigma}}_\gamma
 := \bs{\sigma}_\gamma\circ\gamma;
\)
in particular, the lifting $\Hat{\sigma}$ associated to
$\sigma\in\Sec(E,\pi,B)$ is given via
\(
\Hat{\sigma}\colon\gamma\mapsto \Hat{\sigma}_\gamma
 = \sigma|_{\gamma(J)}\circ\gamma.
\)

	Every derivation $D$ along paths generates a map
\[
\overline{D} \colon \PSec^1(E,\pi,B)\to \PLift^0(E,\pi,B),
\]
which may be called a
\emph{derivation of $C^1$ sections along paths}, such that if
 $\bs{\sigma}\in \PSec^1(E,\pi,B)$, then
\(
\overline{D}\colon\bs{\sigma} \mapsto \overline{D}\bs{\sigma}
 = \overline{D}(\bs{\sigma})
\)
where
$\overline{D}\bs{\sigma}\colon\gamma\mapsto \overline{D}^\gamma\bs{\sigma}$
is a lifting of paths paths defined by
\(
\overline{D}^\gamma\bs{\sigma}\colon s\mapsto
(\overline{D}^\gamma\bs{\sigma})(s):=D_s^\gamma \hat{\bs{\sigma}}
\)
with $\hat{\bs{\sigma}}$ being the lifting generated by $\bs{\sigma}$, \ie
$\gamma\mapsto \hat{\bs{\sigma}}_\gamma:= \bs{\sigma}_\gamma\circ \gamma$.
Notice, if $\gamma\colon J\to B$ has self\ndash intersections points and
$x_0\in\gamma(J)$ is such a point, the map $\gamma(J)\to\pi^{-1}(\gamma(J))$
given by
\(
x \mapsto \{ D_s^\gamma(\hat{\bs{\sigma}}) | \gamma(s)=x,\ s\in J \},
\)
 $x\in\gamma(J)$, is generally multiple\ndash valued at $x_0$ and,
consequently it is not a section of $(E,\pi,B)|_{\gamma(J)}$.

	If $B$ is a $C^1$ manifold and for some $\gamma\colon J\to B$
exists a subinterval $J'\subseteq J$ on which the restricted path
$\gamma|J\colon J'\to B$ is without self\ndash intersections, \ie
 $\gamma(s)\not=\gamma(t)$ for $s,t\in J'$ and $s\not=t$, we can define the
\emph{derivation along $\gamma$ of sections} over $\gamma(J')$ as a map
	\begin{equation}	\label{3-2.17}
\mathsf{D}^\gamma\colon \Sec^1\bigl((E,\pi,B)|_{\gamma(J')}\bigr) \to
         \Sec^0\bigl((E,\pi,B)|_{\gamma(J')}\bigr)
	\end{equation}
such that
	\begin{equation}	\label{3-2.17*}
(\mathsf{D}^\gamma\sigma)(x) := D_{s}^{\gamma}\hat{\sigma}
   \qquad \text{for $x=\gamma(s)$}
	\end{equation}
where $s\in J'$ is unique for a given $x$ and
$\hat{\sigma}\in\PLift\bigl((E,\pi,B)|_{\gamma(J')}\bigr)$ is given by
$\hat{\sigma}=\sigma|_{\gamma(J')}\circ\gamma|_{J'}$. Generally the
map~\eref{3-2.17} defined by~\eref{3-2.17*} is multiple\nobreakdash-valued at
the points of self\ndash intersections of $\gamma$, if any, as
\(
(\mathsf{D}^{\gamma}\sigma)(x)
                := \{ D_{s}^{\gamma}\hat{\sigma} : s\in J,\ \gamma(s)=x \} .
\)
The so\ndash defined map $\mathsf{D}\colon\gamma\mapsto\mathsf{D}^\gamma$ is
called \emph{section\ndash derivation%
\index{section-derivation along paths|defined}
along paths}. As we said, it is single\nobreakdash-valued only along paths
without self\ndash intersections.

	Generally a section along paths or lifting of paths does not define a
(single-value) section of the bundle as well as to a lifting along paths
there does not correspond some (single\ndash value) section along paths.
The last case admits one important special exception: if a lifting
$\lambda$ is such that the lifted path $\lambda_\gamma$
is an `exact topological copy' of the
underlying path $\gamma\colon J\to B$, \ie if there exist $s,t\in J$,
$s\not=t$ for which $\gamma(s)=\gamma(t)$, then
$\lambda_\gamma(s)=\lambda_\gamma(t)$. Such a lifting $\lambda$ generates a
section $\overline{\lambda}\in\PSec(E,\pi,B)$ along paths given by
 $\overline{\lambda}\colon\gamma\mapsto\overline{\lambda}_\gamma$ with
 $\overline{\lambda}\colon\gamma(s)\mapsto\lambda_\gamma(s)$. In the general
case, the mapping $\gamma(s)\mapsto\lambda_\gamma(s)$ for a lifting $\lambda$
of paths is multiple\ndash valued at the points of self\ndash intersection of
$\gamma\colon J\to B$, if any; for injective path $\gamma$ this map is a
section of $(E,\pi,B)|_{\gamma(J)}$. Such mappings will be called
\emph{multiple\ndash valued sections along paths}.

	With every derivation $D$ along paths in $(E,\pi,B)$ can be
associated a derivation $\widetilde{D}$ along paths in $\morf_B(E,\pi,B)$.
For this end every lifting $\PLift(\morf(E,\pi,B))$ should be regarded
as a map
	\begin{equation}	\label{3.30}
A \colon \PLift(E,\pi,B) \to \PLift(E,\pi,B)
	\end{equation}
such that, if $\lambda\in \PLift(E,\pi,B)$, $\gamma\colon J\to B$ and
$s\in J$, then
	\begin{equation}	\label{3.31}
A\colon\lambda\mapsto A(\lambda)	
\colon\gamma\mapsto
(A(\lambda))_\gamma = A_\gamma(\lambda_\gamma),\quad
A_\gamma(\lambda_\gamma)\colon s\mapsto A_\gamma(s) (\lambda_\gamma(s)) .
	\end{equation}

	For every derivation $D$ along paths in $(E,\pi,B)$, we define
	\begin{equation}	\label{3.32}
\widetilde{D}\colon \PLift^1(\morf_B(E,\pi,B)) \to \PLift^0(\morf_B(E,\pi,B))
	\end{equation}
by
	\begin{equation}	\label{6.3}
\widetilde{D}\colon A\mapsto \widetilde{D}(A) := D\circ A
	\end{equation}
where $A\in\PLift^1(E,\pi,B)$ is considered as a map~\eref{3.30}. Putting
	\begin{equation}	\label{3.34}
\widetilde{D}^\gamma(A) := D^\gamma\circ A,\quad
\widetilde{D}_s^\gamma(A) := D_s^\gamma\circ A,
	\end{equation}
it is a trivial verification that the map $\widetilde{D}$ is a derivation
along paths in $\morf_B(E,\pi,B)$. The map $\widetilde{D}$ will be called
\emph{induced (from $D$) derivation along paths}.

	\begin{Defn}	\label{3-Defn2.3}
\index{derivation along paths!generated by linear transport|defined}%
\index{linear transport along paths!derivation along paths generated by|defined[\ff{}]}
        The derivation $D$ along paths generated by a $C^1$ linear
transport $L$ along paths is a map of type~\eref{3-2.19a} assigning to every
path $\gamma\colon J\to B$ a map $D^\gamma$, derivation along $\gamma$
generated by $L$, such that
$D^\gamma\colon s \mapsto D_{s}^{\gamma}$, $s\in J$, is a
map~\eref{3-2.19b}, called derivation along $\gamma$ at $s$ assigned to $L$,
given via
	\begin{equation}	\label{3-2.18}
D_{s}^{\gamma}(\lambda)
   := \lim_{\varepsilon\to0}\Bigl\{\frac{1}{\varepsilon}\bigl[
       L_{s+\varepsilon\to s}^{\gamma}\lambda_\gamma(s+\varepsilon)
       - \lambda_\gamma(s)\bigr]\Bigr\}
	\end{equation}
for every lifting $\lambda\in\PLift^1(E,\pi,B)$
with $\lambda\colon\gamma\mapsto\lambda_\gamma$.
	\end{Defn}

	\begin{Rem}	\label{3-Rem2.3}
The operator $D_{s}^{\gamma}$ is an analogue of the covariant differentiation
assigned to a linear connection;
cf., e.g.,~\cite[p.~139, equation~(12)]{Dombrowski}.
	\end{Rem}

	\begin{Rem}	\label{3-Rem2.3new}
	Notice, if $\gamma$ has self-intersections and $x_0\in\gamma(J)$ is
such a point, the mapping $x\mapsto\pi^{-1}(x)$, $x\in\gamma(J)$ given by
$x\mapsto \{D_s^\gamma(\lambda)|\gamma(s)=x,\ s\in J\}$ is, generally,
multiple\ndash valued at $x_0$.
	\end{Rem}

	Let $L$ be a linear transport along paths in $(E,\pi,B)$. For every
path $\gamma\colon J\to B$ choose some $s_0\in J$ and
$u_0\in\pi^{-1}(\gamma(s_0))$. The mapping
	\begin{equation}	\label{3-2.20new}
\overline{L} \colon\gamma\mapsto \overline{L}_{s_0,u_0}^{\gamma} ,
 \quad  \overline{L}_{s_0,u_0}^{\gamma} \colon J\to E,
 \quad  \overline{L}_{s_0,u_0}^{\gamma} \colon t\mapsto
	\overline{L}_{s_0,u_0}^{\gamma}(t) := L_{s_0\to t}^{\gamma} u_0
	\end{equation}
is, evidently, a lifting of paths.

	\begin{Defn}	\label{3-Defn2.4}
The lifting of paths $\overline{L}$ from $B$ to $E$ in $(E,\pi,B)$ defined
via~\eref{3-2.20new} is called lifting (of paths) generated by the (linear)
transport $L$.
	\end{Defn}

	Equations~\eref{3.2} and~\eref{3.4}, combined with~\eref{3-2.18},
immediately imply
	\begin{gather}		\label{3-2.20}
D_{t}^{\gamma} (\overline{L}) \equiv 0, \qquad t\in J,
			\\	\label{3-2.21}
D_s^\gamma(a\lambda + b\mu)
  = a D_s^\gamma\lambda + b D_s^\gamma\mu,
		\qquad	a,b\in\mathbb{C},
		\quad   \lambda,\mu\in\PLift^1(E,\pi,B),
	\end{gather}
where $s_0\in J$ and $u(s)=L_{s_0\to s}^{\gamma}u_0$ are fixed. In other
words, equation~\eref{3-2.20} means that the lifging $\overline{L}$ is
constant along every path $\gamma$ with respect to $D$.

	Let $\{e_i(s;\gamma)\}$ be a field of smooth bases along
$\gamma\colon J\to B,\ s\in J$. Combining the linearity of $L$
with~\eref{3-2.8} and~\eref{3-2.18}, we find the explicit local action of
$D_s^\gamma$:%
\footnote{%
Here and below we suppose the existence of derivatives like
$\od\lambda_\gamma^i(s)/\od s$, viz.\
 $\lambda_\gamma^i\colon J\to\mathbb{C}$
to be a $C^1$ mapping. This, of course, imposes some smoothness conditions on
$\gamma$ which we assume to hold. Evidently, for the purpose $\gamma$ must
be at least continuous. Without going into details, we notice that the most
natural requirement for $\gamma$, when $B$ is a manifold,
is to admit it to be a $C^1$ map.%
}
	\begin{equation}	\label{3-2.22}
D_s^\gamma\lambda
   = \sum_{i}
     \biggl[ \frac{\od \lambda_\gamma^i(s)}{\od s}
     + \Sprindex[\Gamma]{j}{i}(s;\gamma)\lambda_\gamma^j(s)
     \biggr]  e_i(s;\gamma) .
	\end{equation}
Here the \emph{(2-index) coefficients}%
\index{linear transport along paths!coefficients of|defined[\ff{}]}%
\index{linear transport along paths!coefficients!2-index|defined}%
\index{coefficients!of linear transport along paths|defined[\ff{}]}%
\index{coefficients!2-index of linear transport along paths|defined}
$\Sprindex[\Gamma]{j}{i}$ of the
linear transport $L$ are defined by
	\begin{equation}	\label{3-2.23}
\Sprindex[\Gamma]{j}{i}(s;\gamma)
   := \frac{\Sprindex[L]{j}{i}(s,t;\gamma)}{\pd t}\bigg|_{t=s}
   = - \frac{\Sprindex[L]{j}{i}(s,t;\gamma)}{\pd s}\bigg|_{t=s}
	\end{equation}
and, evidently, uniquely determine the generated by $L$ derivation $D$ along
paths.

	A trivial corollary of~\eref{3-2.21} and~\eref{3-2.22} is the
assertion that the derivation along paths generated by a linear transport is
actually a derivation along paths (see definition~\ref{Defn3.3}).

	If the transport's matrix $\Mat{L}$ has a
representation
	\begin{equation}	\label{3-2.14}
\Mat{L}(t,s;\gamma) = \Mat{F}^{-1}(t;\gamma) \Mat{F}(s;\gamma)
	\end{equation}
for some non-degenerate matrix-valued function $\Mat{F}$, which is a
corollary of~\eref{3.4}, from~\eref{3-2.23}, we get
	\begin{equation}	\label{3-2.25}
\Mat{\Gamma}(s;\gamma):=\bigl[ \Sprindex[\Gamma]{j}{i}(s;\gamma) \bigr]
   = \genfrac{.}{|}{}{}{\pd\Mat{L}(s,t;\gamma)}{\pd t}_{t=s}
   = \Mat{F}^{-1}(s;\gamma)\frac{\od \Mat{F}(s;\gamma)}{\od s} .
	\end{equation}
From here and~\eref{3-2.23}, we see that the change
$\{e_i\}\to\{e_{i}^{\prime}=\sum_{j}A_{i}^{j}e_j\}$
of the local bases along $\gamma$
with a nondegenerate $C^1$ matrix $A:=\bigl[A_{i}^{j}\bigr]$ implies
\[
\Mat{\Gamma}(s;\gamma) = \bigl[ \Sprindex[\Gamma]{j}{i}(s;\gamma) \bigr]
	\mapsto  \Mat{\Gamma}^\prime(s;\gamma)
   = \bigl[ \Sprindex[\Gamma]{j}{\prime\,i}(s;\gamma) \bigr]
\]
with
	\begin{equation}	\label{3-2.26}
\Mat{\Gamma}^\prime(s;\gamma)
  = A^{-1}(s;\gamma) \Mat{\Gamma}(s;\gamma) A(s;\gamma)
    + A^{-1}(s;\gamma)\frac{\od A(s;\gamma)}{\od s}.
	\end{equation}

	It is a fundamental result~\cite{bp-LTP-general,bp-normalF-LTP} that
there exists a bijective correspondence between linear transports along paths
and derivations along paths: a linear transport generates derivation
via~\eref{3-2.18} and, \emph{vice versa}, for every derivation along paths
exists a unique transport generating it by~\eref{3-2.18}. Locally this
correspondence is established by the coincidence of the transport's and
derivation's coefficients.%
\footnote{%
The coefficients (components) of derivation $D$ along paths are defined by
$D_s^\gamma e_i = \sum_{j}\Sprindex[\Gamma]{i}{j}(s;\gamma) e_j(s;\gamma)$.%
}

	Every transport $L$ along paths in a vector bundle $(E,\pi,B)$
generates a linear transport $\lindex[\mspace{-2.3mu}L]{}{\circ}$ along
paths in the bundle $\morf_B(E,\pi,B)$ of point\ndash restricted morphisms
over $B$ in $(E,\pi,B)$. If $\gamma\colon J\to B$, explicitly we
have~\cite[equations~(3.9)\Ndash(3.12)]{bp-TP-morphisms}
\(
\lindex[\mspace{-2.3mu}L]{}{\circ}
	\colon\gamma\mapsto \lindex[\mspace{-2.3mu}L]{}{\circ}^\gamma
	\colon (s,t)\mapsto \lindex[\mspace{-2.3mu}L]{}{\circ}^\gamma_{s\to t}
\)
 $s,t\in J$ with
	\begin{multline}	\label{3.35}
\lindex[\mspace{-2.3mu}L]{}{\circ}^\gamma_{s\to t} (\varphi_{\gamma(s)})
 := L_{s\to t}^{\gamma} \circ \varphi_{\gamma(s)} \circ L_{t\to s}^{\gamma}
\\
\in(\pi_{0}^{B})^{-1} (\gamma(t))
	=\{ \psi|\psi\colon\pi^{-1}(\gamma(t)) \to \pi^{-1}(\gamma(t)) \}
	\end{multline}
for every
$\varphi_{\gamma(s)}\colon\pi^{-1}(\gamma(s)) \to \pi^{-1}(\gamma(s))$. The
transport $\lindex[\mspace{-2.3mu}L]{}{\circ}$ will be called
\emph{associated} to $L$ (in $\morf_B(E,\pi,B)$).

	The generated by $\lindex[\mspace{-2.3mu}L]{}{\circ}$ derivation
along paths in $\morf_B(E,\pi,B)$ will be denoted by
$\lindex[\mspace{-2.3mu}D]{}{\circ}$ and called
\emph{derivation associated to the derivation $D$} generated by $L$. So, if
$A\in\PLift^1(\morf_B(E,\pi,B))$, then
	\begin{equation}	\label{3.36}
\lindex[\mspace{-2.3mu}D]{}{\circ}_{s}^{\gamma}(A)
   := \lim_{\varepsilon\to0}\Bigl\{\frac{1}{\varepsilon}\bigl[
       \lindex[\mspace{-2.3mu}L]{}{\circ}_{s+\varepsilon\to s}^{\gamma}
		A_\gamma(s+\varepsilon)
       -  A_\gamma(s)\bigr]\Bigr\} .
	\end{equation}
If the lift $A_\gamma$ of $\gamma\colon J\to B$ in the bundle space
$E_{0}^{B}$ of $\morf_B(E,\pi,B)$ is \emph{linear} and the matrix of
$A_\gamma$ in $\{e_i(s;\gamma)\}$ is $\Mat{A}_\gamma(s)$,  then
from~\eref{3-2.22}\Ndash\eref{3-2.25}, one finds the explicit matrix of
$\lindex[\mspace{-2.3mu}D]{}{\circ}_{s}^{\gamma}A$ as
	\begin{equation}	\label{3.37}
\pmb{\boldsymbol{[}} \lindex[\mspace{-2.3mu}D]{}{\circ}_{s}^{\gamma}A
		\pmb{\boldsymbol{]}}
 = [\Mat{\Gamma}(s;\gamma),\Mat{A}_\gamma(s)]_{\_}
	+ \frac{\od \Mat{A}_\gamma(s)}{\od s}
	\end{equation}
where $[\cdot,\cdot]_{\_}$ means the commutator of matrices and the equality
$\Mat{L}(s,s;\gamma)=\openone$, $\openone$ being the unit matrix, was used
(see~\eref{3-2.14} or~\eref{3.2}).

	Under some assumptions, the matrix of the induced derivative
$(\widetilde{D}_{s}^{\gamma}A)\lambda=D_{s}^{\gamma}(A(\lambda ))$
is (see~\eref{3.34} and~\eref{3-2.22})
	\begin{equation}	\label{3.38}
\pmb{\boldsymbol{[}} (\widetilde{D}_{s}^{\gamma}A)\lambda
		\pmb{\boldsymbol{]}}
 = \frac{\od \Mat{A}_\gamma(s)}{\od s} \Mat{\lambda}_\gamma(s)
  + \Mat{A}_\gamma(s) \frac{\od\Mat{\lambda}_\gamma(s)}{\od s}
  + \Mat{\Gamma}(s;\gamma) \Mat{A}_\gamma(s) \Mat{\lambda}_\gamma(s)
\end{equation}
where $\lambda\in\PLift^1(E,\pi,B)$ is a $C^1$ lifting along paths in
$(E,\pi,B)$ and  $\Mat{\lambda}_\gamma(s)$ is the matrix of
$\lambda_\gamma(s)$ in $\{e_i(s;\gamma)\}$.

	In our investigation the above-presented general definitions and
results will be applied to the particular case of a Hilbert bundle. Since its
dimension is generically infinite, some problems connected with convergence
of sums (which generally are integrals) or decompositions like
$\sum_{i}\lambda^ie_i$ could arise. We shall comment on these problems
 in Sect.~\ref{V} of the present work.

\section[The Hilbert bundle description of quantum mechanics]
	{The Hilbert bundle description \\of quantum mechanics}
\label{new-I}
\setcounter{equation} {0}

	As we shall see in this investigation, the Hilbert bundles provide a
natural mathematical framework for a geometrical formulation of quantum
mechanics. In it all quantum\ndash mechanical quantities, such as
Hamiltonians, observables, wavefunctions, etc., have an adequate description.
For instance, the evolution of a systems is described as an appropriate
(parallel or, more precisely, linear) transport of system's state
liftings of path or sections along paths. We have to emphasize on the fact
that the new bundle formulation of quantum mechanics and the conventional one
are completely equivalent at the present stage of the theory.

\subsection{Brief literature overview}
	\label{new-I.1}

	Several attempts have been made for a (partial) (re)formulation of
non\ndash relativistic quantum mechanics in terms of bundles. Works
containing such material were mentioned in Sect.~\ref{I}. 
Below  are marked only those of them which directly or indirectly lead to
some essential elements of our approach to quantum mechanics.

	It seems for the first time the appropriate bundle approach to
quantum mechanics was developed in~\cite{Asorey&Carinena&Paramio} where the
single Hilbert space of quantum mechanics is replaced with an infinitely many
copies of it forming a bundle space over the 1\ndash dimensional `time'
manifold (\ie over $\mathbb{R}_+$). In this Hilbert fibre bundle the quantum
evolution is (equivalently) described as a kind of `parallel' transport of
appropriate objects over the bundle's base.

	Analogous construction, a Hilbert bundle over the system's phase
space, is used in the Prugove\v{c}ki's approach to quantum theory (see, e.g.
the references in~\cite{Drechsler/Tuckey-96}).

	The gauge, \ie linear connection, structure in quantum mechanics is
first mentioned~\cite{Wilczek-Zee}.
That structure is pointed to be connected with the system's Hamiltonian.
This observation will find natural explanation in our work
(see~\cite{bp-BQM-equations+observables}).

	Some ideas concerning the interpretation of quantum evolution as a
kind of a `parallel' transport in a Hilbert bundle can also be found
in~\cite{Graudenz-94,Reuter}.

\subsection{Motivation}
	\label{new-I.2}

	Below are present some non-exactly rigorous
ideas and statements whose only purpose is the \emph{motivation} for
applying the fibre bundle formalism  to quantum mechanics. Another excellent
arguments and motives confirming this approach are given
in~\cite{Asorey&Carinena&Paramio}.

	Let $\base$ be a differentiable manifold, representing in our
context the space in which the (nonrelativistic) quantum-mechanical
objects `live', i.e.\  the usual 3-dimensional coordinate space
(isomorphic to $\mathbb{R}^3 $ with the corresponding structures).%
\footnote{\label{footnote-base}%
In the following $\base$ can naturally be considered also as the Minkowski
space-time of special relativity. In this case the below-defined
observer's trajectory $\gamma$ is his world line. But we avoid this
interpretation because only the nonrelativistic case is investigated
here. It is important to be noted that mathematically all of what
follows is valid in the case when by $\base$ is understood an arbitrary
differentiable manifold. The physical interpretation
of these cases will be given elsewhere.
}
Let $\gamma\colon J\to \base$, $J$ being an $\mathbb R$-interval, be
the trajectory of
an observer describing the behaviour of a quantum system at any moment
$t\in J$ by a state vector $\Psi_\gamma(t)$ depending on $t$ and, possibly,
on $\gamma$.%
\footnote{%
In this way we introduce the (possible) explicit dependence of the
description of system's state on the concrete observer with respect to which
it is determined.%
}
For a fixed point $x=\gamma(t)\in \base$ the variety of state vectors
describing a quantum system and corresponding to different observers form a
Hilbert space $\bHil_{\gamma(t)}$ which
\emph{%
depends on $\gamma(t)=x$, but not on~$\gamma$ and~$t$ separately}.%
\footnote{
If there exists a global time, as in the nonrelativistic quantum
mechanics, the parameter $t\in J$ can be taken as such. Otherwise by
$t$ we have to understand the local (`proper' or `eigen-') time of a
concrete observer.%
}

	\begin{Rem}	\label{Rem-base}
	As we said above in footnote~\ref{footnote-base}, the next
considerations are completely valid mathematically if $\base$ is an arbitrary
differentiable manifold and $\gamma$ is a path in it. In this sense $\base$
and $\gamma$ are free parameters in our theory and their concrete choice is
subjected only to \emph{physical reasons}, first of all, ones requiring
adequate physical interpretation of the resulting theory.
(The arbitrariness of $\base$ in a similar construction is mentioned
in~\cite[sect.~I]{Reuter} too.) Typical candidates
for $\base$ are: the 3\ndash dimensional Euclidean space $\mathbb{E}^3$,
$\mathbb{R}^3$, the 4\ndash dimensional Minkowski space $M^4$ of special
relativity or the Riemannian space $V_4$ of general relativity, the system's
configuration or phase space, the `time' manifold
$\mathbb{R}_+:=\{a:a\in\mathbb{R},\ a>0\}$, etc.\ Correspondingly, $\gamma$
obtains interpretation as particle's trajectory, its world line, and so on.
The degenerate case when $\base$ consists of a single point corresponds (up
to an isomorphism - see below) to the conventional quantum mechanics.
Throughout this work, we most often take $\base=\mathbb{R}^3$ as a natural
choice corresponding to the non-relativistic case investigated here but, as we
said, this is not required by necessity. Elsewhere we shall see that
$\base=M^4$ or $\base=V_4$ are natural choices in the relativistic region.
An expanded commend on these problems will be given in
 Sect.~\ref{XII} of this work. 
Here we want only to note that the interpretation of $\gamma$ as an
observer's (particle's) trajectory or world line, as accepted in this work,
is reasonable but not necessary one. Maybe more adequate is to interpret
$\gamma$ as a mean (in quantum-mechanical sense) trajectory of some point
particle but this does not change anything in the mathematical structure of
the bundle approach proposed here.
	\end{Rem}

	The spaces $\bHil_{\gamma(t)}$ must be isomorphic as, from physical
view-point, they simply represent the possible variety of state vectors
from different positions.
	In this way over $\base$ arises a natural bundle structure, viz.\ a
\emph{Hilbert bundle} $\HilB$ with a total space $\bHil$, projection
$\proj\colon \bHil\to \base$ and isomorphic fibres $\proj^{-1}(x):=\bHil_x$.
	Since $\bHil_x$, $x\in \base$ are isomorphic, there exists a Hilbert
space $\Hil$ and (linear) isomorphisms
$l_x\colon \bHil_x\to\Hil,\ x\in \base$.
Mathematically $\Hil$ is the typical (standard) fibre of $\HilB$.
       The maps $\Psi_\gamma\colon J\to\proj^{-1}(\gamma(J))$ can be considered
as sections over any part of $\gamma$ without self-intersections
(see below).

	Now a natural question arises: how the quantum evolution in time in
the bundle constructed is described? There are two almost `evident' ways to do
this. On one hand, we can postulate the conventional quantum mechanics in
every fibre $\bHil_x$, \ie the Schr\"odinger equation for the state vector
$\Psi_\gamma(t)\in\bHil_{\gamma(t)}$ with $\bHil_{\gamma(t)}$ being (an
isomorphic copy of) the system's Hilbert space. But the only thing one gets
in this way is an isomorphic image of the usual quantum mechanics in any
fibre over $\base$. Therefore one can not expect some new results or
descriptions in this direction (see below~\eref{4.3b} and the comments after
it). On the other hand, we can demand the ordinary quantum mechanics to be
valid in the fibre $\Hil$ of the bundle $\HilB$. This means to identify
$\Hil$ with the system's Hilbert space of states and to describe the quantum
time evolution of the system via the vector
	\begin{equation}	\label{4.3}
\psi(t) = l_{\gamma(t)} ( {\Psi}_\gamma(t) ) \in \Hil
	\end{equation}
which evolves according to~(\ref{2.1}) or~(\ref{2.5}). This approach is
accepted in the present investigation. What we intend to do further, is, by
using the basic relation~\eref{4.3}, to `transfer' the quantum mechanics from
$\Hil$ to $\HilB$ or, in other words, to investigate the quantum evolution in
terms of the vector $\Psi_\gamma(t)$  connected with $\psi(t)$
via~\eref{4.3}. Since $l_x,\ x\in \base$ are isomorphisms, both descriptions
are \emph{completely equivalent}. This equivalence resolves a psychological
problem that may arise \emph{prima facie}: the single Hilbert space $\Hil$ of
standard quantum theory~\cite{Dirac-PQM, Fock-FQM, Messiah-QM,
Prugovecki-QMinHS, L&L-3} is replaced with a, generally, infinite number
copies $\bHil_x$, $x\in\base$ thereof (cf.~\cite{Asorey&Carinena&Paramio}).
In the present investigation we shall show that the merit one gains from this
is an entirely geometrical reformulation of quantum mechanics in terms of
Hilbert fibre bundles.

	The evolution of a quantum system will be described in a fibre bundle
$\HilB$ with fixed isomorphisms $\{l_x,\ x\in \base\}$ such that
$l_x\colon \bHil_x\to\Hil$, where $\Hil$ is the Hilbert space in which the
system's
evolution is described through the usual Schr\"odinger picture of motion.

	So, in the Schr\"odinger picture a quantum system is described by a
state vector $\psi$ in $\Hil$. Generally~\cite{Neumann-MFQM}
$\psi$ depends (maybe implicitly) on the observer with
respect to which the evolution is studied%
\footnote{%
Usually this dependence is not written explicitly, but it is always presented
as actually $t$ is the time with respect to a given observer.
}
	and it satisfies the Schr\"odinger
equation~(\ref{2.5}). We shall refer to this representation of quantum
mechanics as a
\emph{Hilbert space description}.
In the new
\emph{(Hilbert fibre) bundle description},
which will be studied below, the linear isomorphisms
$l_x\colon \bHil_x=\proj^{-1}(x)\to\Hil,\ x\in \base$ are supposed
arbitrarily fixed%
\footnote{%
The particular choice of $\{l_x\}$  (and, consequently, of the fibres
$\bHil_x$)
is inessential for our investigation.%
}
and the quantum systems are described by a
\emph{state liftings of paths or sections along paths}
$\Psi$ of a bundle $\HilB$ whose typical fibre is the
Hilbert space $\Hil$ (the same Hilbert space
as in the Hilbert space description).

	Generally, to any vector $\varphi\in\Hil$ there corresponds a
unique (global) section $\overline{\Phi}\in\Sec\HilB$ defined via
	\begin{equation}	\label{4.3b}
\overline{\Phi}\colon x\mapsto\overline{\Phi}_x :=
			l_{x}^{-1}(\varphi)\in \bHil_x,
\qquad x\in \base,\quad \varphi\in\Hil.
	\end{equation}
Consequently to a state vector $\psi(t)\in\Hil$ one can assign the (global)
section $\overline{\Psi}(t)$,
\(
\overline{\Psi}(t)\colon x\mapsto\overline{\Psi}_x(t) =
					l_{x}^{-1}(\psi(t))\in \bHil_x
\)
and thus obtaining in $\bHil_x$ for every $x\in \base$ an isomorphic picture
of (the evolution in) $\Hil$. But in this way one can not expect
significantly new results
as the evolution in $\Hil$ is simply replaced with the
(linearly isomorphic to it) evolution in $\bHil_x$ for every arbitrary fixed
$x\in \base$.%
\footnote{%
The machinery of global sections like~\eref{4.3b} is used in~\cite{Reuter}
for the bundle approach to quantum mechanics contained in this paper.%
}
This reflects the fact that the quantum mechanical description is defined up
to linear isomorphism(s) (see note~\ref{Note4.4} below).  Besides, on the
contrary to the bundle description, in this way one looses the explicit
dependence on the observer. So, in it one cannot get really new results with
respect to the Hilbert space description.

\subsection{Basic ideas and statement of the problems}
	\label{new-I.3}

	Taking into account the (more or less heuristic) arguments from the
previous subsection, we pose the following problem: given a Hilbert bundle
$\HilB$ with the properties described in Subsect.~\ref{III.1} and a path
$\gamma\colon J\to\base$, describe the quantum evolution of some quantum
system in this bundle provided the standard fibre $\Hil$ is the system's
Hilbert space of states. For the moment, we identify the bundle's base
$\base$ with the space\ndash time model used: for definiteness we take for it
the 3\ndash dimensional Euclidean space $\mathbb{E}^3$. The path mentioned
will be interpreted as a trajectory of a certain observer, correspondingly
its parameter will be treated as a (global) time.

	At precisely this point some natural questions arise. First of all,
why should one replace the single conventional Hilbert space of the system
with a Hilbert bundle, \ie with (generally) infinite number of different
`local' Hilbert spaces each of which is associated to a single space point?
And, why the introduced reference path is required? These are basic moments
which \emph{a posteriori} will be justified by the results but \emph{a
priori} their essence is in the following. Conventionally, the system's
evolution is described by \emph{different state vectors, one for every
instant of time, in the unique Hilbert space of the system}. These state
vectors \emph{generically depends on the observer} with respect to which the
system is explored. This dependence is often implicit one in quantum
mechanics but it is always presented: the states might be different because,
e.g., the observers could have different velocities, or be rotated relative
to each other. So, different observers assign, generally, different state
vectors to one and the same quantum system at a given moment and
\emph{these vectors belong to the (initial) Hilbert space of the system}. In
this context the shift to a Hilbert bundle pursuits a twofold goal: the
\emph{explicit observer\ndash dependence} of the `state vectors'%
\footnote{%
Here we use inverted commas as, actually, the right term is bundle state
vector, \ie a state lifting or section at some point; \emph{vide infra} in
this section.%
}
and the \emph{split of the time values of the `state vectors' into different
Hilbert spaces}. We achieve this by describing the system's state at a time
$t\in J$ with respect to observer with trajectory $\gamma\colon J\to\base$
with a `state vector' from the `local' Hilbert space attached to the point
$\gamma(t)$, \ie \emph{from the fibre}
$\bHil_{\gamma(s)}:=\proj^{-1}(\gamma(s))$. Besides through $\gamma$, the
observer\ndash dependence of the `state vectors' is introduced, maybe
implicitly, via the Hamiltonian which does not exists \emph{per ce} but is
always given with respect to some concrete observer. Consequently, if we have
two observers with trajectories $\alpha\colon J_\alpha\to\base$ and
$\beta\colon J_\beta\to\base$ with $J_\alpha\cap J_\beta\not=\varnothing$, at
a moment $t\in J_\alpha\cap J_\beta$ they will describe the state of a system
via some vectors
 $\Psi_\alpha\in\bHil_{\alpha(t)}$ and $\Psi_\beta\in\bHil_{\beta(t)}$. In
particular, if it happens that at the moment $t$ the observers are at one and
the same point $x=\alpha(t)=\beta(t)$, the `state vectors' $\Psi_\alpha$ and
$\Psi_\beta$ will be from a single fibre, the one over $x$, \ie the Hilbert
space $\bHil_x=\proj^{-1}(x)$. But generally these vectors will be different
unless the observers are absolutely identical at the moment $t$.%
\footnote{%
For example, if the observers have non-zero relative acceleration at $x$, it
is quite natural that they will assign different `state vectors' to the
system at the moment $t$.%
}

	At the moment it is not clear what one gains from `unwrapping' the
time evolution from the single Hilbert space $\Hil$ to a collection
 $\{\bHil_{\gamma(t)}|t\in J\}$ of `local' Hilbert spaces along the
observer's trajectory $\gamma$. We shall try to explain this in
Subsect~\ref{new-I.4}. In advance, we want only to state the main merit of
the proposed approach: a self\ndash consistent purely geometrical formulation
of (nor\ndash)relativistic quantum mechanics in terms of Hilbert bundles.

	Now it is time for explicit rigorous statement of the basic assertions
and the problems we are going to solve later in this investigation. Notice,
if the opposite is not explicitly stated, we consider only pure quantum states
that conventionally are described by vectors in a Hilbert space.

	\begin{Post}	\label{Post4.1}
	Let there be given a quantum system and $\Hil$ be its Hilbert space
of states. To this system we assign a $C^1$ compatible Hilbert bundle $\HilB$
with bundle space $\bHil$, projection $\proj\colon\bHil\to\base$, and base
$\base$. Besides, we suppose:

{\bfseries\upshape(i)\hphantom{\bfseries\upshape ii}\hspace{0.5em}}%
The base $\base$ and the bundle space $\bHil$ are $C^1$
differentiable manifolds.

{\bfseries\upshape(ii)\hphantom{\bfseries\upshape i}\hspace{0.5em}}%
The point-trivializing (isometric) isomorphisms
$l_x\colon\proj^{-1}(x)\to\Hil$, $x\in M$, are fixed and of class $C^1$.
Their dependence on $x$ is also required to be of class $C^1$, \ie $\HilB$ is
of class $C^1$.

{\bfseries\upshape (iii)\hspace{0.5em}}%
The (standard) fibre of $\HilB$ is the system's Hilbert
space of states $\Hil$ in which the conventional quantum  mechanics is valid.
	\end{Post}

	\begin{Note}	\label{Note4.1}
	It should be emphasize, here we introduce two parameters which are
left free from the quantum mechanics and are external to it: the base $\base$
and the set of isomorphisms $\{l_x\}$. For the sake of physical
interpretation (see remark~\ref{Rem-base}), we identify $\base$ with the
space(\ndash time) model, in particular with the 3\ndash dimensional
Euclidean space $\mathbb{E}^3$ (or the Minkowski 4\ndash dimensional
space\ndash time $M^4$, or the Riemannian 4\ndash manifold  $V_4$ of general
relativity, etc.). This does not influence the basic scheme which is
valid for arbitrary manifold $\base$. What concerns the set $\{l_x\}$, in
the present work  we consider it as given and its analysis and interpretation
will be given elsewhere. In this connection, we want to notice three things:
(i) The arbitrariness in $\{l_x\}$ reflects the natural one in the choice of
the system's Hilbert space of states which is defined up to isomorphism;
(ii) As we shall see, the mathematical formalism depends on the choice of
$\{l_x\}$ but the physically predictable results (the mean values
(mathematical expectations) of the operators) do not;
(iii) In another investigation we intend to show that on the base of the set
$\{l_x\}$ of isomorphisms is very likely to be achieved a kind of unification
of quantum mechanics and gravity.
	\end{Note}

	\begin{Defn}	\label{Defn4.1}
	The bundle $\HilB$ introduce via postulate~\ref{Post4.1} will be
called Hilbert bundle (of states) of the quantum system, or simply system's
Hilbert bundle (of states).
	\end{Defn}

	\begin{Post}	\label{Post4.2}
	Let $J\subseteq\mathbb{R}$ be real interval representing the period
of time in which a quantum system is investigated, $\HilB$ be its Hilbert
bundle, and $\gamma\colon J\to\base$ be a $C^1$ path in the base $\base$. In
$\HilB$ the state of the system at a moment $t\in J$ is described by a map
$\Psi$ assigning to a pair $(\gamma,t)$ a vector
$\Psi_\gamma(t)\in\proj^{-1}(\gamma(t))=\bHil_{\gamma(t)}$ such that
	\begin{equation}	\label{4.3a}
\Psi_\gamma(t) = l_{\gamma(t)}^{-1}(\psi(t))\in \bHil_{\gamma(t)}
	\end{equation}
where $\psi(t)\in\Hil$ is the conventional state vector in the system's
Hilbert space of states ($\equiv$ the bundle's fibre) describing the system's
state at the moment $t$ in the (usual) quantum mechanics.
	\end{Post}

	\begin{Defn}	\label{Defn4.2}
	The description of a quantum system via the map $\Psi$ (resp.\
$\psi$) in the Hilbert bundle $\HilB$ (resp.\ Hilbert space $\Hil$) will be
called Hilbert bundle (resp.\ Hilbert space) description (of the
quantum mechanics of the system).
	\end{Defn}

	\begin{Note}	\label{Note4.2}
	Since  the maps $l_x\colon\bHil\to\Hil$ are isomorphisms, the
description of the quantum states by $\Psi$ and $\psi$ is completely
equivalent.
	\end{Note}

	\begin{Note}	\label{Note4.3}
	As we said above, the path $\gamma$ will be physically interpreted as
a trajectory (or, possibly, world line) of an observer moving in $\base$ and
with respect to which the quantum system is studied (or `who' investigates
it). So, in the bundle description the `state vector' $\Psi_\gamma(t)$,
representing the system state at a moment $t$, explicitly depends on the
observer which is depicted in the index $\gamma$ in $\Psi_\gamma(t)$.  This
is contrary to the conventional quantum mechanics where this dependence is
implicitly assumed almost everywhere. Thus we come to the above\ndash
mentioned situation: different observers describe the system's state at a
fixed moment by vectors from, generally, different fibres of the bundle;
these vectors belong to one and the same fibre over some point in $\base$ iff
the observers happen to be simultaneously in it but, even in this case, the
vectors need not to coincide, they are generically different unless the
observers are absolutely identical.
	\end{Note}

	\begin{Note}	\label{Note4.4}
The bundle, as well as the
conventional, description of quantum mechanics is defined up to a linear
isomorphism(s). In fact, if $\imath\colon \Hil\to\Hil^\prime$,
$\Hil^\prime$ being a Hilbert space, is a linear isomorphism (which may
depend on the time $t$),
then $\psi^\prime(t)=\imath(\psi(t))$ equivalently describes
the evolution of the quantum system in $\Hil^\prime$.
(Note that in this way, for $\Hil^\prime=\Hil$, one can obtain the
known pictures of motion in quantum mechanics --- see~\cite{Messiah-QM}%
 or Sect.~\ref{VII}
.)
In the bundle case
the shift from $\Hil$ to $\Hil^\prime$ is described by the transformation
$l_x\to l_{x}^{\prime}:=\imath\circ l_x$ which reflects the
arbitrariness in the choice of the typical fibre (now $\Hil^\prime$ instead
of $\Hil$) of $\HilB$. There is also arbitrariness in the choice of the
fibres $\bHil_x=\proj^{-1}(x)$
which is of the same character as the one in the case of
$\Hil$, viz.\ if $\imath_x\colon \bHil_x\to \bHil_{x}^{\prime},\ x\in \base$
are linear
isomorphisms, then the fibre bundle $(\bHil^\prime,\proj^\prime,\base)$ with
$\bHil^\prime:=\bigcup_{x\in \base}\bHil_{x}^{\prime},\
\left.\proj^\prime\right|_{\bHil_{x}^{\prime}}:= \proj\circ\imath_{x}^{-1}$,
typical fibre $\Hil$, and isomorphisms
$l^\prime_x := l_x\circ\imath_{x}^{-1}$
can equivalently be used to describe the state of a quantum
system. In the most general case, we have a fibre bundle
$(\bHil^\prime,\proj^\prime,\base)$ with fibres
$\bHil_{x}^{\prime}=\imath_{x}^{-1}(\bHil_x)$, typical fibre
$\Hil^\prime = \imath(\Hil)$,
and isomorphisms
\(
l_x^\prime :=
\imath\circ l_x\circ\imath_{x}^{-1}\colon \bHil_{x}^{\prime}\to\Hil^\prime.
\)
Further we will not be interested in such generalizations. Thus, we shall
suppose that all of the mentioned isomorphisms are fixed.
	\end{Note}

	Let us now look on the mathematical nature of the map $\Psi$
introduce via postulate~\ref{Post4.2}. From one hand, as the notation
suggests, the mapping $\Psi\colon\gamma\mapsto\Psi_\gamma$ with
$\Psi_\gamma\colon t\mapsto\Psi_\gamma(t)$ is a lifting of paths,
 $\Psi\in\PLift\HilB$, which is a trivial corollary of~\eref{4.3a}. On the
other hand, we can consider $\Psi$ as a multiple\ndash valued section along
paths; for this end one has to put
$\Psi\colon\gamma\mapsto\lindex[\Psi]{\gamma}{}$, $\gamma\colon J\to\base$,
with
\(
\lindex[\Psi]{\gamma}{}\colon x\mapsto
	\{ \Psi_\gamma(t)|\gamma(t)=x,\ t\in J \}
\)
for $x\in\gamma(J)$. If one employs multiple\ndash valued sections along
paths, the basic problem is how exactly the values corresponding to some
`time' value $t$ are chosen and how the transition between different
`time' values is depicted;  of course, this problem arises at the points of
self\ndash intersections of $\gamma$, if any. Mathematically the work with
multiple-valued maps is considerably more difficult than the treatment of
single valued ones. The correct regorous treatment of $\Psi$ as a section
requires additional rules describing, besides the correspondence
$\gamma(t)\mapsto\lindex[\Psi]{\gamma}{}(\gamma(t))$, the mapping
$t\mapsto\Psi_\gamma(t)$ which is equivalent to the consideration of $\Psi$
as a lifting of paths. By this reason, in the general case, we shall look on
$\Psi$ as a lifting of paths. There is one important special case when both
approaches to $\Psi$ are transparently equivalent: when only paths $\gamma$
without self\ndash intersections are employed. This is a consequence from the
fact that now the map $\gamma\colon J\to\gamma(J)$ is bijective as
$\gamma\colon J\to\base$ is injective. In particular, if for given $\gamma$
there is a subinterval $J'\subset J$ such that the restricted path
$\gamma|_{J'}$ is injective, the maps $\Psi_{\gamma|J'}$ and
 $\lindex[\Psi]{\gamma|J'}{}$ are completely equivalent representations of
$\Psi$ along $\gamma|J'$.

	The physical preference to interpret $\Psi$ as a section of lifting
depends on the concrete choice of $\base$ and the corresponding
interpretation of $\gamma$. For example, if $\base$ is the space\ndash time
of special or general relativity and $\gamma$ is the world line of  (real
point\ndash like) observer, then $\gamma$ is without self\ndash
intersections and $\Psi$ along $\gamma$ can naturally be interpreted as
section in $\Sec\HilB|_{\gamma(J)}$. On the other hand, if
$\base=\mathbb{E}^3$ is the Euclidean space of classical mechanics and
$\gamma\colon J\to\mathbb{E}^3$ is the trajectory of some point\ndash like
object, treated as an observer, then $\gamma$ could have self\ndash
intersections and, correspondingly, $\Psi$ is more easily treated as lifting
of paths.

	\begin{Defn}	\label{Defn4.3}
	The unique lifting of paths $\Psi$ or (multiple-valued) section along
paths $\Psi$ corresponding to the state vector $\psi$ from conventional
quantum mechanics will be called state lifting (of paths) or state section
(along paths) respectively.
	\end{Defn}

	For brevity, we call, by abuse of the language, a particular value of
$\Psi$, say $\Psi_\gamma(t)$, a \emph{bundle state vector} (at a moment $t$,
or, more precisely, at the (space) point $\gamma(t)$ and at the instant $t$,
\ie at a space\ndash time point $(\gamma(t),t)$ if $\base$ is treated as a
space\ndash time model).

	Since the entering in~\eref{4.3a} mappings $l_x$, $x\in\base$, are
isomorphisms, the correspondences
	\begin{multline}	\label{4.2new}
\text{\scshape state vector} \boldsymbol{\iff}
\text{\scshape state lifting of paths}
\\
\boldsymbol{\iff}
\text{\scshape state section along paths}
	\end{multline}
are bijective (isomorphisms).%
\footnote{%
In~\eref{4.2new} the state sections are, generally, multiple-values sections
along paths.%
}
Hence, the description of a quantum system via state vectors, or liftings of
paths, or sections along paths are equivalent.

	On the base of postulates~\ref{Post4.1} and~\ref{Post4.2}, the
formalism of conventional quantum mechanics concerning solely the wave
function (state vector)  $\psi$ can be transferred, equivalently, in the
Hilbert bundle description in terms  of the state lifting $\Psi$ of paths.
Equation~\eref{4.3a} plays the major role in this reformulation. In this
direction, our first goal is the Hilbert bundle description of the quantum
evolution, \ie the change of the state liftings/sections $\Psi$ in time. In
Sect.~\ref{IV} we shall see that the bundle evolution of a quantum system is
represented by a suitable linear transport along paths in the system's
Hilbert bundle. In
the next, second, part
of the present investigation the corresponding (bundle) equations of motion
will be derived.

	We completely understand that even at this early, introductory,
stage of our work, a lot of concrete problems arise. They are  connected with
the transferring and/or interpretation of particular results of conventional
quantum mechanics in the case of its bundle (re)formulation. These questions
are, as a rule, out of the subject of the present work, devoted to the
general formalism, and have to be considered separately of it. Regardless of
this, we want to pay attention to one such problem which may lead to
methodological difficulties.

	It is well known~\cite{Dirac-PQM,Fock-FQM,Messiah-QM}, in most
situations, the wave function (vector) of a particle is not localized at a
single  space point, but it is spread over some space region that could be
even the entire space, as in the case of momentum eigenstate.
\emph{Prima facie} a superficial conclusion can be made that such a state is
included in the `local' Hilbert space at some point, \ie in the fibre over
it. Such a conclusion is generally entirely wrong (unless we are dealing with
a state localized at a single point or the base $\base$ consists of a single
point)! Suppose $\psi\in\Hil$ is the wave function of some quantum system
with respect to some observer and $\psi(x,t)$ is its value at a space point
$x$ at time $t\in J$. Take the particular choice $\base=\mathbb{R}^3$ and
let $\gamma\colon J\to\mathbb{R}^3$ be the observer's trajectory.%
\footnote{%
Other choices, such as $\base=V_4,M^4,U^4,\ldots$, do not change anything in
the next conclusions. The same concerns the interpretation of $\gamma$.%
}
Since the mappings $l_x$, $x\in\base$ are (isometric) isomorphisms,
from~\eref{4.3a} follows that the bundle state vector $\Psi_\gamma(t)$ is
non\ndash zero if and only if the state vector $\psi(x,t)$ is non\ndash zero.
Let $W_t=\Supp\psi(\cdot,t)\subset\mathbb{R}^3=\base$  be the support of
$\psi(\cdot,t)$, \ie $\psi(x,t)\not=0$ for $x\in W_t$
and
$\psi(x,t)=0$ for $x\not\in W_t$ (if $W_t\not=\base$).
The above said implies $\Psi_\gamma(t)\not=0$ iff $\gamma(t)\in W_t$, in
other words
$\Psi_\gamma(t)\not=0$ iff $\proj\bigl(\Psi_\gamma(t)\bigr)\in W_t$.
Consequently, the non\ndash zero bundle state vectors are spread over the
same region (of space) as the `original' non\ndash zero state vectors.
Besides, the non\ndash zero bundle state vectors are in the `local' Hilbert
spaces attached to the corresponding points in $W_t$, \viz
$\Psi_\gamma(t)\not=0$ is in the fibre
$\proj^{-1}(\gamma(t))\subset\proj^{-1}(W_t)$. In conclusion, the state
liftings of paths are localized, \ie are non\ndash zero, in the same space
region as the conventional wave functions. Analogous result can be obtained
if we take for $\base$ other space(\ndash time) models, such as $V_4$, $M^4$,
etc.

	\begin{Rem}	\label{Rem4.4}
	The above interpretation of the case $\base=\mathbb{R}^3$ (or
$\base=V_4$, etc.) is quite more natural than the one of conventional quantum
mechanics: the non\ndash zero values of the state liftings/sections are
situated in the fibres just above the points at which the wave function is
non\ndash zero, while its values belong to an abstract Hilbert space which
is highly non\ndash local object, associated with the whole space(\ndash time)
rather than with some particular point in it. Mathematically our theory is
valid if $\base$ is arbitrary manifold, but if $\base$ is not a space(\ndash
time) model, the above (and other) `nice' interpretation(s) could be lost. For
example, if $\base$ consists of a single point, $\base=\{x\}$ we have
$\bHil=\bHil_x=l_x^{-1}(\Hil)$ and, according to note~\ref{Note4.4}, we
obtain an isomorphic copy in $\bHil$ of the standard quantum mechanics. Now,
generally, for $\gamma\colon J\to\{x\}$ is hard to be found a `good'
interpretation but, if, e.g., $x$ is in $\mathbb{R}^3$ (or in $V_4$, etc.),
then $\gamma$ can be interpreted as trajectory (world line) of an observer
situated at a space point $x$ during the whole period of `observation'.
	\end{Rem}

	From one hand, as mentioned earlier, the postulate~\ref{Post4.1}
and~\ref{Post4.2} are enough for the bundle reformulation of the state vector
(wave function) formalism. In particular, the probabilistic interpretation
of quantum mechanics is retained: since
	\begin{equation}	\label{4.3new}
\langle \psi(t)| \psi(t)\rangle
 = \langle \Psi_\gamma(t)| \Psi_\gamma(t)\rangle_{\gamma(t)} ,
	\end{equation}
which is a corollary of~\eref{4.8} and~\eref{4.3a}), the bundle state vector
$\Psi_\gamma(t)$
can be interpreted as a probability amplitude. Form other hand, these
postulates do not allow us to transfer in the bundle description the
predictions of quantum mechanics concerning the observables. For this end,
new initial assertions are required. They will be presented in
 Sect.~\ref{VI} 
of this investigation in which the exploration of the observables in the
bundle approach begins. In short, their essence is: in the bundle approach
the observables are described via (Hermitian) liftings of paths or
(multiple\ndash valued) morphisms along paths (in the bundle of restricted
morphisms of the system's Hilbert bundle or in the Hilbert bundle of states
respectively) and their mean values (mathematical expectations) are such that
they coincide with the mean values of the corresponding (Hermitian) operators
representing the same observables in the Hilbert space quantum mechanics. On
the ground of these assertions, the whole machinery of quantum mechanics (of
pure states) can be reformulated in terms of fibre bundles. This will be
done in the next
 sections 
 of our work. Due to the just mentioned coincidence of the mean values of the
operators and liftings corresponding to observables, \emph{the predictions
of Hilbert space and Hilbert bundle quantum mechanics are absolutely
identical, \ie these are different representations of a single theory, the
quantum mechanics}. For the bundle description of mixed states, additional
postulates are required. They will be presented further. As we shall see, in
the bundle approach the mixed states are represented via density liftings
of paths (or multiple\ndash valued density sections along paths) such that
the mean values of the liftings (or sections) corresponding to
observables coincide with the mean values of the corresponding to them
Hermitian operators (in the Hilbert space description) computed by means of
the ordinary density operator (matrix). Consequently, as in the case of pure
states, now we have also a complete coincidence of the predictions of Hilbert
space and Hilbert bundle versions of quantum mechanics.

	Beginning with the next section, following the above lines, the
purpose of this work is the bundle formulation of the general formalism of
quantum mechanics.

\subsection{Preliminary recapitulation}
	\label{new-I.4}

	The summary and discussion of the bundle version of quantum mechanics
will be presented in the concluding part of this work. Below we give a short
abstract of them with the hope that it will help for the better understanding
of our investigation. It also serves as a partial motivation for the present
work.

	The bundle formulation of quantum mechanics is a purely geometrical
version of conventional quantum mechanics to which it is completely
equivalent; hence these are simply different `faces' of a single theory, the
quantum mechanics.
The proposed geometric formulation of quantum mechanics is \emph{dynamical}
in a sense that all geometrical structures employed for the description of
a quantum system depend on and are determined form the dynamical
characteristics of the system.
The new form of the theory has three free parameters:
the bundle's base $\base$,
the set $\{l_x|x\in\base$ of point trivializing isometric isomorphisms,
and the path $\gamma\colon J\to\base$. The choice of these
objects is external to quantum mechanics and is subjected to reasons like the
physical interpretation of the theory and its connection with other physical
theories. As a working hypothesis, we suggest to interpret $\base$ as a
space(\ndash time) model and $\gamma$ as a trajectory (world line) of an
observer along which the quantum evolution is studied.

	In the Hilbert bundle description the system Hilbert space is
replaced with a suitable Hilbert bundle. In it the system state is
represented via appropriate state lifting of paths in the case of pure states
or density liftings of paths if the state is mixed. In
both cases, the quantum evolution in time is characterized by a linear
transport along  $\gamma$ of the state lifting or density liftings in the
system Hilbert bundle or in the bundle of its point\ndash restricted
morphisms over the base respectively. The corresponding equations of motion
are derived. The probabilistic interpretation of quantum theory remains
valid.

	 In the new bundle approach, the observables are describe via
liftings of paths in the bundle of restricted morphisms over the
base in the Hilbert bundle of states. They are so\ndash defined that their
mean values coincide with the mean values of the corresponding Hermitian
operators representing observables in conventional quantum theory. Therefore
the physical predictions of the Hilbert space and Hilbert bundle versions of
quantum mechanics are identical. The bundle equations of motion, governing
the time evolution of observables are derived.

	From bundle's view-point, an observable is integral of motion iff
it is a constant lifting of paths, \viz iff it is linearly transported
in the bundle of restricted morphisms over the base in the Hilbert bundle of
states with respect to the linear transport induced in this bundle by
the evolution transport of the state liftings.

	We also pay attention to the bundle version of the different pictures
of motion. The corresponding equations of motion for the state liftings (or
density liftings) and observables are considered in the bundle pictures of
motion. We point to an interesting result: in terms of local frames, the
bundle Heisenberg picture of motion corresponds to the choice of a suitable
normal frame, \ie a frame in which the matrix of the evolution transport of
the state liftings is unit. Since the normal frames are the mathematical
objects corresponding to the physical concept of an inertial frame, the above
means that the (bundle) Heisenberg picture of quantum mechanics is something
like a `quantum mechanics in a (bundle) inertial frame'.

	At last, we consider problems concerning the role of observers,
physical interpretation, and possible generalizations of bundle quantum
mechanics. In these directions the new form of the theory admits a lot of
developments which is due to the afore mentioned three free parameters in it.
We point that the presented formalism can be transferred in the relativistic
region too.

\section {The (bundle) evolution transport}
\label{IV}
\setcounter{equation} {0}

	The purpose of this section is the Hilbert bundle description of the
quantum evolution of a quantum system. More precisely, we want to find the
time\ndash dependence of the state liftings of paths (or sections along
paths) of a system provided the time\ndash dependence of its (conventional)
wave function (state vector) is known.%
\footnote{%
The corresponding bundle equations of motion will be derived in
the second part
of this investigation.%
}
We shall prove that this is achieved via a suitable linear transport along
paths, called evolution transport, in the system's Hilbert bundle.

	According to postulate~\ref{Post4.1}, assertion~(iii), the evolution
of a system in the fibre $\Hil$ of the system's Hilbert bundle $\HilB$ is
given via the evolution operator $\ope{U}$ (see Sect.~\ref{II}). This
operator has a `transport like' properties, similar
to~\eref{3.1}--\eref{3.linear}. Indeed, using~(\ref{2.1}), we get
\(
\psi(t_3)=\ope{U}(t_3,t_2)\psi(t_2) =
\ope{U}(t_3,t_2) [ \ope{U}(t_2,t_1)\psi(t_1) ],\
\psi(t_3)=\ope{U}(t_3,t_1)\psi(t_1),\ \mathrm{and}\
\psi(t_1)=\ope{U}(t_1,t_1)\psi(t_1)
\)
for every moments $t_1,t_2,t_3$ and arbitrary state vector $\psi$. Hence
	\begin{align}
\ope{U}(t_3,t_1) &= \ope{U}(t_3,t_2)  \circ \ope{U}(t_2,t_1), \label{4.1} \\
\ope{U}(t_1,t_1) &= \id_{\Hil}.			\label{4.2}
\\
\intertext{Besides, by definition,
$\ope{U}(t_2,t_1)\colon \Hil\to\Hil$ is a linear unitary operator, i.e.\ for
\( \lambda_i\in\mathbb{C} \) and
\( \psi_i(t_1)\in\Hil \), $i=1,2$,
we have:}
\ope{U}(t_2,t_1) \biggl(\sum_{i=1,2}^{} \lambda_i\psi_i(t_1)\biggr) &=
\sum_{i=1,2}^{}\lambda_i \ope{U}(t_2,t_1) \psi_i(t_1),	\label{4.2a} \\
\ope{U}^\dag(t_1,t_2) &= \ope{U}^{-1}(t_2,t_1). 		\label{4.2b}
\\
\intertext{From~(\ref{4.1}) and~(\ref{4.2}), evidently, follows}
\ope{U}^{-1}(t_2,t_1) &= \ope{U}(t_1,t_2)		\label{4.2c}
\intertext{and consequently}
\ope{U}^\dag(t_1,t_2) &= \ope{U}(t_1,t_2).		\label{4.2d}
	\end{align}

	If one takes as a primary object the Hamiltonian $\Ham$,
these facts are direct consequences of~(\ref{2.8}).

	Thus the properties of the evolution operator are very similar
to the ones defining a ((flat) Hermitian) linear transport along paths
in a Hilbert bundle. In fact, below we show that the bundle analogue of the
evolution operator is a kind of such transport.

	Along any path $\gamma$, we define the bundle analogue of the
evolution operator $\ope{U}(t,s)\colon \Hil\to\Hil$ as a linear mapping
\(
\mor{U}_\gamma(t,s)\colon \bHil_{\gamma(s)} \to \bHil_{\gamma(t)},\ s,t\in J
\)
such that
	\begin{equation}	\label{4.4}
\Psi_\gamma(t) = \mor{U}_\gamma(t,s) \Psi_\gamma(s)
	\end{equation}
for every instants of time $s,t\in J$.  Hence $\mor{U}_\gamma$ connects the
different time values of the bundle state vectors.
Analogously to~(\ref{4.1}) and~(\ref{4.2}), now we have:
	\begin{align}				\label{4.5}
\mor{U}_\gamma(t_3,t_1) &= \mor{U}_\gamma(t_3,t_2)\circ
\mor{U}_\gamma(t_2,t_1), \qquad t_1,t_2,t_3\in J,	\\ \label{4.6}
\mor{U}_\gamma(t,t) &= \id_{\bHil_{\gamma(t)}}, \qquad t\in J.
	\end{align}

	Comparing~(\ref{4.4}) with~(\ref{2.1}) and using~(\ref{4.3a}),
we find
	\begin{alignat}{2}	\label{4.7}
\mor{U}_\gamma(t,s) &=
		l_{\gamma(t)}^{-1}\circ \ope{U}(t,s) \circ l_{\gamma(s)},
&\qquad s,t&\in J
\\ \intertext{or}
			\label{4.7'}
\ope{U}(t,s)  &=
l_{\gamma(t)} \circ \mor{U}_\gamma(t,s) \circ l_{\gamma(s)}^{-1},
&\qquad s,t&\in J.
	\end{alignat}
This shows the equivalence of the description of evolution
of a quantum systems via $\ope{U}$ and~$\mor{U}_\gamma$.

	A trivial corollary of~(\ref{4.7}) is the \emph{linearity} of
$\mor{U}_\gamma$ and
	\begin{equation}	\label{4.7a}
\mor{U}_\gamma^{-1}(t,s) = \mor{U}_\gamma(s,t) .
	\end{equation}

	As $l_x\colon \bHil_x\to\Hil,\ x\in \base$ are linear isomorphisms,
from~(\ref{4.5})--(\ref{4.7}) follows that
$\mor{U}\colon \gamma\mapsto \mor{U}_\gamma$
with
\(
\mor{U}_\gamma\colon (s,t)\mapsto \mor{U}_\gamma(s,t)
=:\mor{U}_{t\to s}^{\gamma} \colon  \bHil_{\gamma(t)} \to \bHil_{\gamma(s)}
\)
is a \emph{linear transport along paths} in $\HilB$.%
\footnote{%
\label{L-transport:evolution-transport}%
In the context of quantum mechanics it is more natural to define
$\mor{U}_\gamma(s,t)$ from $\bHil_{\gamma(t)}$ into $\bHil_{\gamma(s)}$
instead from
$\bHil_{\gamma(s)}$ into $\bHil_{\gamma(t)}$, as is the map
\(
\mor{U}_{s\to t}^{\gamma}=\mor{U}_\gamma(t,s) \colon
		\bHil_{\gamma(s)} \to \bHil_{\gamma(t)}
\).
The latter notation is better in the general theory of transports along
paths~\cite{bp-normalF-LTP,bp-LTP-general}. Consequently, when applying
results from~\cite{bp-normalF-LTP,bp-LTP-general}, we have to remember that
they are valid for the maps $\mor{U}_{s\to t}^{\gamma}$
(or $\mor{U}^\gamma\colon (s,t)\mapsto \mor{U}_{s\to t}^{\gamma}$).
That is why for the
usage of some results  concerning general linear transports along paths
from~\cite{bp-normalF-LTP,bp-LTP-general}
for $\mor{U}_\gamma(s,t)$ or $\mor{U}_\gamma$ one has to write them for
$\mor{U}_{s\to t}^{\gamma}$ (or $\mor{U}^\gamma$) and then to use the
connection
\(
\mor{U}_{s\to t}^{\gamma} = \mor{U}_\gamma(t,s)=\mor{U}_{\gamma}^{-1}(s,t)
\)
(or $\mor{U}^\gamma=\mor{U}_{\gamma}^{-1}$).
Some results for $\mor{U}_{s\to t}^{\gamma}$ and $\mor{U}_\gamma(s,t)$
coincide but
this is not always the case. In short, the results for linear
transports along paths are transferred to the considered in this work
case by replacing $L_{s\to t}^{\gamma}$ with
$\mor{U}_\gamma(t,s)=\mor{U}_{\gamma}^{-1}(s,t)$.%
}
 This transport is
\emph{Hermitian} (see Sect.~\ref{III}).
	In fact, applying~(\ref{4.12}) to $\mor{U}_\gamma(t,s)$ and
using~(\ref{4.7}), we get
	\begin{equation}	\label{4.13}
\mor{U}_{\gamma}^{\ddag}(t,s) =
l_{\gamma(t)}^{-1} \circ \ope{U}^\dag(s,t) \circ l_{\gamma(t)}.
	\end{equation}
So, using~(\ref{4.2d}), once again~(\ref{4.7}), and~(\ref{4.2c}), we find
	\begin{equation}	\label{4.14}
\mor{U}_{\gamma}^{\ddag}(t,s) = \mor{U}_\gamma(t,s) =
				\mor{U}_{\gamma}^{-1}(s,t).
	\end{equation}

	Hence $\mor{U}_\gamma(t,s)$ is simultaneously Hermitian and unitary
operator, as it should be for any Hermitian or unitary transport along paths
in a Hilbert bundle (see Sect.~\ref{III}). Consequently,  $\mor{U}$ is an
isometric transport along paths.

	Above we defined the transport $\mor{U}$ by~\eref{4.4} from
which~\eref{4.4}--\eref{4.14} follow. It is a simple exercise to prove that
if $\mor{U}$ is defined via~\eref{4.7}, the remaining equations
of~\eref{4.4}--\eref{4.14}  are fulfilled. Consequently,~\eref{4.4}
and~\eref{4.7} are equivalent definitions of the transport $\mor{U}$ along
paths.

	\begin{Defn}	\label{Defn5.1}
	The isometric linear transport $\mor{U}$ along paths, defined
via equation~\eref{4.4} or~\eref{4.7}, in the system's Hilbert bundle $\HilB$
of states is called evolution transport (of the system) or bundle evolution
operator.
	\end{Defn}

	In this way, we see that the evolution transport $\mor{U}$ is a
Hermitian (and hence unitary) linear transport along paths in
$\HilB$. Consequently, to any unitary evolution operator $\ope{U}$
in the Hilbert space $\Hil$ there corresponds a unique isometric linear
transport $\mor{U}$ along paths, the evolution transport, in the Hilbert
bundle $\HilB$ and vice versa.

	Let us summarize. In the Hilbert bundle description, the time
evolution of a quantum system is represented by means of the evolution
transport along paths in the system's Hilbert bundle. It connects the
different time values of the state liftings according to~\eref{4.4} along the
reference path $\gamma$. Equation~\eref{4.7} is the link between the
evolution transport and evolution operator; it is equivalent to~\eref{4.3a}
provided~\eref{4.4} is postulated.

\section {The bundle equations of motion}
\label{V}
\setcounter{equation} {0}

	In conventional quantum mechanics, the time-dependence of the state
vector $\psi\in\Hil$ of a quantum system is governed via the Schr\"odinger
equation~\eref{2.5}. It is natural to expect the existence of an analogous
equation for the state lifting $\Psi$ replacing $\psi$ by~\eref{4.3a} in the
bundle description of quantum mechanics. The derivation of this equation (or
of its variants), which should be only in bundle terms, is the major purpose
of the present section. Regardless of some technical problems, the idea is
quite simple: using~\eref{4.3a} and~\eref{4.4} or~\eref{4.7}, one should
transform the Schr\"odinger equation in `pure' bundle terms. A realization of
such a procedure is given below. The resulting (invariant) bundle equation of
motion has an amazingly transparent geometrical meaning: it expresses the
fact that the state liftings/sections are linearly transported along the
reference path along which the quantum evolution is explored.

\subsection{Derivation of the equations}
	\label{V.1}

	If we substitute~(\ref{4.7'}) into~(\ref{2.5})--(\ref{2.8}),
we `get' the `bundle' analogues of~(\ref{2.5})--(\ref{2.8}).
But they will be wrong! This is due to the fact that they will contain
partial derivatives like
\(
\partial l_{\gamma(t)}/\partial t,\
\partial \Psi_\gamma(t)/\partial t,\) and
\(
\partial \mor{U}_\gamma(t,t_0)/\partial t,
\)
which are not defined at all.
For instance, for in the first case we must have
\(
\partial l_{\gamma(t)}/\partial t =
\lim_{\varepsilon\to0}\left(	\frac{1}{\varepsilon}
( l_{\gamma(t+\varepsilon)} - l_{\gamma(t)} )	\right),
\)
but the `difference' in this limit is not defined (for $\varepsilon\not=0$)
because $l_{\gamma(t+\varepsilon)}$ and $l_{\gamma(t)}$ act on different
spaces, viz.\ on
$\bHil_{\gamma(t+\varepsilon)}$ and $\bHil_{\gamma(t)}$ respectively.
The same is the situation with $\partial \mor{U}_\gamma(t,t_0)/\partial t$.
The most obvious is the contradiction in the following relation
\(
\partial \Psi_\gamma(t)/\partial t =
\lim_{\varepsilon\to0}\left(	\frac{1}{\varepsilon}
( \Psi_\gamma(t+\varepsilon) - \Psi_\gamma(t) )\right),
\)
because $\Psi_\gamma(t+\varepsilon)$ and $\Psi_\gamma(t)$ belong to different
(for $\varepsilon\neq0$) vector spaces.

	One can go through this difficulty by defining, for example,
$\partial \Psi_\gamma(t)/\partial t$ like
$l_{\gamma(t)}^{-1}\partial \psi_\gamma(t)/\partial t$
(cf.~(\ref{4.3})) but this does not lead to some important and new results.

	To overcome this problem, we are going to introduce local bases (or
coordinates) and to work with the matrices of the corresponding
operators and vectors in them.

	Let $\{e_a(x),\quad a\in\Lambda\}$ be a basis in
$\bHil_x=\proj^{-1}(x),\ x\in \base$.
The indices $a,b,c,\ldots\in\Lambda$ may
take discrete, or continuous, or both values. More precisely, the set
$\Lambda$ has a decomposition $\Lambda=\Lambda_d\bigcup\Lambda_c$ where
$\Lambda_d$ is a union of (a finite or countable) subsets of
$\mathbb{N}$ (or, equivalently, of
$\mathbb{Z}$) and $\Lambda_c$ is union of subsets of $\mathbb{R}$ (or,
equivalently, of $\mathbb{C}$). Note that $\Lambda_d$ or  $\Lambda_c$,
but not both, can be empty.
	This is why sums like%
\footnote{%
Here and henceforth in this work, we use the Einstein rule for summation over
indices repeated on different levels.%
}
$\lambda^a e_a(x)$ or $\lambda_a\mu^a$ for
$a\in\Lambda$ and $\lambda^a,\mu_a\in\mathbb{C}$ must be understood as a sum
over the discrete (enumerable) part(s) of $\Lambda$, if any, plus the
(Stieltjes or Lebesgue) integrals over the continuous part(s) of $\Lambda$,
if any. For instance:
\(
\lambda^ae_a(x):=\sum_{a\in\Lambda} \lambda^ae_a(x) :=
\sum_{a\in\Lambda_d} \lambda^ae_a(x) +
\int_{a\in\Lambda_c} \lambda^ae_a(x) da.
\)
	By this reason it is better to write
\(
\int \!\!\!  \!\!\!  \sum_{a\in\Lambda}:=
\sum_{a\in\Lambda_d} + \int_{a\in\Lambda_c} da
\)
	instead of \( \sum_{a\in\Lambda}  \),
but we shall avoid this complicated notation by using the assumed
summation convention on indices repeated on different levels.%
\footnote{%
For details concerning infinite dimensional matrices see, for
instance,~\cite{Neumann-MFQM} and~\cite[chapter~VII, \S~18]{Messiah-QM}.
A comprehensive presentation of the theory of infinite matrices is given
in~\cite{Cooke-infty-mat}; this book is mainly devoted to infinite discrete
matrices but it contains also some results on continuous infinite matrices
related to Hilbert spaces.%
}

	The matrices corresponding to vectors or operators in a given
field of bases will be denoted with the same symbol but in
\textbf{boldface}, for example:
$\mmor{U}_\gamma(t,s) :=
\left[ \left( \mor{U}_\gamma(t,s) \right)_{{\ }b}^{a}  \right]$
and
$\boldsymbol{\Psi}_\gamma(s) :=
\left[ \Psi_\gamma^{a}(s)  \right]$, where
\(
\mor{U}_\gamma(t,s)\left(e_b(\gamma(s))\right) =:
\left( \mor{U}_\gamma(t,s) \right)_{{\ }b}^{a} e_a(\gamma(t))
\)
and
\(
\Psi_\gamma(s) =: \Psi_\gamma^{a}(s) e_a(\gamma(s)).
\)%
\footnote{%
The matrices $\mope{U}(t,s)$ and $\mmor{U}(t,s)$ are closely related to
propagator functions~\cite{Bjorken&Drell-1}, but we will not need these
explicit connections.  For explicit calculations and construction of
$\mope(t,s)$, see~\cite[\S~21, \S22]{Bjorken&Drell-1}%
}

	Analogously, we suppose in $\Hil$ to be fixed a basis
$\{f_a(t),\quad a\in\Lambda\}$ with respect to which we shall use the same
bold-faced matrix notation, for instance:
$\mope{U}(t,s)=\left[ \ope{U}_{{\ }a}^{b}(t,s) \right]$,
\(
\ope{U}(t,s)\left(f_a(s)\right) =:
\left( \ope{U}(t,s) \right)_{{\ }a}^{b} f_b(t)
\),
\(
\boldsymbol{\psi}(t) = \left[ \psi^a(t) \right],\
\psi(t) =: \psi^a(t) f_a(t)
\)
and the `two-point' matrix
\(
\boldsymbol{l}_x(t) = \bigl[ \left(l_x\right)_{{\ }a}^{b}\!(t)  \bigr]
\)
is defined via
\(
l_x\left(e_a(x)\right) =: \left(l_x\right)_{{\ }a}^{b}\!(t) f_b(t).
\)
	Generally $\boldsymbol{l}_x(t)$ depends on $x$ and $t$, but if
$x=\gamma(s)$ for some $s\in J$, we put $t=s$ as from physical reasons
is clear that $\bHil_{\gamma(t)}$ corresponds to $\Hil$ at the `moment'
$t$, i.e.\ the components of $l_{\gamma(s)}$ are with respect to
$\{e_a(\gamma(s))\}$ and
$\{f_a(s)\}$. The same remark concerns `two-point' objects like
$\mor{U}_\gamma(t,s)$
and $\ope{U}(t,s)$
whose components will be taken with respect to pairs of bases like
$( \{e_a(\gamma(t))\} , \{e_a(\gamma(s))\} )$
and
$( \{f_a(t)\} , \{f_a(s)\} )$ respectively.

	Evidently, the equations~(\ref{4.3}), (\ref{4.4})--(\ref{4.7})
remain valid \emph{mutatis mutandis} in the introduced matrix notation:
the kernel letters have to be made bold-faced, the operator composition
(product) must be replaced by matrix multiplication, and the identity
map $\id_{\bHil_x}$ has
to be replaced by the unit matrix
\(
\openone_{\bHil_x} := \left[ \delta_{a}^{b} \right] :=
\left[ \left(\id_{\bHil_x}\right)_{{\ }a}^{b} \right]
\)
of $\bHil_x$ in $\{e_a(x)\}$. Here
$\delta_{a}^{b}=1$ for $a=b$ and
$\delta_{a}^{b}=0$ for $a\neq b$, which means that
\(
e_a(x) = \delta_{a}^{b} e_b(x).
\)
	For instance, using the above definitions, one verifies
that~(\ref{4.7}) is equivalent to
	\begin{equation}	\label{5.01}
\mmor{U}_\gamma(t,s) =
\boldsymbol{l}_{\gamma(t)}^{-1}(t)
\mope{U}(t,s)\boldsymbol{l}_{\gamma(s)}(s).
	\end{equation}

	Let
$\boldsymbol{\Omega}(x):=\left[\Omega_{a}^{{\ }b}(x)\right]$
and
$\boldsymbol{\omega}(t):=\left[\omega_{a}^{{\ }b}(t)\right]$
be nondegenerate matrices. The changes
\[
\{e_a(x)\} \to \{ e_a^\prime (x):=\Omega_{a}^{{\ }b}(x)e_b(x) \} 
,\qquad
\{f_a(t)\} \to \{ e_a^\prime (t):=\omega_{a}^{{\ }b}(t)e_b(t) \}
\]
of the bases in $\bHil_x$ and $\Hil$, respectively, lead to the
transformation of the matrices of the components of $\Phi_x\in \bHil_x$ and
$\phi\in\Hil$, according to
	\begin{equation}	\label{5.00}
\boldsymbol{\Phi}_x\mapsto \boldsymbol{\Phi}_{x}^{\prime} =
\bigl(\boldsymbol{\Omega}^\top(x)\bigr)^{-1}\boldsymbol{\Phi}_x,\quad
\boldsymbol{\phi}\mapsto \boldsymbol{\phi}^{\prime} =
\bigl(\boldsymbol{\omega}^\top(t)\bigr)^{-1}\boldsymbol{\phi}.
	\end{equation}
Here the super script $\top$ means matrix transposition, for example
\(
\boldsymbol{\Omega}^\top(x) :=
\left[ \left({\Omega}^\top(x)\right)_{{\ }b}^{a} \right]
\)
with
\(
\left({\Omega}^\top(x)\right)_{{\ }b}^{a} :=
{\Omega}_{b}^{{\ }a} (x) .
\)
  One easily verifies the transformation
	\begin{equation}	\label{5.02}
\boldsymbol{l}_x(t)\mapsto  \boldsymbol{l}_{x}^{\prime}(t) =
\bigl( \boldsymbol{\omega}^\top(t) \bigr)^{-1}
\boldsymbol{l}_x(t)
\boldsymbol{\Omega}^\top(x)
	\end{equation}
of the components of the linear isomorphisms $l_x\colon \bHil_x\to\Hil$ under
the above changes.

	For any operator $\ope{A}(t)\colon\Hil\to\Hil$ we have
	\begin{equation}	\label{5.02a}
\mope{A}(t)\mapsto\mope{A}^\prime(t)
	= \bigl(\boldsymbol{\omega}^\top(t)\bigr)^{-1}
	  \mope{A}(t)
	  \boldsymbol{\omega}^\top(t).
	\end{equation}

	Analogously, if  $\mor{A}(t)$  is a morphism of $\HilB$, i.e.\ if
$\mor{A}\colon\bHil\to\bHil$ and $\pi\circ\mor{A}=\id_\base$, and
$\mor{A}_x:=\left.\mor{A}(t)\right|_{\bHil_x}$, then
	\begin{equation}	\label{5.02b}
\mmor{A}_x(t)\mapsto\mmor{A}^\prime_x(t)
	= \bigl(\boldsymbol{\Omega}^\top(t)\bigr)^{-1}
	  \mmor{A}_x(t)
	  \boldsymbol{\Omega}^\top(t).
	\end{equation}

	Note that the components of $\ope{U}(t,s)$, when referred to a pair
of bases $\{e_a(t)\}$ and $\{e_a(s)\}$, transform according to
	\begin{equation}	\label{5.03}
\mope{U}(t,s) \mapsto \mope{U}^\prime(t,s) =
\bigl( \boldsymbol{\omega}^\top(t) \bigr)^{-1}
\mope{U}(t,s)
\boldsymbol{\omega}^\top(s).
	\end{equation}
Analogously, the change
\( \{e_a(\gamma(t))\} \to \{ e_a^\prime (t;\gamma) :=
\Omega_{a}^{{\ }b}(t;\gamma)e_b(\gamma(t)) \} \),
with a nondegenerate matrix
$\boldsymbol{\Omega}(t;\gamma):=\left[ \Omega_{a}^{{\ }b}(t;\gamma) \right]$
along $\gamma$, implies
\footnote{%
Cf.~\cite[equation~\eref{+:2.10}]{bp-normalF-LTP}
or~\cite[equation~(4.10)]{bp-LTP-general}, where the notation
$ H(t,s;\gamma)=\mmor{U}_\gamma(s,t;\gamma) $
and
$ \mor{A}(t)=\boldsymbol{\Omega}^\top(t;\gamma)$
is used.%
}
	\begin{equation}	\label{5.04}
\mmor{U}_\gamma(t,s) \mapsto \mmor{U}_\gamma^\prime(t,s) =
\bigl( \boldsymbol{\Omega}^\top(t;\gamma)\bigr)^{-1}
\mmor{U}_\gamma(t,s)
\boldsymbol{\Omega}^\top(s;\gamma).
	\end{equation}

	Substituting $\psi(t)=\psi^a(t)f_a(t)$ into~(\ref{2.5}), we get the
\emph{matrix Schr\"odinger equation}
	\begin{equation}	\label{5.05}
\frac{d\boldsymbol{\psi}(t)}{dt} =
\mHam(t) \boldsymbol{\psi}(t)
	\end{equation}
where
	\begin{equation}	\label{5.06}
\mHam(t) :=
\HamM(t) - \ih \boldsymbol{E}(t)
	\end{equation}
is the \emph{matrix Hamiltonian} (in the Hilbert space description).
	Here $\boldsymbol{E}(t)=\left[E_{a}^{{\ }b}(t)\right]$
determines the expansion of
%
%
${d f_a(t)}/{dt}$ over $\{f_a(t)\}\subset\Hil$, that is
${d f_a(t)}/{dt}=E_{a}^{{\ }b}(t) f_b(t)$; if $f_a(t)$ are independent
of $t$, which is the usual case,
we have $\boldsymbol{E}(t)=0$. In the last case
$\mHam = \HamM$.
It is important to be noted that $\mHam$ is
independent of
$\boldsymbol{E}(t)$. In fact, applying~(\ref{2.7}) to the basic vector
$f_a(t)$, we get
\(
\Ham(t) f_a(t) =
\ih [ (\frac{\partial}{\partial t} \ope{U}(t,t_0) ) f_b(t_0) ]
\ope{U}_{a}^{{\ }b}(t_0,t) =
\ih [ \frac{\partial}{\partial t} ( f_c(t) \ope{U}_{b}^{{\ }c}(t,t_0) ) ]
\ope{U}_{a}^{{\ }b}(t_0,t)
\),
so that
	\begin{equation}	\label{5.07}
\HamM(t) =
\ih \frac{\partial \mope{U}(t,t_0)}{\partial t}
\mope{U}(t_0,t) + \ih \boldsymbol{E}(t)
	\end{equation}
which leads to
	\begin{equation}	\label{5.08}
\mHam(t) =
\ih \frac{\partial \mope{U}(t,t_0)}{\partial t}
\mope{U}(t_0,t).
	\end{equation}

	Substituting the matrix form of~(\ref{4.3}) into~(\ref{5.05}), we
find the \emph{matrix\ndash bundle Schr\"odinger equation}

	\begin{equation}	\label{5.1}
\ih\frac{d \boldsymbol{\Psi}_\gamma(t) }{dt} =
\mbHam_{\gamma}(t) \boldsymbol{\Psi}_\gamma(t)
	\end{equation}
where the
\emph{matrix-bundle Hamiltonian}
is
	\begin{equation}	\label{5.2}
\mbHam_{\gamma}(t)  :=
\boldsymbol{l}_{\gamma(t)}^{-1}(t)
\HamM(t) \boldsymbol{l}_{\gamma(t)}(t)
 -
\ih \boldsymbol{l}_{\gamma(t)}^{-1}(t)
\left(
\frac{d \boldsymbol{l}_{\gamma(t)}(t) }{dt} +
\boldsymbol{E}(t) \boldsymbol{l}_{\gamma(t)}(t)
\right).
	\end{equation}

	Combining~(\ref{5.06}) and~(\ref{5.2}), we find the following
connection between the conventional and bundle matrix Hamiltonians:
	\begin{equation}	\label{5.2'}
\mbHam_{\gamma}(t)  =
\boldsymbol{l}_{\gamma(t)}^{-1}(t)
\HamM^{\mathbf{m}}(t) \boldsymbol{l}_{\gamma(t)}(t)
-
\ih \boldsymbol{l}_{\gamma(t)}^{-1}(t)
\frac{d \boldsymbol{l}_{\gamma(t)}(t)}{dt}.
	\end{equation}

	\begin{rem}{5.1}
	Choosing $e_a(x)=l_x^{-1}(f_a)$ for
$df_a(t)/dt=0$, we get
$\boldsymbol{l}_x(t)=\left[ \delta_{a}^{b} \right]$. Then
$\bHamM_{\gamma}(t) = \HamM(t)$. So, as
$\Ham^\dag=\Ham$, we have
\(
\left(\mbHam_{\gamma}(t)\right)^\dag =
\HamM^\dag(t) =
\HamM(t) = \mbHam_{\gamma}(t)
\)
where we use the dagger ($\dag$) to denote also matrix Hermitian
conjugation.
	Here $\mbHam_{\gamma}(t)$
is a Hermitian matrix in the chosen basis,
but in other bases it may not be such (see below~(\ref{5.12})).
Analogously, choosing $\{f_a(t)\}$ such that $\boldsymbol{E}(t)=0$,
we see that
$\mHam(t)=\HamM(t)$
is a Hermitian matrix, otherwise it may not be such.
	\end{rem}

	\begin{rem}{5.2}
	Note, due to~(\ref{5.2'}), the transition
$\mHam\to\mbHam_{\gamma}$
is very much alike a gauge (or connection)
transformation~\cite{Konopleva&Popov}
(see also below equations~(\ref{5.10})--(\ref{5.12})).
	\end{rem}

	Because of~(\ref{5.1}) and~(\ref{4.4}) there is a bijective
correspondence between $\mmor{U}_\gamma$ and $\mbHam_{\gamma}$
expressed through the initial-value problem (cf.~(\ref{2.6}))
	\begin{equation}	\label{5.3}
\ih \frac{\partial\mmor{U}_\gamma(t,t_0)}{\partial t} =
\mbHam_{\gamma}(t) \mmor{U}_\gamma(t,t_0), \qquad
\mmor{U}_\gamma(t_0,t_0) = \openone_{\bHil_{\gamma(t_0)}},
	\end{equation}
or via the equivalent to it integral equation
	\begin{equation}	\label{5.4}
\mmor{U}_\gamma(t,t_0) = \openone_{\bHil_{\gamma(t_0)}} +
\iih
\int\limits_{t_0}^{t}
\mbHam_{\gamma}(\tau)
\mmor{U}_\gamma(\tau,t_0)
d\tau.
	\end{equation}

	So, if $\mbHam_{\gamma}$ is given, we have
(cf.~(\ref{2.8}))
	\begin{equation}	\label{5.5}
\mmor{U}_\gamma(t,t_0) =
\Texp
\int\limits_{t_0}^{t}
\iih
\mbHam_{\gamma}(\tau)
d\tau
	\end{equation}
and, conversely, if $\mmor{U}_\gamma$ is given, then (cf.~(\ref{2.7})
and~(\ref{5.08}))
\footnote{%
Expressions like
$ ({\partial\ope{U}(t,t_0)}/{\partial t} ) \ope{U}(t_0,t)$,
$( {\partial\mmor{U}_\gamma(t,t_0)}/{\partial t} )
\mmor{U}_\gamma^{-1}(t,t_0)$,
and
$\ope{U}{(t,t_0)}\ope{U}(t_0,t_1)$
are independent of $t_0$
due to~\cite[propositions~\ref{+:Prop2.1}
		and~\ref{+:Prop2.4}]{bp-normalF-LTP}
or~\cite[propositions~2.1 and~2.4]{bp-LTP-general}
(see also~(\ref{3.4}), \eref{3-2.14}, and~\cite[lemma~3.1]{bp-TP-general}).%
}
	\begin{equation}	\label{5.6}
\mbHam_{\gamma}(t) =
\ih
\frac{\partial\mmor{U}_\gamma(t,t_0)}{\partial t}
\mmor{U}_\gamma^{-1}(t,t_0) =
\frac{\partial\mmor{U}_\gamma(t,t_0)}{\partial t}
\mmor{U}_\gamma^{}(t_0,t).
	\end{equation}

	The next step is to write the above matrix equations into an
invariant, i.e.\  basis-independent, form. For this purpose we shall use the
introduce in~\cite{bp-normalF-LTP,bp-LTP-general} derivation along paths
uniquely corresponding to any linear transport along paths in a vector
bundle.

	According to definitions~\ref{Defn3.3} and~\ref{3-Defn2.3} the
derivation along paths corresponding to the bundle evolution transport
$\mor{U}$ is a linear mapping
\[
D\colon \PLift^1\HilB \to \PLift^0\HilB,
\]
$\PLift^k\HilB$ being the set of $C^k$ liftings of paths from $\base$ to
$\bHil$, such that for every $C^1$ lifting $\lambda$ of paths and every path
$\gamma\colon J\to\base$, we have
$D\colon\lambda\mapsto D(\lambda)=D\lambda$
and $D\lambda\colon\gamma\mapsto D^{\gamma}(\lambda)=(D\lambda)_\gamma$ is
defined by
$D^\gamma\lambda\colon s\mapsto D_{s}^{\gamma}\lambda\in\bHil_{\gamma(s)}$
with
	\begin{equation}	\label{5.7}
D_{s}^{\gamma}(\lambda)
   := \lim_{\varepsilon\to0}\Bigl\{\frac{1}{\varepsilon}\bigl[
       U_\gamma(s,s+\varepsilon)\lambda_\gamma(s+\varepsilon)
       - \lambda_\gamma(s)\bigr]\Bigr\}
	\end{equation}
where $\lambda\colon\gamma\mapsto\lambda_\gamma$.

	By~\eref{3-2.22} (see
also~\cite[equation~\eref{+:2.22}]{bp-normalF-LTP}
or~\cite[proposition~4.2]{bp-LTP-general}) the explicit local form
of~(\ref{5.7}) in a frame $\{e_i(\cdot,\gamma)\}$ along $\gamma$ is
	\begin{equation}	\label{5.8}
\mor{D}_{s}^{\gamma}\lambda =
\left(
\frac{d\lambda^a_\gamma(s)}{ds} +
\Gamma_{{\ }b}^{a}(s;\gamma)\lambda^b_\gamma(s)
\right)
e_a(s;\gamma)
	\end{equation}
where the \emph{coefficients}
$\Gamma_{{\ }a}^{b}(s;\gamma)$ of $\mor{U}$ are defined by
(cf.~\eref{3-2.23})
	\begin{equation}	\label{5.9}
\Gamma_{{\ }a}^{b}(s;\gamma) :=
\left.
\frac{\partial\left(\mor{U}_\gamma(s,t)\right)_{{\ }a}^{b}}{\partial t}
\right|_{t=s} =
- \left.
\frac{\partial\left(\mor{U}_\gamma(t,s)\right)_{{\ }a}^{b}}{\partial t}
\right|_{t=s} .
	\end{equation}

	Using~(\ref{4.6}) and~(\ref{5.6}), both for $t_0=t$, we see that
	\begin{equation}	\label{5.10}
\boldsymbol{\Gamma}_\gamma(t) :=
\bigl[ \Gamma_{{\ }a}^{b}(t;\gamma) \bigr] =
- \iih \mbHam_{\gamma}(t)
	\end{equation}
which expresses a fundamental result:
\emph{up to a constant the matrix-bundle Hamiltonian coincides with the
matrix of coefficients of the bundle evolution transport}
(in a given field of bases). Let us recall, using another arguments,
analogous result was obtained in~\cite[sect.~5]{bp-BQM-preliminary}.

	There are two invariant operators corresponding to the Hamiltonian
$\Ham$ in $\Hil$: the evolution transport $\mor{U}$ and the
corresponding to
it derivation along paths $\mor{D}$. Equations~(\ref{5.1})--(\ref{5.10}), as
as well as the general results of~\cite[\S~\ref{+:Sect2}]{bp-normalF-LTP}
and~\cite[\S~4]{bp-LTP-general}, imply that these
three operators, namely $\Ham$, $\mor{U}$, and $\mor{D}$, are equivalent in a
sense that  if one of them is given, then the remaining ones are uniquely
determined.

	\begin{exmp}{5.0}
Let $\{e_a(x)\}$  be fixed by $e_a(x)=l_{x}^{-1}(f_a)$ for $df(t)/dt=0$.
Then
$\mbHam_{\gamma}(t)$ is a Hermitian matrix (see
remark~\ref{rem5.1}). Consequently, in this case,
$\boldsymbol{\Gamma}_\gamma(t)$ is anti-Hermitian, i.e.\
\(
\left(\boldsymbol{\Gamma}_\gamma(t)\right)^\dag =
- \boldsymbol{\Gamma}_\gamma(t).
\)
Note that for other choices of the bases this property may not hold.
	\end{exmp}

	\begin{exmp}{5.1}
	Let $\Ham$ be given and independent of $t$, i.e.\
$\partial \Ham(t)/\partial t =0$, and $\{e_a(x)\}$  be fixed by
$e_a(x)=l_{x}^{-1}(f_a)$ for $df(t)/dt=0$. Then
$\boldsymbol{l}_x(t)=\left[\delta_{a}^{b}\right]$ with
$\delta_{a}^{b}$ defined above.
Equations~(\ref{5.2}) and~(\ref{5.10}) yield
$\mbHam_{\gamma}(t)=\HamM(t)$ and
$\boldsymbol{\Gamma}_\gamma(t)=-\HamM(t)/\ih$.
Finally, now the solution of~(\ref{5.3}) is
\(
\mmor{U}_\gamma(t,t_0) =
\exp \left( \HamM(t)(t-t_0)/\ih \right)
\)
(cf.~(\ref{5.5})).
	\end{exmp}

	According to~\cite[equation~\eref{+:2.26}]{bp-normalF-LTP}
(or~\cite[equation~(4.11)]{bp-LTP-general}) and
footnote~\ref{L-transport:evolution-transport} on
page~\pageref{L-transport:evolution-transport}, if the basis
$\{e_a(t;\gamma)\}$ along $\gamma$ is change to
$\{ e_{a}^{\prime}(t;\gamma) =
\Omega_{a}^{{\ }b}(t;\gamma) e_b(\gamma(t)) \}$
with $\det\boldsymbol{\Omega}(t;\gamma)\not=0$,
\(
\boldsymbol{\Omega}(t;\gamma) :=
\left[\Omega_{a}^{{\ }b}(t;\gamma)\right],
\)
then $\boldsymbol{\Gamma}_\gamma(t)$ transforms into%
\footnote{%
In~\cite{bp-normalF-LTP,bp-LTP-general} the matrix
$\mor{A}(t)=\boldsymbol{\Omega}^\top(t;\gamma)$
instead of $\boldsymbol{\Omega}(t;\gamma)$.%
}
(see~\eref{3-2.26})
	\begin{equation}	\label{5.11}
\boldsymbol{\Gamma}_\gamma^\prime(t) =
(\boldsymbol{\Omega}^{\top}(t;\gamma))^{-1}
\boldsymbol{\Gamma}_\gamma(t)
\boldsymbol{\Omega}^\top(t;\gamma) +
(\boldsymbol{\Omega}^{\top}(t;\gamma))^{-1}
\frac{d \boldsymbol{\Omega}^\top(t;\gamma)}{dt}.
	\end{equation}
This result is also a corollary of~(\ref{5.03}) and~(\ref{5.9}).

	Hence (see~(\ref{5.10})), the matrix-bundle Hamiltonian undergoes
the change
$\mbHam_{\gamma}(t) \mapsto \,^{\prime}\!\mbHam_{\gamma}(t)$
where
	\begin{equation}	\label{5.12}
\,^{\prime}\!\mbHam_{\gamma}(t) =
(\boldsymbol{\Omega}^\top(t;\gamma))^{-1}
\mbHam_{\gamma}(t)
\boldsymbol{\Omega}^\top(t;\gamma)
-\ih
(\boldsymbol{\Omega}^\top(t;\gamma))^{-1}
\frac{d \boldsymbol{\Omega}^\top(t;\gamma)}{dt}.
	\end{equation}
This result can be deduced from~(\ref{5.2'}) too.

	Now we are able to write into an invariant form the matrix-bundle
Schr\"odinger equation~(\ref{5.1}). Substituting~(\ref{5.10}) into~(\ref{5.1})
and using~(\ref{5.8}), we find that~(\ref{5.1}) is equivalent to
	\begin{equation}	\label{5.13}
\mor{D}_{t}^{\gamma}\Psi = 0
	\end{equation}
or, as $t\in J$ is arbitrary, to
	\begin{equation}	\label{5.13new}
\mor{D}^{\gamma}\Psi = 0.
	\end{equation}
Since $\gamma\colon J\to\base$ is arbitrary, the last equation can be
rewritten as
	\begin{equation}	\label{5.24new}
D\Psi = 0.
	\end{equation}
This is the (invariant) \emph{bundle Schr\"odinger equation}
(for the state liftings). Since it coincides with the
\emph{linear transport equation}~\cite[definition~5.2]{bp-LTP-appl}
for the evolution transport,
it has a very simple and fundamental geometrical meaning.
By~\cite[proposition~5.4]{bp-LTP-appl} this
is equivalent to the statement that
\emph{$\Psi_\gamma$ is a (linearly) transported along $\gamma$ lifting}
with respect to the evolution transport (expressed in other terms
via~(\ref{4.4}); see~\cite[definition~2.2]{bp-TP-general}).
Note that~(\ref{5.13}) and~(\ref{4.4}) are compatible
as~\cite[equation~(4.5)]{bp-LTP-general} is fulfilled
(see also~\cite[equation~\eref{+:2.20}]{bp-normalF-LTP}):
\(
\mor{D}_{t}^{\gamma} ( \overline{\mor{U}} ) \equiv0,\ t\in J
\)
where $\overline{\mor{U}}\in\PLift\HilB$  is the lifting of paths generated
by $\mor{U}$ (see definition~\ref{3-Defn2.4}).
	Moreover, if $\mor{D}$ is given (independently of $\mor{U}$, e.g.
through~(\ref{5.8})),
from~\cite[proposition~5.4]{bp-LTP-appl} follows that $\mor{U}$ is the unique
solution of the (invariant) initial-value problem%
\footnote{%
In fact,~\eref{5.14} is the inversion of~(\ref{5.7}) with respect to
$\mor{U}$.%
}
	\begin{equation}	\label{5.14}
\mor{D}_{t}^{\gamma} (\overline{\mor{U}}) = 0
\qquad \overline{\mor{U}}_\gamma(t_0,t_0) = \id_{\bHil_{\gamma(t_0)}}
	\end{equation}
for fixed $t_0\in J$. Since here $t\in J$  and $\gamma\colon J\to\base$ are
arbitrary, the equation in this initial\ndash value problem is equivalent to
	\begin{equation}	\label{5.14new}
\mor{D}^{\gamma} (\overline{\mor{U}}) = 0
	\end{equation}
or to
	\begin{equation}	\label{5.14new-1}
\mor{D} (\overline{\mor{U}}) = 0 .
	\end{equation}
This is the \emph{bundle Schr\"odinger equation for the evolution transport}
$\mor{U}$.

	\begin{Rem}	\label{Rem5.3}
	Mathematically equation~\eref{5.24new} (or~\eref{5.13}) is a trivial
corollary of~\eref{4.4} and~\eref{3-2.20}. But this derivation
of~\eref{5.24new} leaves open the problem for its relation (equivalence) with
the Schr\"odinger's one. Besides, such a `quick' derivation of~\eref{5.24new}
leaves hidden the above\ndash pointed properties of the matrix Hamiltonians,
in particular the fundamental relation~\eref{5.10}.
	\end{Rem}

\subsection{Inferences}
	\label{V.2}

	Thus we see that there are two equivalent ways for
describing the unitary evolution of a quantum system:
(i) by means of the evolution transport $\ope{U}$ (see~(\ref{2.1})) or by the
Hermitian Hamiltonian $\Ham$ (see~(\ref{2.5})) in the Hilbert space $\Hil$
(which is the typical fibre in the bundle description) and
(ii) via the evolution transport $\mor{U}$ (see~(\ref{4.4})),
which is a Hermitian (and unitary) transport along paths, or the derivation
$\mor{D}$ along paths (see~(\ref{5.13})) in the Hilbert bundle $\HilB$.
In the bundle description $\mor{U}$ corresponds to $\ope{U}$
(see~(\ref{4.7})) and $\mor{D}$ to $\Ham$ (see~(\ref{5.8}) and~(\ref{5.10})).

	We derived the bundle Schr\"odinger equation~\eref{5.24new} from the
`classical' Schr\"odinger equation~\eref{2.5}; the equivalence of the two
equations is evident from the above considerations.

	Now we have at our disposal all tools required for pure bundle
description of the evolution of a quantum system.

	Given a system characterized by a derivation $D$ along paths. If the
system's bundle state victor $\Psi_{\gamma}^{0}$ is know along
$\gamma\colon J\to\base$ at a point $t_0\in J$, the state lifting $\Psi$
of paths is a solution of the bundle Schr\"odinger equation~\eref{5.13}
under the initial condition
	\begin{equation}	\label{5.24new-1}
\Psi_\gamma(t_0) = \Psi_{\gamma}^{0} .
	\end{equation}
By virtue of~\eref{5.8}, equation~\eref{5.13} and the
condition~\eref{5.24new-1} from a standard initial\ndash value problem for a
first order system of ordinary differential equations (with respect to the
time $t$) which has solutions along  $\gamma$~\cite{Hartman}.%
\footnote{%
This initial-value problem is analogous (and equivalent) to the one for the
Schr\"odinger equation~\eref{2.5} and condition~\eref{2.5a}.%
}
This solution is
\[
\Psi_\gamma(t) = \mor{U}_\gamma(t,t_0)\Psi_\gamma^0
\]
where the evolution transport $\mor{U}$ could be found as the unique solution
of the initial\ndash value problem~\eref{5.14}

	Above we supposed the system to be described via a derivation $D$
along paths instead by a Hamiltonian $\Ham$. These are equivalent approaches.
Actually, in a local field of bases along $\gamma$, the matrix of $\Ham$ and
the one of the coefficients of $D$ are connected by~\eref{5.10}
and~\eref{5.2} and, hence, can uniquely be expressed through each other.
Consequently, if the Hamiltonian $\Ham$ is known, one can construct from it
the derivation $D$ and \emph{vice versa}. In the next section we shall see
that to the Hamiltonian $\Ham$, as an observable, in the bundle description
corresponds, besides $D$, a suitable lifting $\mor{H}$ of paths or
(multiple\ndash valued) section along paths, the bundle Hamiltonian.

	Now we shall derive a new form of the bundle Schr\"odinger equation
in terms of the derivation $\widetilde{D}$ along paths in $\morf_\base\HilB$
induced by the derivation $\mor{D}$ along paths generated by the evolution
transport $\mor{U}$.%
\footnote{%
For the notation and corresponding definitions, see Subsect.~\ref{III.3}, in
particular, equations~\eref{3.30}\Ndash\eref{3.34}.%
}

	Applying equation~\eref{5.8}, we can find the explicit (matrix of
the) action of
$\widetilde{D}_t^\gamma(C):=D_t^\gamma\circ C$,
$C\in \PLift^1(\morf_\base\HilB)$, on a state lifting $\Psi$ provided the
lift  $C_\gamma$ is \emph{linear}.

	Let
\(
\pmb{\boldsymbol{[}} X \pmb{\boldsymbol{]}}
\)
be the matrix of a vector or an operator $X$ in $\{e_a\}$.
	Due to~(\ref{5.8}), we have
\[
	\pmb{\boldsymbol{[}}
( \tilde{\mor{D}}_{t}^{\gamma}(C)) \Psi
	\pmb{\boldsymbol{]}}
=
\left(
\frac{d}{dt} \boldsymbol{C}_\gamma(t)
\right)
\boldsymbol{\Psi}_\gamma(t)
+
\boldsymbol{C}_\gamma(t)
\left( \frac{d}{dt} \boldsymbol{\Psi}_\gamma(t) \right) +
\boldsymbol{\Gamma}_\gamma(t)
\boldsymbol{C}_\gamma(t)
\boldsymbol{\Psi}_\gamma(t),
\]
which is a special case of~\eref{3.38}.
	Substituting here \( \frac{d}{dt} \boldsymbol{\Psi}_\gamma(t) \)
from~(\ref{5.1}) and using~(\ref{5.10}), we obtain the matrix equation
	\begin{equation}	\label{6.4}
	\pmb{\boldsymbol{[}}
( \tilde{\mor{D}}_{t}^{\gamma}(C) ) \Psi
	\pmb{\boldsymbol{]}}
=
\left( \frac{d}{dt} \boldsymbol{C}_\gamma (t) \right)
	\boldsymbol{\Psi}_\gamma(t)
+
\left[  \boldsymbol{\Gamma}_\gamma(t),\boldsymbol{C}_\gamma (t) \right]_{\_}
	\boldsymbol{\Psi}_\gamma(t),
	\end{equation}
where $[\cdot,\cdot]_{\_}$ denotes the commutator
of matrices, or
	\begin{equation}	\label{6.5}
\pmb{\boldsymbol{[}} \tilde{\mor{D}}_{t}^{\gamma} (C) \pmb{\boldsymbol{]}} =
\frac{d}{dt} \boldsymbol{C}_t +
\left[
\boldsymbol{\Gamma}_\gamma(t),\boldsymbol{C}_\gamma (t)
\right]_{\_} .
	\end{equation}

	Comparing this equation with~\eref{3.37}, we get
\(
\pmb{\bs{[}} \widetilde{D}_{t}^{\gamma} (C) \pmb{\bs{]}}
 = \pmb{\bs{[}}
	\lindex[\mspace{-2.3mu}D]{}{\circ}_{t}^{\gamma} (C) \pmb{\bs{]}}
\)
where $\lindex[\mspace{-2.3mu}D]{}{\circ}$ is the derivation along paths in
$\morf_\base\HilB$ associated to $D$ according to~\eref{3.36}.
Therefore the invariant bundle form of~\eref{6.5} is
	\begin{equation}	\label{6.5new}
\widetilde{D}(C) = \lindex[\mspace{-2.3mu}D]{}{\circ} (C),
	\end{equation}
where $C\in\PLift^1(\morf_\base\HilB)$ acts \emph{only on state liftings}
according to~\eref{3.31} and $C_\gamma$ is \emph{linear}, or, equivalently,
we can write
	\begin{equation}	\label{6.5new-1}
\widetilde{D}|_{\mathcal{O}}
 = \lindex[\mspace{-2.3mu}D]{}{\circ}|_{\mathcal{O}}
	\end{equation}
with $\mathcal{O}$ being the set of just-described liftings $C$.

	We derived~(\ref{6.5new}) under the assumption that
 $C_\gamma$ is linear and $C$
acts on state liftings, \ie on ones satisfying the matrix\ndash bundle
Schr\"odinger
equation~(\ref{5.1}). Conversely, if we apply~(\ref{6.5}) to some vector
\( \Phi_\gamma(t)\in \bHil_{\gamma(t)} \) and compare the result with the
one for
\( \left( \mor{D}_{t}^{\gamma}(C) \right) ( \Phi ) \)
obtained through~(\ref{5.8}) (see above), we see that \(\Phi_\gamma(t)\)
satisfies~(\ref{5.1}). Consequently, equation~(\ref{6.5new}) with linear
$C_\gamma$ is valid if and only if $C$ is applied on liftings representing
the evolution of a quantum system. Hence \( \Psi \) is a state lifting, i.e.\
it satisfies, for instance, the bundle Schr\"odinger
equation~(\ref{5.24new}), iff the equation
	\begin{equation}	\label{6.5new-2}
(\widetilde{D}(C)) \Psi = ( \lindex[\mspace{-2.3mu}D]{}{\circ} (C) ) \Psi,
	\end{equation}
is valid for every lifting $C$ in the bundle of restricted morphisms such
that $C_\gamma$ is linear for every $\gamma$.  In particular~(\ref{6.5new-2})
is valid for the (Hermitian) liftings (of paths) corresponding to observables
(see further Sect.~\ref{VI}) and \( \Psi \) satisfying the bundle
Schr\"odinger equation~(\ref{5.24new}).

	The over-all above discussion shows the equivalence
of~(\ref{6.5new-2}) (for every $C$ with $C_\gamma$ linear) with the
Schr\"odinger equation (in anyone of its (equivalent) forms mentioned until
now). That is why~(\ref{6.5new-2}) can be called \emph{matrix\ndash lifting
Schr\"odinger equation.}

	We want to point to a substantial difference between, from one hand,
the bundle Schr\"odinger equation~\eref{5.24new}, or~\eref{6.5new-2},
or~\eref{5.14new-1} and, from other hand, the initial conditions for it
(see~\eref{5.24new-1} or~\eref{5.14}) or the conventional Schr\"odinger
equation~\eref{2.5} and the initial conditions~\eref{2.5a} for it. The bundle
Schr\"odinger equations are absolutely invariant in a sense that they do not
depend on some coordinates, space(\ndash time) points, or reference paths
like $\gamma$ and hence, in our interpretation, are observer\ndash
independent. In this attitude, the bundle Schr\"odinger equations are
analogous to the covariant equations in general relativity which, due to
their tensorial character, have similar properties. In contrast to the
mentioned observation, the initial bundle conditions depend on the reference
path $\gamma$, \ie are observer\ndash dependent as, e.g., the conventional
Hamiltonian $\Ham$ is such.%
\footnote{%
For instance, suppose two point-like free particles 1 and 2 have masses $m_a$
and momentum operators $p_a$, $a=1,2$ with respect to some observer. The
particle's Hamiltonians are  $\Ham_a=p_a^2/2m$, $a=1,2$. The Hamiltonian of
the second particle with respect to the first one is $\Ham_{1,2}=p^2/2m$
(after the elimination of the center of mass movement) with
$p:=(m_2p_1-m_1p_2)/m_1m_2$ and $m:=m_1m_2/(m_1+m_2)$. For details,
see~\cite[chapter~IX, \S\S11,~12]{Messiah-1}.%
}
Consequently in the Hilbert bundle description the observer\ndash
dependence, \ie the dependence on the reference path $\gamma$, is `moved'
from the equations of motion to the initial conditions for them. It is clear,
this dependence cannot be removed completely due to the equivalence
between the Hilbert space and Hilbert bundle descriptions of quantum
mechanics.

	Since now we have in our disposal the machinery required for analysis
of~\cite{Asorey&Carinena&Paramio}, we, as promised in
Sect.~\ref{I}, want to make some comments on it.
In~\cite[p.~1455, left column, paragraph~4]{Asorey&Carinena&Paramio} is
stated ``that in the Heisenberg gauge (picture) the Hamiltonian is the null
operator''. If so, all eigenvalues of the Hamiltonian vanish and, as they are
picture\ndash independent, they are null in any picture of quantum mechanics.
Consequently, form here one deduces the absurd conclusion that the `energy
levels of any system coincide and correspond to one and the same energy equal
to zero'. Since the paper~\cite{Asorey&Carinena&Paramio} is mathematically
completely correct and rigorous, there is something wrong with the physical
interpretation of the mathematical scheme developed in it. Without
going into details, we describe below the solution of this puzzle which
simultaneously throws a bridge between~\cite{Asorey&Carinena&Paramio} and
the present investigation.

	In~\cite{Asorey&Carinena&Paramio} the system's Hilbert space
$\mathfrak{H}$ is replace by a differentiable Hilbert bundle
$\mathfrak{E}(\mathbb{R}_+,\mathfrak{H})$
(in our terms $(\mathfrak{E},\pi,\mathbb{R}_+)$ with a fibre $\mathfrak{H}$),
$\mathbb{R}_+:=\{t :  t\in\mathbb{R}, t\ge0\}$,
which is an associated Hilbert bundle of the principle fibre bundle
$\mathfrak{P}\bigl(\mathbb{R}_+,\mathfrak{U}(\mathfrak{H})\bigr)$ of
orthonormal bases of $\mathfrak{H}$ where $\mathfrak{U}(\mathfrak{H})$ is
the unitary group of (linear) bounded invertible operators in
$\mathfrak{H}$ with bounded inverse. Let
$p:\mathfrak{U}(\mathfrak{H})\to GL(\mathbb{C}, \dim\mathfrak{H})$
be a (linear and continuous) representation of
$\mathfrak{U}(\mathfrak{H})$ into the general
linear group of $\dim\mathfrak{H}$\ndash dimensional matrices. An obvious
observation is that~\cite[equation~(4.6)]{Asorey&Carinena&Paramio} under $p$
transforms, up to notation, to our equation~\eref{5.12}
(in~\cite{Asorey&Carinena&Paramio} is taken $\hbar=1$). Thus we see that
what in~\cite{Asorey&Carinena&Paramio} is called Hamiltonian is actually
the (analogue of the) matrix\ndash bundle Hamiltonian $\mbHam_\gamma(t)$,
not the Hamiltonian $\Ham$ itself. This
immediately removes the above\ndash pointed conflict: as we shall see later
(see Sect.~\ref{VII}, equation~\eref{7.3}) along any $\gamma$ (or,
over $\mathbb{R}_+$ in the notation of~\cite{Asorey&Carinena&Paramio} - see
below),  we can choose a field of frames (bases) in which $\mbHam_\gamma(t)$
identically vanishes but, due to~\eref{5.2}, this does not imply the
vanishment of the Hamiltonian at all. This particular choice of the frame
along $\gamma$  corresponds to the `Heisenberg gauge'
in~\cite{Asorey&Carinena&Paramio}, normally known as Heisenberg picture.

	Having in mind the above, we can
describe~\cite{Asorey&Carinena&Paramio} as follows. In it we have
$\bHil=\mathfrak{E}$, $\base=\mathbb{R}_+$, $\Hil=\mathfrak{H}$ (the conventional
system's Hilbert space), $J=\mathbb{R}_+$, $\gamma=\id_{\mathbb{R}_+}$
(other choices of $\gamma$  correspond to reparametrization of the time), and
$\frac{\pd}{\pd t},\ t\in\mathbb{R}_+$ is the analog of $D^+$
in~\cite{Asorey&Carinena&Paramio}. As we already pointed, the
matrix\ndash bundle Hamiltonian $\mbHam_\gamma(t)$ represents the operator
$A(t)$ of~\cite{Asorey&Carinena&Paramio}, incorrectly identified there with
the `Hamiltonian' and the choice of a field of bases over
$\gamma(J)=\mathbb{R}_+=\base$ corresponds to an appropriate `choice of the
gauge' in~\cite{Asorey&Carinena&Paramio}. Now, after its correspondence
between~\cite{Asorey&Carinena&Paramio} and the present work is set, one can
see that under the representation $p$ the main results
of~\cite{Asorey&Carinena&Paramio}, expressed
by~\cite[equations~(4.5), (4.6) and~(4.8]{Asorey&Carinena&Paramio},
correspond to our equations~\eref{5.13} (see also~\eref{5.8}), \eref{5.12}
and~\eref{5.02b} respectively.

	Ending with the comment on~\cite{Asorey&Carinena&Paramio}, we note
two things. First, this paper uses a rigorous mathematical base, analogous to
the one in~\cite{Prugovecki-QMinHS}, which is not a goal of our work. And,
second, the ideas of~\cite{Asorey&Carinena&Paramio} are a very good motivation
for the present investigation and are helpful for its better understanding.

\section {The bundle description of observables}
\label{VI}
\setcounter{equation} {0}

	In quantum mechanics is accepted that to any dynamical variable,
say $\dyn{A}$, there corresponds a unique observable, say
$\ope{A}(t)$, which is a Hermitian linear operator in the Hilbert space
$\Hil$, i.e.\ $\ope{A}(t)\colon \Hil\to\Hil$, $\ope{A}(t)$ is linear, and
$\ope{A}^\dag=\ope{A}$~\cite{Messiah-QM,Prugovecki-QMinHS,Fock-FQM}.

	The mean value of an observable $\ope{A}$ in a state with state vector
$\psi\in\Hil$ with finite norm is calculated according to~\eref{2.9}. It
is interpreted as an observed (mean) value of the dynamical variable $\dyn{A}$
at a state $\psi$. This assumption and the probabilistic interpretation of
the wave function $\psi$ are the main tools for predicting experimentally
observable results in quantum mechanics. As we said earlier in
Subsect~\ref{new-I.3}, the latter of these tools is transferred in
Hilbert bundle quantum mechanics in an evident way. The bundle version of the
former one is the main task of this section. Below will be shown that the
proper bundle analogue of $\ope{A}$ is a suitable lifting of paths (in the
bundle of restricted morphisms of the Hilbert bundle of states) or a
(generally multiple\ndash valued) morphism along paths (in the system's
Hilbert bundle).

\subsection{Heuristic introduction}
	\label{VI.1}

	Let $\psi^{(\lambda)}(t)\in\Hil$ be an eigenvector of $\ope{A}(t)$
with eigenvalue $\lambda$ ($\in\mathbb{R}$), i.e.\
$\ope{A}(t)\psi^{(\lambda)}(t) = \lambda \psi^{(\lambda)}(t)$.
According to~(\ref{4.3a}) to $\psi^{(\lambda)}(t)$ corresponds the vector
\(
\Psi_\gamma^{(\lambda)}(t) =
l_{\gamma(t)}^{-1} \psi^{(\lambda)}(t) \in \bHil_{\gamma(t)}
\)
in the bundle description. But the Hilbert space and Hilbert bundle
descriptions of a quantum evolution should be fully equivalent.
Hence to $\ope{A}(t)$ in $\bHil_{\gamma(t)}$
should correspond certain operator which we denote by
$\mor{A}_{\gamma}(t)$. We define this operator by
demanding every
$\Psi_\gamma^{(\lambda)}(t)$
to be its eigenvector with eigenvalue $\lambda$, i.e.\
\(
\left( \mor{A}_{\gamma}(t) \right)
				\Psi_\gamma^{(\lambda)}(t) :=
\lambda \Psi_\gamma^{(\lambda)}(t).
\)
Combining this equality with the preceding two, we easily verify that
\(
\bigl( \mor{A}_{\gamma}(t) \circ
			l_{\gamma(t)}^{-1} \bigr)
\psi^{(\lambda)}(t) =
\bigl(l_{\gamma(t)}^{-1} \circ \ope{A}(t) \bigr) \psi^{(\lambda)}(t)
\)
where the linearity of $l_x$ has been used. Admitting that
$\{\psi^{(\lambda)}(t)\}$ is a complete set of vectors, i.e.\ a basis of
$\Hil$, we find
	\begin{equation}	\label{6.0}
\mor{A}_{\gamma}(t) =
l_{\gamma(t)}^{-1} \circ \ope{A}(t) \circ l_{\gamma(t)}
\colon  \bHil_{\gamma(t)}\to\bHil_{\gamma(t)}.
	\end{equation}

	More `physically', the same result is derivable from~(\ref{2.9}) too.
The mean value
\( \langle \ope{A} \rangle_{\psi}^{t} \) of \(\ope{A}\)
	at a state \(\psi(t)\) is given
by~(\ref{2.9}) and the mean value of
\( \mor{A}_{\gamma}(t) \) at a state \(\Psi_\gamma(t) \)
is
	\begin{equation}	\label{6.1'}
\left\langle \mor{A}_{\gamma}(t) \right\rangle_{\Psi_\gamma(t)} :=
\left\langle \mor{A}_{\gamma}(t) \right\rangle_{\Psi_\gamma}^{t} :=
\frac{
\langle \Psi_\gamma(t) | \mor{A}_\gamma(t) \Psi_\gamma(t) \rangle_{\gamma(t)}
}
{
\langle \Psi_\gamma(t) | \Psi_\gamma(t) \rangle_{\gamma(t)}
},
	\end{equation}
i.e.\ it is given via~(\ref{2.9}) in which the scalar
product $\langle\cdot | \cdot\rangle_x$, defined by~(\ref{4.8}), is used
instead of
$\langle\cdot | \cdot\rangle$.
	We define
$\mor{A}_{\gamma}(t)$ by demanding
	\begin{equation}	\label{6.1''}
\langle \ope{A}(t)\rangle_{\psi}^{t} =
\langle \mor{A}_{\gamma}(t) \rangle_{\Psi_\gamma}^{t}.
	\end{equation}
Physically this condition is quite natural as it means that the observed
values of the dynamical variables are independent of the way we
calculate them. From this equality, (\ref{4.3}), and~(\ref{4.8}), we get
\(
\langle \psi(t) | \ope{A}(t)\psi(t) \rangle =
\langle \psi(t) |
l_{\gamma(t)}^{}
\circ \mor{A}_{\gamma}(t) \circ
l_{\gamma(t)}^{-1} \psi(t)
\rangle
\)
which, again, leads to~(\ref{6.0}). Thus we have also proved the equivalence
of~(\ref{6.0}) and~(\ref{6.1''}).

	The above considerations lead to the idea that to every observable
$\ope{A}$ at a moment $t$ there should correspond an operator
$\mor{A}_\gamma(t)$, given by~\eref{6.0}, in the fibre
$\bHil_{\gamma(t)}=\pi^{-1}(\gamma(t))$. It is almost evident, if
$\gamma\colon J\to\base$ is without self\ndash intersections, the collection
of maps $\{\mor{A}_\gamma(t)|t\in J\}$ forms a morphism over $\gamma(J)$ of
the restricted on $\gamma(J)$ system's Hilbert bundle.

\subsection{Rigorous considerations}
	\label{VI.2}

	As it was mentioned earlier (see Subsect.~\ref{new-I}),
postulates~\ref{Post4.1} and~\ref{Post4.2} are not enough for the bundle
description of observables. The contents of Subsect.~\ref{VI.1} confirms this
opinion. Relying on the above not quite rigorous results, we formulate the
missing section of the chain as the next postulate.

	\begin{Post}	\label{Post6.1}
	Let $\HilB$ be the Hilbert bundle of a quantum system,
$\gamma\colon J\to\base$, and $t\in J$. In the bundle description of
quantum mechanics, every dynamical variable $\dyn{A}$ characterizing the
system is represented by a map $\mor{A}$ assigning to the pair $(\gamma,t)$
a map $\mor{A}_\gamma(t)\colon\pi^{-1}(\gamma(t))\to\pi^{-1}(\gamma(t))$ such
that
	\begin{equation}	\label{6.1new}
\mor{A}_\gamma(t) = l_{\gamma(t)}^{-1} \circ\ope{A}(t)\circ l_{\gamma(t)}
	\end{equation}
where $\ope{A}(t)\colon\Hil\to\Hil$ is the linear Hermitian operator (in the
system's Hilbert space $\Hil$) representing $\dyn{A}$ in the conventional
quantum mechanics. If at a moment $t\in J$ the system is in a state
characterized by a bundle state vector $\Psi_\gamma(t)$ with a finite norm
(in $\bHil_{\gamma(t)}$), the observed value of $\dyn{A}$ (or of $\mor{A}$)
with respect to $\gamma$ at a moment $t$
is equal to the mean value of $\mor{A}_\gamma(t)$ in $\Psi_\gamma(t)$ which,
by definition, is given by~\eref{6.1'}.
	\end{Post}

	From~\eref{6.1new}, \eref{6.1'}, \eref{4.3a}, and~\eref{2.9},
we derive~\eref{6.1''}. This simple result has a fundamental meaning: the
observed values of a dynamical variable are (and must be!) independent of the
way they are calculated. This assertion may be called
`\emph{principle of invariance of the observed (mean) values}' and its
essence is the independence of the physically measurable quantities of the
mathematical way we describe them. In our context, it means the coincidence of
the observed values of a dynamical variable calculated in the Hilbert bundle
and Hilbert space descriptions. In other words, we can express the same by
saying that the predictions of both, conventional and bundle, versions of
quantum mechanics are absolutely identical regardless of the existence of
three free parameters (the base $\base$, the set $\{l_x|x\in\base\}$, and the
path $\gamma$) in the bundle case.

	Let us now clarify the mathematical nature of the mapping $\mor{A}$
introduced via postulate~\ref{Post6.1}. First of all, the maps
$\mor{A}_\gamma(t)$ are linear as $\ope{A}$ and $l_{\gamma(t)}$ are such
(see~\eref{6.1new}). If we define $\mor{A}$ as a map
$\mor{A}\colon\gamma\mapsto\mor{A}_\gamma$ with
$\mor{A}_\gamma\colon t\mapsto\mor{A}_\gamma(t)$, we see that
$\mor{A}_\gamma\colon J\to\bHil_{0}^{\base}$, where
	\begin{multline*}
\bHil_{0}^{\base} :=
\{
  \varphi_x \,|\, \varphi_x\colon\bHil_x\to\bHil_x,\ x\in\base
\}
\\
=
\{
\varphi_x \,|\, \varphi_x=\varphi|_{\bHil_x},\
		x\in\base,\ \varphi\in\Morf_B\HilB
\}
	\end{multline*}
is the bundle space of the bundle $\morf_\base\HilB$ of restricted morphisms
over $\base$ (see Subsect.~\ref{new-I.1}). Since the bundles $\HilB$
and $\morf_\base\HilB$ have a common base, the manifold $\base$, we conclude
that $\mor{A}_\gamma$ is a lifting of $\gamma\colon J\to\base$ in
 $\morf_\base\HilB$ (not in $\HilB$!). Consequently, the map $\mor{A}$, as
considered above, is a \emph{lifting of paths in the bundle of restricted
$\base$\ndash morphisms of the system's Hilbert bundle of states},
	\begin{equation}	\label{6.1new-1}
\mor{A}\in\PLift\bigl( \morf_\base\HilB \bigr) .
	\end{equation}
The linear maps
$\mor{A}_\gamma(t)\colon\bHil_{\gamma(t)}\to\bHil_{\gamma(t)}$
are Hermitian. Indeed, using~\eref{4.11} and~\eref{4.12} for
$y=x=\gamma(t)$ and $\mor{A}_{x\to x}=\mor{A}_\gamma(t)$, and~\eref{6.1new},
we get
	\begin{equation}	\label{6.1new-2}
\mor{A}_{\gamma}^{\ddag}(t) = \mor{A}_{\gamma}
	\end{equation}
where the Hermiticity of $\ope{A}$ was used. A lift $\mor{A}_\gamma$ in
$\morf_\base(\HilB)$ of $\gamma\colon J\to\base$ is called Hermitian
if~\eref{6.1new-2} holds for every $t\in J$; we denote this by writing
symbolically $\mor{A}_{\gamma}^{\ddag}=\mor{A}_\gamma$. Respectively, a
lifting $\mor{A}$ in $\PLift(\morf_\base\HilB)$ is Hermitian,
$\mor{A}^\ddag=\mor{A}$, if $\mor{A}\colon\gamma\mapsto\mor{A}_\gamma$ and
$\mor{A}_{\gamma}^{\ddag}=\mor{A}_\gamma$  for every path
$\gamma\in\Path(\base)$ in $\base$.

	Let us summarize. In the bundle description a dynamical variable
$\dyn{A}$ is represented by a Hermitian lifting $\mor{A}$ of paths in the
bundle of restricted morphisms over the base in the Hilbert bundle of states.
For $\mor{A}$ equations~\eref{6.1new} holds and its mean value along $\gamma$
at a moment $t$ for a system with state lifting $\Psi$ is
	\begin{equation}	\label{6.1new-3}
\langle \mor{A} \rangle _{\Psi}^{t,\gamma}
 := \langle \mor{A}_\gamma(t) \rangle _{\Psi_\gamma}^{t}
	\end{equation}
with the r.h.s.\ of this equality given by~\eref{6.1'}.

	The map $\mor{A}$, provided via postulate~\ref{Post6.1}, can also be
considered as a (multiple\ndash valued) morphism along paths of the Hilbert
bundle of states.%
\footnote{%
Cf.\ the analogous situation concerning state liftings and sections in
Subsect.~\ref{new-I.3}.%
}
From one hand, define
$\mor{A}\colon\gamma\mapsto\lindex[\mor{A}]{\gamma}{}$ with
\(
\lindex[\mor{A}]{\gamma}{}\colon x\mapsto
	\{ \mor{A}_\gamma(t)|\gamma(t)=x,\ t\in J \}
\)
for $x\in\gamma(J)$. If $\gamma$ is without self\ndash intersections, then
$\lindex[\mor{A}]{\gamma}{}$ is in $\Morf_{\gamma(J)}\HilB|_{\gamma(J)}$ (see
Subsect.~\ref{new-I.1}). From other hand, we can defined
 $\mor{A}\colon\gamma\mapsto\lindex[\mor{A}]{\gamma}{}$ as a map
 $\lindex[\mor{A}]{\gamma}{}\colon \pi^{-1}(\gamma(J))\to\pi^{-1}(\gamma(J))$
with
\(
\lindex[\mor{A}]{\gamma}{}|_{\pi^{-1}(x)}
 = \{ \mor{A}_\gamma(t) | \gamma(t)=x,\ t\in J \} .
\)
In this case, if $\gamma$ is without self\ndash intersections,
 $\lindex[\mor{A}]{\gamma}{}\in \Morf_{\gamma(J)}\HilB|_{\gamma(J)}$, \ie up
to a  bijective map $\lindex[\mor{A}]{\gamma}{}$ is in
 $\Sec\bigl( \morf_{\gamma(J)}\HilB|_{\gamma(J)} \bigr)$.
Recalling that a morphism $\varphi$ over $\base$ along paths of a bundle
$(E,\pi,B)$ is a map
\(
\varphi\colon\gamma\mapsto
	\varphi_\gamma\in \Morf_{\gamma(J)}\HilB|_{\gamma(J)}
\)
for  every path $\gamma\in\Path(B)$, we see that $\mor{A}$ is a morphism over
$\base$ along paths without self\ndash intersections. But if $\gamma$ is not
injective, the map $\mor{A}\colon\gamma\mapsto\lindex[\mor{A}]{\gamma}{}$ is,
generally, multiple\ndash valued morphism (over $\base$) along paths of
$\HilB$ and it gives an alternative description of the map $\mor{A}$
introduced via postulate~\ref{Post6.1}. If the multiplicity of $\mor{A}$ as a
morphism along paths is really presented, this description will rarely be
employed; if $\mor{A}$ as a morphism is single\ndash valued, it is somewhat
`simpler' to consider $\mor{A}$ as a morphism than as a lifting of paths
and, respectively, this interpretation will be preferred.

	\begin{Defn}	\label{Defn6.1}
	The unique Hermitian lifting of paths in the bundle of restricted
morphisms (over the base of the Hilbert bundle of states) corresponding to a
dynamical variable will be called observable lifting (of paths)
Respectively, the corresponding (multiple\ndash valued) morphism (over the
base) along paths of the Hilbert bundle of states will be called observable
morphism (along paths).
	\end{Defn}

	By virtue of~\eref{6.1new-2}, the observable morphisms along paths
are Hermitian,
	\begin{equation}	\label{6.2}
\mor{A}^\ddag=\mor{A},
	\end{equation}
which is also a corollary of~\eref{6.1new} and~\eref{4.15}.

	Generally, to every operator $\ope{A}\colon \Hil\to\Hil$ there
corresponds a unique (global) morphism
$\overline{\mor{A}}\in\Morf\HilB$ given by
	\begin{equation}	\label{6.1}
\overline{\mor{A}}_x = \left.\overline{\mor{A}}\right|_{\bHil_{x}} =
l_{x}^{-1} \circ \ope{A} \circ l_{x}, \qquad x\in \base,
\quad \ope{A}\colon \Hil\to\Hil.
	\end{equation}
Consequently to an observable $\ope{A}(t)$  can be assigned the (global)
morphism $\overline{\mor{A}}(t)$,
 $\overline{\mor{A}}(t)|_{\bHil_x}=l_{x}^{-1}\circ \ope{A}(t)\circ l_x$.
But this morphism $\overline{\mor{A}}(t)$ is almost useless for our goals as
it simply gives in any fibre $\bHil_x$ a linearly isomorphic image of the
initial observable $\ope{A}(t)$ (see Sect.~\ref{new-I}).

	Notice that $\mor{A}_\gamma(t)$ generally
depends explicitly on $t$ even if $\ope{A}$ does not.
In fact, from~(\ref{6.0}) we get
	\begin{equation}	\label{6.6}
\frac{\partial \mmor{A}_\gamma(t) } {\partial t} =
\left[
\boldsymbol{g}_\gamma(t), \mmor{A}_\gamma(t)
\right]_{\_}
+
\boldsymbol{l}_{\gamma(t)}^{-1}(t)
\frac{\partial \mope{A}(t) }{\partial t}
\boldsymbol{l}_{\gamma(t)}(t),
	\end{equation}
where $[\cdot,\cdot]_{\_}$ denotes the commutator of corresponding
quantities, and
	\begin{equation}	\label{6.7}
\boldsymbol{g}_{\gamma}(t) := -  \boldsymbol{l}_{\gamma(t)}^{-1}(t)
\frac{d \boldsymbol{l}_{\gamma(t)}(t) }{d t} .
	\end{equation}

	In particular, to the Hamiltonian $\Ham$ in $\Hil$ there
corresponds the \emph{bundle Hamiltonian} $\mor{H}$ given by
	\begin{equation}	\label{6.2'}
\bHam_\gamma(t) :=
l_{\gamma(t)}^{-1} \circ \Ham(t) \circ l_{\gamma(t)}.
	\end{equation}
It is an observable lifting of paths or morphism along paths.

	From~(\ref{6.2'}), using~(\ref{2.7}) and~(\ref{4.7}), we find
	\begin{equation}	\label{6.2'a}
\bHam_\gamma(t) = \ih l_{\gamma(t)}^{-1} \circ
\frac{\partial \ope{U}(t,t_0)}{\partial t} \circ l_{\gamma(t_0)} \circ
\mor{U}_\gamma(t_0,t).
	\end{equation}
	From here a relationship between the matrix-bundle Hamiltonian and
the bundle Hamiltonian can be obtained. For this purpose, we
write~(\ref{6.2'a}) in a matrix form and using~(\ref{5.6}) and
\( {d f_a(t)}/{dt}=E_{a}^{{\ }b}f_b(t)\),
we get:
	\begin{equation}	\label{6.2''}
\bHamM_{\gamma}(t)  = \mbHam_{\gamma}(t) +
\ih \boldsymbol{l}_{\gamma(t)}^{-1}(t)
\left( \frac{d \boldsymbol{l}_{\gamma(t)} (t) }{dt} +
\boldsymbol{E}(t)\boldsymbol{l}_{\gamma(t)}(t)
\right).
	\end{equation}
Substituting here~(\ref{5.2}), we obtain
	\begin{equation}	\label{6.10'}
\bHamM_\gamma(t) =
\boldsymbol{l}_{\gamma(t)}^{-1}(t)
\HamM(t)\boldsymbol{l}_{\gamma(t)}(t)
	\end{equation}
which is simply the matrix form of~(\ref{6.2'}). Combining~(\ref{6.2''})
with~(\ref{5.2'}), we find the following connection between the matrix of
the bundle Hamiltonian and the matrix Hamiltonian:
	\begin{equation}	\label{6.10''}
\bHamM_\gamma(t) =
\boldsymbol{l}_{\gamma(t)}^{-1}(t)
\mHam(t)\boldsymbol{l}_{\gamma(t)}(t) +
\ih \boldsymbol{l}_{\gamma(t)}^{-1}(t)
\boldsymbol{E}(t)\boldsymbol{l}_{\gamma(t)}(t) .
	\end{equation}

    	Notice, due to~(\ref{6.1}) as well as to~(\ref{6.0}), to the
identity map of $\Hil$ there corresponds a morphism along paths equal to the
identity map of $\bHil$:
	\begin{equation}	\label{6.2'''}
\id_\Hil \longleftrightarrow \id_\bHil .
	\end{equation}

\subsection{Functions of observables}
	\label{VI.3}

	The results expressed by~(\ref{6.0}) and~(\ref{6.1}) enable functions
of observables in $\Hil$ to be transferred into ones of liftings of
paths (morphisms along paths) or morphisms of $\HilB$, respectively. We will
illustrate this in the case of, e.g., two variables.

	Let
\(
\ope{G}\colon (\ope{A}(t),\ope{B}(t)) \mapsto
 \ope{G}(\ope{A}(t),\ope{B}(t))\colon  \Hil\to\Hil
\)
be a function of the observables
$\ope{A}(t),\ope{B}(t)\colon \Hil\to\Hil$.
It is natural to define the bundle analogue $\mor{G}$ of $\ope{G}$ by
\[
\mor{G}\colon(\mor{A},\mor{B})\mapsto\mor{G}(\mor{A},\mor{B})\colon
\gamma\mapsto\mor{G}_\gamma(\mor{A},\mor{B})\colon
\proj^{-1}(\gamma(J))\to\proj^{-1}(\gamma(J)),
\]
where
$\mor{G}_\gamma(\mor{A},\mor{B})$ is a lifting of $\gamma$ and
	\begin{multline}	\label{6.071}
\left. G_\gamma(\mor{A},\mor{B}) \right|_{t} :=
l_{\gamma(t)}^{-1}\circ\ope{G}(\ope{A}(t),\ope{B}(t))\circ l_{\gamma(t)} \\
=
l_{\gamma(t)}^{-1}\circ
\ope{G}(l_{\gamma(t)}\circ\mor{A}_\gamma(t)\circ l_{\gamma(t)}^{-1} ,
	    l_{\gamma(t)}\circ\mor{B}_\gamma(t)\circ l_{\gamma(t)}^{-1})
\circ l_{\gamma(t)}.
	\end{multline}
Thus $G(\mor{A},\mor{B})$ is an observable lifting of paths.
This definition becomes evident in the cases when $\ope{G}$ is a
polynom or if it is expressible as a convergent power series; in both
cases the multiplication has to be understood as an operator composition.
If we are dealing with one of these cases, the definition~(\ref{6.071})
follows from the fact that for any observable liftings
$\mor{A}_1,\ldots,\mor{A}_k$, $k\in\mathbb{N}$
of paths, the equality
	\begin{equation}	\label{6.072}
\mor{A}_{1,\gamma}(t)\circ
\mor{A}_{2,\gamma}(t)\circ \cdots \circ \mor{A}_{k,\gamma}(t) =
l_{\gamma(t)}^{-1}\circ (\ope{A}_1(t)\circ\ope{A}_2(t)\circ\cdots
\circ\ope{A}_k(t)) \circ l_{\gamma(t)}
	\end{equation}
holds due to~(\ref{6.0}). In these cases $\mor{G}(\mor{A},\mor{B})$ depends
only on $\mor{A}$ and $\mor{B}$ and it is explicitly independent on the
isomorphisms $l_x$, $x\in\base$.

	The commutator of two operators is a an important operator
function in quantum mechanics. In the Hilbert space and bundle
descriptions it is defined, respectively, by
$[\ope{A},\ope{B}]_{\_}:=\ope{A}\circ\ope{B}-\ope{B}\circ\ope{A}$
and
$[\mor{A},\mor{B}]_{\_}:=\mor{A}\circ \mor{B} -\mor{B}\circ \mor{A}$,
where (see~\eref{6.071})
\(
(A\circ B)\colon\gamma\mapsto (A\circ B)_\gamma
 		t\colon\mapsto (A\circ B)_\gamma(t)
 = A_\gamma(t) \circ B_\gamma(t).
\)
The relation
	\begin{equation}	\label{6.073}
[\mor{A}_\gamma(t),\mor{B}_\gamma(t)]_{\_} =
l_{\gamma(t)}^{-1} \circ [\ope{A},\ope{B}]_{\_} \circ l_{\gamma(t)}
	\end{equation}
is an almost evident corollary of~(\ref{6.0}). It can also be
considered as a special case of~(\ref{6.071}). In particular, to
commuting observables (in~$\Hil$) there correspond commuting observable
liftings or morphisms:
	\begin{equation}	\label{6.074}
[\ope{A},\ope{B}]_{\_}=0   \iff   [\mor{A},\mor{B}]_{\_}=0.
	\end{equation}

	A little more general is the result, following
from~(\ref{6.073}), that to observables whose commutator is a c-number
there correspond observable liftings with the same c-number as a
commutator:
	\begin{equation}	\label{6.075}
[\ope{A},\ope{B}]_{\_}=c \id_\Hil   \iff   [\mor{A},\mor{B}]_{\_}
 =c \id_\bHil.
	\end{equation}
for some $c\in\mathbb{C}$. In particular, the bundle analogue of the
famous relation $[\ope{Q},\ope{P}]_{\_}=\ih\id_\Hil$
between a coordinate $\ope{Q}$ and the conjugated to it momentum
$\ope{P}$ is $[\mor{Q},\mor{P}]_{\_}=\ih\id_\bHil$.

	A bit more complicated is the case for operators and liftings of
paths at different `moments'.
Let $\gamma\colon J\to\base$ and $r,s,t\in J$.
If
$\Breve{\ope{G}}_{s,t}\colon(\ope{A},\ope{B})\mapsto
\ope{G}(\ope{A}(s),\ope{B}(t))$,
we define the bundle analogue $\Breve{\mor{G}}_{s,t}$ of
$\Breve{\ope{G}}_{s,t}$ by
\[
\Breve{\mor{G}}_{s,t}\colon
(\mor{A},\mor{B})\mapsto \Breve{\mor{G}}_{s,t}(\mor{A},\mor{B})\colon
\gamma \mapsto\Breve{\mor{G}}_{\gamma;s,t}(\mor{A},\mor{B})\colon
\proj^{-1}(\gamma(J)) \to\proj^{-1}(\gamma(J)),
\]
where
	\begin{multline} 	\label{6.081}
\left.{\Breve{\mor{G}}}_{\gamma;s,t} (\mor{A},\mor{B})\right|_{r}
:=
l_{\gamma(r)}^{-1}\circ \ope{G}(\ope{A}(s),\ope{B}(t)) \circ l_{\gamma(r)}
\\
=
l_{\gamma(r)}^{-1}\circ
\ope{G}(
l_{\gamma(r)}\circ\Breve{\ope{A}}_{\gamma;s}(r)\circ l_{\gamma(r)}^{-1} ,
l_{\gamma(r)}\circ\Breve{\ope{B}}_{\gamma;t}(r)\circ l_{\gamma(r)}^{-1}
)
\circ l_{\gamma(r)}
\colon \bHil_{\gamma(r)} \to \bHil_{\gamma(r)}. 
	\end{multline}
Here
	\begin{equation}	\label{6.082}
\Breve{\ope{A}}_{\gamma;t}(r) :=
l_{\gamma(r)}^{-1}\circ\ope{A}(t)\circ l_{\gamma(r)} =
l_{t\to r}^{\gamma}\circ\mor{A}(t)\circ l_{r\to t}^{\gamma}
\colon \bHil_{\gamma(r)} \to \bHil_{\gamma(r)},
	\end{equation}
where~\eref{6.0} has been used and
 $l_{s\to t}^{\gamma}:=l_{\gamma(s)\to\gamma(t)}$ is the (flat) linear
transport (along paths) from $\gamma(s)$ to $\gamma(t)$ assigned to the
isomorphisms $l_x$, $x\in\base$ (see equation~\eref{4.12h}).%
\footnote{%
According to~\cite[sections~2 and~3]{bp-TP-morphisms} the observable lifting
 $\Breve{\mor{A}}_{\gamma;t}(r)$ along $\gamma$
is obtained via linear transportation of $\mor{A}_\gamma(t)$ along $\gamma$ by
means of the induced by $l_{s\to t}^{\gamma}$ linear transport along paths in
the bundle  $\morf_\base{\HilB}$ of restricted morphisms over $\base$ of
$\HilB$
(see also subsection~\ref{VII.1}).%
}
Now the analogue of~\eref{6.072} is
	\begin{multline}	\label{6.083}
\Breve{\ope{A}}_{1;\gamma;t_1}(r)\circ
\Breve{\ope{A}}_{2;\gamma;t_2}(r)\circ \cdots \circ
\Breve{\ope{A}}_{k;\gamma;t_k}(r)
\\
=
l_{\gamma(r)}^{-1}\circ
( \ope{A}_1(t_1)\circ \ope{A}_2(t_2)\circ\cdots\circ \ope{A}_k(t_k) )
		\circ l_{\gamma(r)}.
	\end{multline}
So, if $\ope{G}$ is a polynom or a convergent power series,
the observable lifting
$\Breve{\mor{G}}_{\gamma;s,t}(\mor{A},\mor{B})$
along  $\gamma$ depends only on
 $\Breve{\mor{A}}_{\gamma;s}(r)$ and $\Breve{\mor{B}}_{\gamma;t}(r)$.

	In particular for $\ope{G}(\cdot,\cdot)=[\cdot,\cdot]_{\_}$, have
	\begin{equation}\label{6.084}
\bigl[ \Breve{\mor{A}}_{\gamma;s}(r) ,
		\Breve{\mor{B}}_{\gamma;t}(r) \bigr]_{\_}
=
l_{\gamma(r)}^{-1}\circ [\ope{A}(s),\ope{B}(t)]_{\_}\circ l_{\gamma(r)}
	\end{equation}
which for $s=r=t$ reduces to~(\ref{6.073}). In this case the analogues
of~(\ref{6.074}) and~(\ref{6.075}) are
	\begin{align}
	\label{6.085}
 [\ope{A}(s),\ope{B}(t)]_{\_} =0 &\iff
\left[
\Breve{\mor{A}}_{\gamma;s}(r) , \Breve{\mor{B}}_{\gamma;t}(r)
\right]_{\_} =0,
\\
	\label{6.086}
 [\ope{A}(s),\ope{B}(t)]_{\_} = c \id_\Hil &\iff
\left[
\Breve{\mor{A}}_{\gamma;s}(r) , \Breve{\mor{B}}_{\gamma;t}(r)
\right]_{\_} = c \id_{\bHil_{\gamma(r)}},
	\end{align}
respectively.

	The above considerations can \emph{mutatis mutandis}, e.g.\ by
replacing $\gamma(t)$ with $x$,  $\ope{A}(t)$ with $\ope{A}$,  $\mor{A}$ with
$\overline{\mor{A}}$, etc., be transferred to global morphisms of $\HilB$, but
this is not required for the present investigation.

\section {Pictures of motion from  bundle view-point}
\label{VII}
\setcounter{equation} {0}

	Well-known are the different pictures (or representations) of
motion of a quantum system~\cite[ch.~VIII,~\S\S~9, 10, 14]{Messiah-QM},
\cite[ch.~III,~\S~14]{Fock-FQM}, \cite[\S~27, \S~28]{Dirac-PQM}:
the Schr\"odinger's, Heisenberg's, interaction's, and other
`intermediate' ones. Nevertheless that they are equivalent from the
view\ndash point of physically predictable results, these special
representations of the quantum\ndash mechanical formalism reflect its
different sides. Correspondingly, the choice of a concrete picture depends on
the particular physical problem under investigation. Bellow we consider
certain general problems connected with these special pictures of motion of a
quantum system from the fibre bundle view-point on quantum mechanics proposed
in this investigation.

\subsection {Schr\"odinger picture}
\label{VII.1}
\setcounter{equation} {0}

	In fact, the bundle Schr\"odinger picture of motion of a
quantum system is the way of its description we have been dealing
until now~\cite{bp-BQM-introduction+transport,bp-BQM-equations+observables}.
Its basic assertions will be summarize in this subsection.

	The states of a quantum system form a Hilbert bundle $\HilB$ whose
base $\base$ is a $C^1$ manifold, interpreted as a space(\ndash time) model.
The system state is described by a lifting $\Psi$ of paths over $\HilB$,
	\begin{equation}	\label{7.01}
\Psi\in\PLift\HilB ,
	\end{equation}
which also admits equivalent interpretation as a, generally, multiple-valued
section along paths in the same bundle. Along every path
$\gamma\colon J\to\base$, interpreted as a trajectory (world line) of some
observer, the different time\ndash values of the bundle state vectors
$\Psi_\gamma(t)$ are connected via equation~\eref{4.4} in which
$\mor{U}_\gamma(t,s)$ is the evolution transport along $\gamma$ from
$s$ to $t$, $s,t\in J$. The state lifting $\Psi$ is generically a variable
in time quantity evolving according to the bundle Schr\"odinger
equation~\eref{5.24new}. This equation, together with some initial condition,
is equivalent to the Schr\"odinger equation (initial\ndash value
problem)~\eref{5.14} for the evolution transport $\mor{U}$.

	In the bundle description to a dynamical variable $\dyn{A}$
corresponds a unique lifting of paths $\mor{A}$ in the bundle of restricted
morphisms of the system's Hilbert bundle of states,
	\begin{equation}	\label{7.02}
A\in\PLift(\morf_\base\HilB)
	\end{equation}
The observable lifting $\mor{A}$ admits also a treatment as a, generally,
multiple\ndash value  section along paths of $\HilB$. With respect to a
reference path $\gamma\colon J\to \base$ at some moment $t\in J$ an
observable lifting $\mor{A}$ reduces to a map
$\mor{A}_\gamma(t)\colon\bHil_{\gamma(t)}\to\bHil_{\gamma(t)}$ which,
generally, is time\ndash dependent regardless that in the Hilbert space
description the corresponding observable $\ope{A}$ may happen to be
time\ndash independent. It (or its evolution) is explicitly given
by~\eref{6.0}.

	Geometrically the maps $\mor{A}_\gamma(t)$ `live' in the bundle space
	\begin{subequations}	\label{7.1}
	\begin{multline}	\tag{\ref{7.1}a}	\label{7.1a}
\bHil_{0}^{\base} :=
\{
  \varphi_x \,|\, \varphi_x\colon\bHil_x\to\bHil_x,\ x\in\base
\}
\\
=
\{
\varphi_x \,|\, \varphi_x=\varphi|_{\bHil_x},\
		x\in\base,\ \varphi\in\Morf_B\HilB
\}
	\end{multline}
of the bundle of point-restricted morphisms of $\HilB$ whose projection is
	\begin{equation}	\tag{\ref{7.1}b}	\label{7.1b}
\proj_{0}^{\base}\colon \bHil_{0}^{\base} \to\base ,
\quad
\proj_{0}^{\base}(\varphi) = x_\varphi
	\end{equation}
	\end{subequations}
for $\varphi\in\bHil_{0}^{\base}$, where $x_\varphi\in\base$ is the unique
element of $\base$ such that
$\varphi\colon\bHil_{x_\varphi}\to\bHil_{x_\varphi}$.

	Since equations~\eref{4.3new} and~\eref{6.1''} are valid, the
probabilistic interpretation of conventional quantum mechanics is retained
and the predictions of Hilbert bundle and Hilbert space versions of quantum
mechanics are identical.

	Summing up, in the bundle Schr\"odinger picture, both the state
liftings and observable liftings generically change in time in the
corresponding bundles as described above.

\subsection {Heisenberg picture}
\label{VII.2}

	The Heisenberg picture is suitable for analyzing some quantum
properties of the systems, as well as for the comparison between classical
and quantum mechanics. In it the time\ndash dependence  is entirely shift to
the dynamical variables, \ie to the observables representing them, while the
state vectors remain constant in time. In this subsection will be proved that
analogous transformation is available in the bundle version of quantum
mechanics too.

	Below we present two different ways for introduction of bundle
Heisenberg picture leading, of course, to one and the same result. The
first one is based  entirely on the bundle approach and reveals its natural
geometric character. The second one is a direct analogue of the usual way in
which one arrives to this picture.

\subsubsection{Hilbert bundle introduction}
	\label{VII.2.1}

	According to~\cite[sect.~\ref{+:Sect4}]{bp-normalF-LTP}
or~\cite[sect.~3]{bp-LTP-general}
\emph{%
every linear transport along paths is locally Euclidean},
i.e.\ (see~\cite[sect.~\ref{+:Sect4}]{bp-normalF-LTP} for details and
rigorous results)
along every path there is a field of
(generally multiple-valued~\cite[remark~4.2]{bp-normalF-LTP})
bases, called normal, in which its matrix is unit.
Such a collection of bases is called a \emph{normal frame} along
the corresponding path.
In particular, along $\gamma\colon  J\to \base$ there exists a frame
\(
\left\{
\{ \tilde e_{a}^{\gamma}(t)\}\mathrm{\ -\ basis\ in\ } \bHil_{\gamma(t)}
\right\}
\)
in which the matrix of the evolution transport $\mor{U}_\gamma(t,s)$ is
\(
   \widetilde{\mmor{U}}_\gamma(t,s)=\openone.
\)
	Explicitly we can put
	\begin{equation}	\label{7.3new}
   \tilde e_{a}^{\gamma}(t) = \mor{U}_\gamma(t,t_0) e_{a}^{\gamma}(t_0),
	\end{equation}
where $t,t_0\in J$, $\gamma$ is not a summation index, and the basis
$\{ e_{a}^{\gamma}(t_0) \}$ in $\bHil_{\gamma(t_0)}$ is
fixed~\cite[proof of proposition~3.1]{bp-LTP-general}
(cf.~\cite[equation~\eref{+:4.2}]{bp-normalF-LTP}).%
\footnote{%
The so-defined field of bases are not uniquely defined at the points of
self-intersection, if any, of $\gamma$. Evidently, they are unique on
any `part' of $\gamma$ without self-intersections. The last case
covers the interpretation of $\gamma$ as an observer's world line, in
which it cannot have self-intersections.
See~\cite[sect.~4]{bp-normalF-LTP} for details.%
}
Because of~(\ref{4.6}), (\ref{5.9}), and~(\ref{5.10}) this class of
frames normal along $\gamma$ for the evolution transport is uniquely defined
by any one of the (equivalent) equalities:
	\begin{equation}	\label{7.3}
\widetilde{\mmor{U}}_\gamma(t,t_0) = \openone,	\quad
\widetilde{\boldsymbol{\Gamma}}_\gamma(t) = \boldsymbol{0},	\quad
\widetilde{\mbHam_\gamma}(t) = \boldsymbol{0}.
	\end{equation}

	So, the matrix-bundle Hamiltonian vanishes in such a special frame
and, consequently (see~(\ref{5.1})), the components of the
bundle state vectors remain constant in time $t$, i.e.\
\( \widetilde{\boldsymbol{\Psi}}_\gamma(t)=\const, \)
but the vectors themselves are not necessary such as the normal frames along
$\gamma$  are generally time\ndash dependent.

	In the normal frame
\( \{  \tilde e_{a}^{\gamma}(t) \}, \)
defined above by~\eref{7.3new}, the components of $\mor{A}_\gamma(t)$ are
	\begin{align*}
\widetilde{(\mor{A}_\gamma(t))}_{ab} &=
\bigl\langle
\tilde e_{a}^{\gamma}(t) |
\bigl( \left.\mor{A}_\gamma\right|_{\bHil_{\gamma(t)}}\bigr)
			\tilde e_{b}^{\gamma}(t)
\bigr\rangle_{\gamma(t)}
\\ &=
\langle
\mor{U}_\gamma(t,t_0) e_{a}^{\gamma}(t_0) |
\mor{A}_\gamma(t) \mor{U}_\gamma(t,t_0) e_{b}^{\gamma}(t_0)
\rangle_{\gamma(t)}
\\ &=
\langle
e_{a}^{\gamma}(t_0) |
\mor{U}_\gamma^{-1}(t,t_0)\mor{A}_\gamma(t)
		\mor{U}_\gamma(t,t_0) e_{b}^{\gamma}(t_0)
\rangle_{\gamma(t)} =
\left( \mor{A}_{\gamma,t}^{\bHam}(t_0) \right)_{ab},
	\end{align*}
where
	\begin{equation}	\label{7.4}
\mor{A}_{\gamma,t}^{\mathrm{H}}(t_0) :=
\mor{U}_\gamma^{-1}(t,t_0)\circ \mor{A}_\gamma(t)\circ \mor{U}_\gamma(t,t_0)
\colon  \bHil_{\gamma(t_0)}\to  \bHil_{\gamma(t_0)}.
	\end{equation}
Hence the matrix elements of $\mor{A}_\gamma(t)$ in
$\{\tilde e_{a}^{\gamma}(t)\}$
coincide with those of $\mor{A}_{\gamma,t}^{\mathrm{H}}(t_0)$ in
$\{e_{a}^{\gamma}(t_0)\}.$
	Consequently, due to~(\ref{2.9}), (\ref{6.1'}), (\ref{6.1''}),
and~(\ref{4.14}),
the mean value of $\mor{A}$ (along $\gamma$) is
	\begin{equation*}\begin{split}
\langle \mor{A}_\gamma(t)\rangle_{\Psi_\gamma}^{t} &=
\left(\widetilde{\mor{A}_\gamma(t)}\right)_{ab}
\widetilde{\Psi}_{\gamma}^{a}(t) \widetilde{\Psi}_{\gamma}^{b}(t) /
\langle
{\Psi}_{\gamma}(t) | {\Psi}_{\gamma}(t)
\rangle_{\gamma(t)}
\\ &=
\Bigl(  \mor{A}_{\gamma,t}^{\mathrm{H} (t_0)}  \Bigr)_{ab}
\widetilde{\Psi}_{\gamma}^{a}(t) \widetilde{\Psi}_{\gamma}^{b}(t) /
\langle
{\Psi}_{\gamma}(t_0) | {\Psi}_{\gamma}(t_0)
\rangle_{\gamma(t)} .
	\end{split}\end{equation*}
But
\(
\widetilde{\boldsymbol{\Psi}}_\gamma(t) =
\widetilde{\boldsymbol{\Psi}}_\gamma(t_0),
\)
hereout
	\begin{equation}	\label{7.5}
\langle \mor{A}_\gamma(t) \rangle_{\Psi_\gamma}^{t} =
\langle \mor{A}_{\gamma,t}^{\mathrm{H}}(t_0) \rangle_{\Psi_\gamma}^{t_0}.
	\end{equation}

	So, the mean value of $\mor{A}_\gamma(t)$ in a state $\Psi_\gamma(t)$
is equal to the mean value of $\mor{A}_{\gamma,t}^{\mathrm{H}}(t_0)$ in the
state $\Psi_\gamma(t_0)$. Taking into account that the only measurable
(observable) physical quantities are the mean
values~\cite{Dirac-PQM,Messiah-QM,Neumann-MFQM}, we infer that the
descriptions of a quantum system along $\gamma$ at a moment $t$ through
either one of the pairs
$( \Psi_\gamma(t),\mor{A}_\gamma(t) )$
and
$( \Psi_\gamma(t_0),\mor{A}_{\gamma,t}^{\mathrm{H}}(t_0) )$
are fully equivalent. The former one is the bundle Schr\"odinger picture of
motion along $\gamma$, reviewed above in Sect.~\ref{VII.1}. The latter
one is the \emph{ bundle Heisenberg picture of motion} of the quantum system
along $\gamma$.%
\footnote{%
Notice, the bundle Heisenberg picture is with respect to some (reference)
path $\gamma$. We shell comment on this fact in Subsect.~\ref{VII.2.3}.%
}
In it the time dependence of the bundle state vectors is entirely
shifted to the observables in conformity with~(\ref{7.4}). In this
description the bundle state vectors are constant and do not evolve in
time. On the contrary, in it the observables depend on time and act on one
and the same fibre of $\HilB$, the one to which belongs the (initial) bundle
state vector. Their evolution is governed by the Heisenberg form of the
bundle Schr\"odinger equation~(\ref{5.13}) which can be derived in the
following way.

	Substituting~(\ref{4.7}) and~(\ref{6.0}) into~(\ref{7.4}),
we get
	\begin{equation}	\label{7.6}
\mor{A}_{\gamma,t}^{\mathrm{H}}(t_0) =
l_{\gamma(t_0)}^{-1}\circ\ope{A}_{t}^{\mathrm{H}}(t_0)\circ l_{\gamma(t_0)}
\colon \bHil_{\gamma(t_0)}\to \bHil_{\gamma(t_0)},
	\end{equation}
where (cf.~(\ref{7.4}))
	\begin{equation}	\label{7.7}
\ope{A}_{t}^{\mathrm{H}}(t_0) :=
\ope{U}(t_0,t)\circ\ope{A}(t)\circ\ope{U}(t,t_0)\colon \Hil\to\Hil
	\end{equation}
is the Heisenberg operator corresponding to $\ope{A}(t)$ in the Hilbert space
description (see below).

	A simple verification shows that
	\begin{equation}	\label{7.8}
\ih\frac{\partial \ope{A}_t^{\mathrm{H}}(t_0)}{\partial t} =
\left[  \ope{A}_t^{\mathrm{H}}(t_0),\Ham_t^{\mathrm{H}}(t_0)  \right] _{\_} +
\ih\Bigl(
\frac{\partial\ope{A}}{\partial t}
\Bigr)_{\!t}^{\!\mathrm{H}} (t_0).
	\end{equation}
Here
\( \left({\partial\ope{A}}/{\partial t}\right)_t^{\mathrm{H}} (t_0) \)
is obtained from~(\ref{7.7}) with ${\partial\ope{A}}/{\partial t}$ instead of
$\ope{A}$ and
	\begin{equation}	\label{7.9}
\Ham_t^{\mathrm{H}}(t_0) =
\ope{U}^{-1}(t,t_0) \Ham(t) \ope{U}(t,t_0) =
\ih\ope{U}^{-1}(t,t_0)\frac{\partial\ope{U}(t,t_0)}{\partial t}
	\end{equation}
(cf.~(\ref{7.7})) with $\Ham(t)$ being the usual Hamiltonian in $\Hil$
(see~(\ref{2.7})), i.e.\ $\Ham_t^{\mathrm{H}}(t_0)$ is the Hamiltonian in the
Heisenberg picture.

	Finally, from~(\ref{7.6}) and~(\ref{7.8}), we obtain
	\begin{equation}	\label{7.10}
\ih\frac{\partial \mor{A}_{\gamma,t}^{\mathrm{H}}(t_0)}{\partial t} =
\left[
\mor{A}_{\gamma,t}^{\mathrm{H}}(t_0), H_{\gamma,t}^{\mathrm{H}}(t_0)
\right] _{\_} +
\ih\left(
\frac{\partial\ope{A}}{\partial t}
\right)_{\!\gamma,t}^{\!\mathrm{H}} (t_0)
	\end{equation}
in which all quantities with subscript $\gamma$ are defined according
to~(\ref{7.6}). This is the \emph{bundle equation of motion (for the
observables) in the Heisenberg picture} of motion of a quantum system. It
determines the time evolution of the observables in this description.

\subsubsection{Hilbert space introduction}
	\label{VII.2.2}

	Now we shall outline briefly how the above results can be obtained by
transferring the conventional Heisenberg picture of motion from the Hilbert
space $\Hil$ to its analogue in the Hilbert bundle $\HilB$.

	The mathematical expectation of an observable $\ope{A}(t)$ in a
state characterized by a bundle state vector $\psi(t)$ with a finite norm is
(see~(\ref{2.9}), (\ref{2.4}), and~(\ref{4.2b}))
\[
\langle\ope{A}(t)\rangle_\psi^t 	=
\frac{\langle\psi(t) | \ope{A}(t)\psi(t)\rangle}
{\langle\psi(t) | \psi(t)\rangle} =
\frac{\langle\psi(t_0) |
	\ope{U}^{-1}(t,t_0)\ope{A}(t)\ope{U}(t,t_0)\psi(t_0)\rangle}
{\langle\psi(t_0) | \psi(t_0)\rangle}	.
\]
	Combining this with~(\ref{7.7}), we find:
	\begin{gather}	\label{7.11}
\langle\ope{A}(t)\rangle_\psi^t 	=
\langle\ope{A}_t^{\mathrm{H}}(t_0)\rangle_{\psi}^{t_0} 	=
\langle\ope{A}_t^{\mathrm{H}}(t_0)\rangle_{\psi_t^{\mathrm{H}}}^{t_0},
\\ \label{7.11a}
\psi_t^{\mathrm{H}}(t_0) := \psi(t_0).
	\end{gather}

	Thus the pair $(\psi(t),\ope{A}(t))$ is equivalent to the pair
$(\psi(t_0),\ope{A}_t^{\mathrm{H}}(t_0))$ from the view-point of observable
quantities. The latter one realizes the Heisenberg picture in $\Hil$, \ie in
the Hilbert space description of quantum mechanics. In it the state vectors
are constant while the observables, generally, change in time according to
the Heisenberg form~(\ref{7.8}) of the equation of motion.

	In the Hilbert bundle description to $\ope{A}$ and
$\ope{A}_t^{\mathrm{H}}(t_0)$ correspond the quantities (see~(\ref{6.0})),
respectively,
\(
\mor{A}_\gamma(t)=l_{\gamma(t)}^{-1}\circ\ope{A}(t)\circ l_{\gamma(t)}
\)
and (see~(\ref{7.7}) and~(\ref{4.7}))
	\begin{equation}	\label{7.12}
\mor{A}_{\gamma,t}^{\mathrm{H}}(t_0)	=
l_{\gamma(t_0)}^{-1}\circ\ope{A}_t^{\mathrm{H}}(t_0)\circ l_{\gamma(t_0)} =
\mor{U}_\gamma^{-1}(t,t_0)\circ \mor{A}_\gamma(t)\circ \mor{U}_\gamma(t,t_0).
	\end{equation}
Hence to the Heisenberg operator $\ope{A}_t^{\mathrm{H}}$
corresponds exactly the above\ndash introduced by~(\ref{7.4}) (Heisenberg)
map
$\mor{A}_{\gamma,t}^{\mathrm{H}}(t_0)$.
In particular, to the Hamiltonian $\Ham(t)$ and its Heisenberg form
$\Ham_t^{\mathrm{H}}(t_0)$, given by~(\ref{7.7}) for $\ope{A}=\Ham$ or
by~(\ref{7.9})
(cf.~(\ref{2.7})), correspond the mappings
(see~(\ref{6.0}) and~(\ref{6.2'}))
\(
\bHam_\gamma(t)	=
l_{\gamma(t)}^{-1}\circ\Ham(t)\circ l_{\gamma(t)}
\)
and (cf.~(\ref{7.6}) and~(\ref{7.9}))
	\begin{equation}	\label{7.13}
\bHam_{\gamma,t}^{\mathrm{H}}(t_0)	=
l_{\gamma(t_0)}^{-1}\circ\Ham_t^{\mathrm{H}}(t_0)\circ l_{\gamma(t_0)}	=
\mor{U}_\gamma^{-1}(t,t_0)\circ H_\gamma(t)\circ \mor{U}_\gamma(t,t_0)
	\end{equation}
the latter of which is exactly the one entering in~(\ref{7.10}).

	Now it is a trivial verification that the mappings
$\mor{A}_{\gamma,t}^{\mathrm{H}}(t_0)$ satisfy the bundle Heisenberg
equation of motion~(\ref{7.10}).

	Thus the both approaches, Hilbert bundle and Hilbert space ones, are
self-consistent and lead to one and the same final result, the bundle
Heisenberg picture of motion.

\subsubsection{Summary and inferences}
	\label{VII.2.3}

	According to the above results, in the bundle Heisenberg picture the
state of a quantum system is represented by a time\ndash independent bundle
state vector
	\begin{equation}	\label{7.13a}
\Psi_{\gamma,t}^{\mathrm{H}}(t_0) =
			\Psi_{\gamma}(t_0) \in \bHil_{\gamma(t_0)}
	\end{equation}
and every dynamical variable $\dyn{A}$ is described by a time-depending
mapping
	\begin{equation}	\label{7.15new}
\mor{A}_{\gamma,t}^{\mathrm{H}}(t_0) :=
\mor{U}_\gamma^{-1}(t,t_0)\circ \mor{A}_\gamma(t)\circ \mor{U}_\gamma(t,t_0)
\in (\proj_{0}^{\base})^{-1}(\gamma(t_0))
	\end{equation}
from the fibre over $\gamma(t_0)$ of the bundle
$\morf_\base\HilB=(\bHil_0^\base,\proj_{0}^{\base},\base)$ of
point\ndash restricted morphisms over $\base$ of $\HilB$.%
\footnote{%
For the notation and mathematical details, see subsections~\ref{VII.2.1}
and~\ref{III.1}.%
}
Here $\gamma\colon J\to\base$ is a path in the base $\base$, $t\in J$ is
arbitrary, and $t_0\in J$ is arbitrarily fixed and interpreted as an initial
moment at which the initial conditions determining the system's state and
dynamical variables are supposed to be known.

	By virtue of~\eref{7.5}, \eref{7.13a}, and~\eref{6.1''}, the mean
values of a dynamical variable $\dyn{A}$ are independent of the way of their
calculation:
	\begin{equation}	\label{7.15new-1}
\langle \mor{A} \rangle_{\Psi}^{t,\gamma} =
\langle \mor{A}_{\gamma,t}^{\mathrm{H}}(t_0)
		\rangle_{\Psi_\gamma^{\mathrm{H}}}^{t_0} =
\langle \ope{A} \rangle_{\Psi}^{t,\gamma} =
\langle \ope{A}_{t}^{\mathrm{H}}(t_0)
		\rangle_{\Psi_\gamma^{\mathrm{H}}}^{t_0} .
	\end{equation}
Hence the predictions of quantum mechanics are identical in the Hilbert
bundle and Hilbert space descriptions, as well as in their presentations in
the Schr\"odinger and Heisenberg pictures.

	We want to emphasize on three features of the bundle Heisenberg
picture as introduced above. First, in it the states are not represented
via state liftings as in the Schr\"odinger picture, but by a particular
bundle state vector corresponding to a concrete value of the state lifting of
the reference path $\gamma$ in the Schr\"odinger picture. Second, in it the
dynamical variables are described via (time\ndash dependent) mappings whose
domain and range is the fibre over the same fixed point of the reference path
$\gamma$  in the system's Hilbert bundle, while in the Schr\"odinger picture
the corresponding objects are liftings of paths in the bundle of restricted
morphisms of the Hilbert bundle of states. Third, the bundle Heisenberg
picture, as formulated above, is explicitly observer\ndash dependent in a
sense that it is always defined with respect to some reference path $\gamma$.%
\footnote{%
Such dependence exists also in the conventional Hilbert space description of
quantum mechanics, but it is so deeply hidden that it seem not to be
mentioned until now.%
}
This fact is in contrast to the Schr\"odinger picture which is formulated in
an observer\ndash independent way, only in terms of liftings of paths and
transports along paths in suitable bundles and in which the observer\ndash
independence is introduced via the initial conditions. Below an analogous
description for the Heisenberg picture will be found too.

	An interesting interpretation of the Heisenberg picture can be given
in the bundle $\morf_\base{\HilB}=(\bHil_0,\proj_0,\base)$ of
point\ndash restricted morphisms over $\base$ of $\HilB$
(see subsection~\ref{VII.1} or~\ref{III.1}). Since $\mor{U}$ is a
transport along paths in the bundle $\HilB$ of states, then, according
to~\eref{3.35} (see also~\cite[equation~(3.12)]{bp-TP-morphisms}),
it induces a transport $^\circ{\!}\mor{U}$ along paths in
$\morf_\base{\HilB}$ whose action on a map
$\mor{A}_\gamma(s)\colon\pi^{-1}(\gamma(s)) \to \pi^{-1}(\gamma(s))$
along $\gamma\colon J\to\base$ is
	\begin{equation}	\label{7.13b1}
^\circ{\!}\mor{U}_\gamma(t,s) (\mor{A}_\gamma(s)) :=
\mor{U}_\gamma(t,s)\circ \mor{A}_\gamma(s)\circ \mor{U}_\gamma(s,t)
\in(\proj_{0}^\base)^{-1}(\gamma(t)).
	\end{equation}

	Comparing, from one hand, this definition with~\eref{7.4}
and, from other hand,~\eref{7.13a} with~\eref{4.4}, we obtain respectively
	\begin{align}	\label{7.13b2}
\mor{A}_{\gamma,t}^{\mathrm{H}}(t_0)
 &= {^\circ{\!}\mor{U}}_\gamma(t_0,t) (\mor{A}_\gamma(t)) ,
\\
	\label{7.17new}
\Psi_{\gamma,t}^{\mathrm{H}}(t_0)
 & = \mor{U}_\gamma(t_0,t)\Psi_\gamma(t) .
	\end{align}
Consequently the pair of transports $(\mor{U},{^\circ{\!}\mor{U}})$ along
paths is just the mapping which maps the bundle Schr\"odinger picture into
the bundle Heisenberg picture.

	A simple corollary of~\eref{7.13b2} and~\eref{3.1} is that the
Heisenberg operators $\mor{A}_{\gamma,t}^{\mathrm{H}}(t_0)$ are connected by
	\begin{equation}	\label{7.13b3}
\mor{A}_{\gamma,t}^{\mathrm{H}}(t_1)  =
{^\circ{\!}}\mor{U}_{\gamma,t}(t_1,t_0) \mor{A}_{\gamma,t}^{\mathrm{H}}(t_0)
	\end{equation}
for every initial moments $t_1,t_0\in J$. Obviously, the map
\(
\mor{A}_{\gamma,t}^{\mathrm{H}}\colon t_0\mapsto
	\mor{A}_{\gamma,t}^{\mathrm{H}}(t_0)
\)
for every $t_0\in J$ is a lifting of $\gamma$ from $\base$ to the bundle
space of the bundle $\morf_\base\HilB$. Therefore the mapping
$\mor{A}^{\mathrm{H}}\colon\gamma\mapsto \mor{A}_{\gamma,t}^{\mathrm{H}}$ is
a lifting of paths in $\morf_\base\HilB$,
	\begin{equation}	\label{7.18new}
\mor{A}^{\mathrm{H}} \in \PLift(\morf_\base\HilB),
	\end{equation}
which, by virtue of~\eref{7.13b3} is $^\circ{\!}\mor{U}$-transported along
every path $\gamma$. Comparing~\eref{7.13b2} and~\eref{3-2.20new}, we infer
that $\mor{A}^{\mathrm{H}}$ coincides with the lifting
$^\circ{\!}\overline{\mor{U}}\in\PLift(\morf_\base\HilB)$ generated by
$^\circ{\!}\mor{U}$ (see definition~\ref{3-Defn2.4}). This observation
allows to be found a new form of the equations of motion for the observables
in the Heisenberg picture which replaces the Schr\"odinger equation in it.

	Let $^\circ{\!}\mor{D}$ be the derivation along paths generated by
the induced transport $^\circ{\!}\mor{U}$ (see
also~\cite{bp-normalF-LTP,bp-LTP-general}). In conformity with~\eref{3.35},
we have
\(
   {^\circ{\!}\mor{D}}\colon\gamma\mapsto{{^\circ{\!}\mor{D}}^\gamma}
   \colon s\mapsto {{^\circ{\!}\mor{D}}_s^\gamma}
\)
with
	\begin{equation}	\label{7.13b4}
{^\circ{\!}\mor{D}}_s^\gamma (\mor{A}) :=
\lim_{\varepsilon\to0}
\Bigl\{
\frac{1}{\varepsilon}
\left[
{^\circ{\!}\mor{U}}(s,s+\varepsilon)
\left( \mor{A}_\gamma(s+\varepsilon) \right) - \mor{A}_\gamma(s)
\right]
\Bigr\}  .
	\end{equation}

	A simple calculation shows that in a local field of bases the matrix
of $^\circ{\!}\mor{D}_s^\gamma (\mor{A})$, in accordance with~\eref{3.37} is
	\begin{equation}	\label{7.13b5}
\pmb{\boldsymbol{[}}
^\circ{\!}\mor{D}_s^\gamma (\mor{A})
\pmb{\boldsymbol{]}} =
-\left[
\mmor{A}_\gamma(s),\boldsymbol{\Gamma}_\gamma(s)
\right]_{\_}
+
\frac{\partial \mmor{A}_\gamma(s)}{\partial s}
	\end{equation}
where
\(
\boldsymbol{\Gamma}_\gamma(s) :=
\left[\Gamma_{\ a}^{b}(s;\gamma)\right]:=
\left.\partial\mmor{U}_\gamma(s,t) /\partial t \right|_{t=s}
\)
is the matrix of the coefficients of $\mor{U}$
(not of $^\circ{\!}\mor{U}$!).
From here, using~\eref{5.10} and~\eref{6.2''}, one, after some matrix
algebra, finds the explicit form of~\eref{7.13b4}:
	\begin{equation}	\label{7.13b6}
^\circ{\!}\mor{D}_t^\gamma (\mor{A}) =
\iih [ \mor{A}_\gamma(t) , \mor{H}_\gamma(t) ]_{\_} +
\left(\frac{\partial \ope{A}}{\partial t}\right)_{\!\gamma(t)},
	\end{equation}
where the last term is defined via~\eref{6.0} and $\mor{H}$ is the bundle
Hamiltonian, given by~\eref{6.2'}.

	The last result, together with~\eref{7.4}, shows that the Heisenberg
equation of motion~\eref{7.10} is equivalent to
	\begin{equation}	\label{7.13b7}
\frac{\partial \mor{A}_{\gamma,t}^{\mathrm{H}}(t_0) } {\partial t} =
\mor{U}_\gamma(t_0,t)\circ
\left(
^\circ{\!}\mor{D}_t^\gamma ( \mor{A} )
\right)
\circ \mor{U}_\gamma(t,t_0).
	\end{equation}
By the way, this equation is also an almost trivial corollary
of~\eref{7.13b4}, \eref{7.13b2}, and~\eref{3.1}. But such a `quick'
derivation leaves the problem for the relation (equivalence) between
equation~\eref{7.13b7} and {7.10} open.

	Now the analogue of~\eref{5.14} is
	\begin{equation}	\label{7.13b8}
^\circ{\!}\mor{D}_t^\gamma \circ ( \overline{^\circ{\!}\mor{U}} ) = 0,
\quad
\overline{ ^\circ{\!}\mor{U} }^\gamma(t_0,t_0)
	= \id_{\proj_{0}^{-1}(\gamma(t_0))}.
	\end{equation}
From here and~\eref{7.13b3}, we derive the equation of motion as
	\begin{equation}	\label{7.13b9}
^\circ{\!}\mor{D}_{t_0}^\gamma
\left( \mor{A}^{\mathrm{H}} \right) =0
	\end{equation}
which is another equivalent form of~\eref{7.10} or~\eref{7.13b7}.

	 Since $\gamma\colon J\to\base$ and $t\in J$ are arbitrary, the last
equation is equivalent to
	\begin{equation}	\label{7.24new}
^\circ{\!}\mor{D}(\mor{A}^{\mathrm{H}}) = 0.
	\end{equation}
This is the \emph{bundle Heisenberg equation of motion} (for the observables)
which replaces the bundle Schr\"odinger equation~\eref{5.24new} in the
Heisenberg picture. It does not depend on the reference path $\gamma$ and, in
this sense, is observer\ndash independent. As in the Schr\"odinger picture
(see Subsect.~\ref{VII.1}), here the observer\ndash dependence is introduced
by the initial conditions at some moment $t_0\in J$. This is clearly seen
from~\eref{7.13b8}  regardless of the fact that the equation of this
initial\ndash value problem can be rewritten as
	\begin{equation}	\label{7.24new-1}
^\circ{\!}\mor{D}( \overline{\mor{U}} ) = 0
		\end{equation}
which is independent of the reference path $\gamma$.

	The Heisenberg bundle state vector~\eref{7.17new} admits a treatment
analogous to the one of $\mor{A}_{\gamma,t}^{\mathrm{H}}(t_0)$. Indeed,
define a lifting of paths
	\begin{equation}	\label{7.18new-1}
\Psi^{\mathrm{H}}\in\PLift\HilB
	\end{equation}
by $\Psi^{\mathrm{H}}\colon\gamma\mapsto\Psi_{\gamma,t}^{\mathrm{H}}$
with
\(
\Psi_{\gamma,t}^{\mathrm{H}}\colon t_0\mapsto
	\Psi_{\gamma,t}^{\mathrm{H}}(t_0),
\)
 $t_0\in J$. By virtue of~\eref{7.13b2}, this lifting is
$\mor{U}$\ndash transported along every path $\gamma$, coincides with the
lifting $\overline{\mor{U}}$ generated by the evolution transport $\mor{U}$
(see definition~\ref{3-Defn2.4}), and, in conformity with~\eref{3-2.20}
satisfies the equation
	\begin{equation}	\label{7.24new-2}
\mor{D}(\Psi^{\mathrm{H}}) = 0
	\end{equation}
with $\mor{D}$ being the derivation along paths generated by $\mor{U}$.
\emph{Pro forma} the last equation coincides with the bundle Schr\"odinger
equation~\eref{5.24new} but its meaning is completely different: in the
Heisenberg picture the system's state is described by a solution
of~\eref{7.24new-2} at a \emph{single} point of some path, while in the
Schr\"odinger picture the state is represented via the solution
of~\eref{5.24new} along a whole path, \ie in the former case the state is
given by a fixed bundle state vector, while in the latter one via a lifting
of paths.

	Now a brief comment on the beginning of Subsect.~\ref{VII.2.1} is in
order. It was shown that in a normal frame~\eref{7.3new}, described via some
of the conditions~\eref{7.3}, the matrix elements of an observable lifting
of paths in
the Schr\"odinger picture coincide with the ones in the Heisenberg picture,
\(
\widetilde{ (\mor{A}(t)) }_{ab}
 =
( \mor{A}_{\gamma,t}^{\mathrm{H}}(t_0) )_{ab}.
\)
In this frame the components of a bundle state vector $\Psi_\gamma(t)$ are
\(
\widetilde{\Psi}_{\gamma}^{a}(t)
 = \Sprindex[\bigl( \mor{U}_{\gamma}^{-1}(t,t_0) \bigr)]{b}{a}
	\Psi_{\gamma}^{b}(t)
 = \Sprindex[\bigl( \mor{U}_{\gamma}(t_0,t) \bigr)]{b}{a}
	\Psi_{\gamma}^{b}(t)
 = \Psi_{\gamma}^{b}(t_0)
 = \bigl( \Psi_{\gamma,t}^{\mathrm{H}}(t_0) \bigr)^{a}.
\)
These results can be expressed by the assertion that in a normal frame the
Schr\"odinger picture of motion is identical with the Heisenberg one.

\subsection {`General' picture}
\label{VII.3}

	The Schr\"odinger and Heisenberg pictures for describing a quantum
system are not the only possible ones. Any transformation of the state
vectors and observables preserving the scalar products leads to a new
`picture' . For investigating different problems, different such pictures may
turn to be suitable. Below we present the general scheme by means of which
such special representations of the quantum\ndash mechanical motion are
generated.

\subsubsection{Introduction}
	\label{VII.3.1}

	The idea of a particular picture of motion is the simultaneous
transformation of the (bundle) state vectors and the observables (observable
liftings) in such a way that the scalar products remain unchanged. As a
consequence of this, the physically predictable results of the theory are
identical with the ones before the transformation. Formally one should
proceed as follows.

	Let $\mor{V}$ be a `two-point' lifting of paths in
$\morf_\base\HilB$, \ie for every $\gamma\colon J\to\base$, we have
$\mor{V}\colon\gamma\mapsto\mor{V}_\gamma$ with
$\mor{V}_\gamma\colon(s,t)\mapsto\mor{V}_\gamma(s,t)$ where
$\mor{V}_\gamma(s,t)\colon\bHil_{\gamma(s)}\to\bHil_{\gamma(t)}$.%
\footnote{%
Example of such map $\mor{V}$ is a transport along paths in
$\morf_\base\HilB$.%
}
Suppose the maps
\(
\mor{V}_\gamma(t,s)\colon \bHil_{\gamma(s)}\to \bHil_{\gamma(t)},\ s,t\in J
\)
are linear, of class $C^1$, and unitary, i.e.\ (see~(\ref{4.12d}))
$\mor{V}_{\gamma}^{\ddag}(t,s)=\mor{V}_{\gamma}^{-1}(s,t)$,
where $\mor{V}_{\gamma}^{-1}(s,t)$ is the \emph{left} inverse of
$\mor{V}_{\gamma}(s,t)$.%
\footnote{%
Every unitary (and hence Hermitian) linear transport along paths in
$\Morf_\base\HilB$ provides an example of $\mor{V}$ with the required
properties. In particular, for $\mor{V}$ can be taken the transport
${^\circ}{\!}L$ associated to some unitary linear transport $L$ along paths
in $\HilB$. The choice $L=\mor{U}$, $\mor{U}$ being the evolution transport,
returns us to the Heisenberg picture --- \emph{vide infra}.%
}
A simple calculation shows that
	\begin{align}	\label{7.14new}
\langle\Psi_\gamma(t) | \Psi_\gamma(t)\rangle_{\gamma(t)}
 &=
\langle\Psi_{\gamma,t}^{\mor{V}}(t_1) |
		\Psi_{\gamma,t}^{\mor{V}}(t_1)\rangle_{\gamma(t_1)}
\\	\label{7.14}
\langle \mor{A}_\gamma(t) \rangle_{\Psi_\gamma}^{t}
 &=
\langle \mor{A}_{\gamma,t}^\mor{V}(t_1)
				\rangle_{\Psi_{\gamma,t}^\mor{V}}^{t_1} ,
	\end{align}
where~(\ref{4.11}) was used, $t_1\in J$, and
	\begin{align}	\label{7.15}
\Psi_{\gamma,t}^\mor{V}(t_1) &:=
\mor{V}_\gamma(t_1,t)\Psi_\gamma(t)\in \bHil_{\gamma(t_1)},
\\				\label{7.16}
\mor{A}_{\gamma,t}^\mor{V}(t_1) &:=
\mor{V}_\gamma(t_1,t) \circ \mor{A}_\gamma(t) \circ
\mor{V}_\gamma^{-1}(t_1,t)\colon
\bHil_{\gamma(t_1)} \to \bHil_{\gamma(t_1)}.
	\end{align}

	Thereof the pairs $(\Psi_\gamma(t),\mor{A}_\gamma(t))$
and $(\Psi_{\gamma,t}^\mor{V}(t_1),\mor{A}_{\gamma,t}^\mor{V}(t_1))$
provide a completely equivalent description of a given quantum system as the
physical predictable results on their base are identical. The latter way of
describing a quantum system will be called the \emph{$\mor{V}$-picture} or
\emph{general picture} of motion. For $t_1=t$ and
$\mor{V}_\gamma(t,t)=\id_{\bHil_{\gamma(t)}}$ it coincides with the
Schr\"odinger picture and for $t_1=t_0$ and
$\mor{V}_\gamma(t_0,t)=\mor{U}_\gamma(t_0,t)$
it reproduces the Heisenberg picture.

	The analogues of~(\ref{7.14}), (\ref{7.15}), and~(\ref{7.16}) in the
Hilbert space space description in the Hilbert space $\Hil$, which is the
typical fibre of the Hilbert bundle $\HilB$ of states, are respectively:
	\begin{align}	\label{7.17}
\langle \ope{A}(t) \rangle_{\psi}^{t} &=
\langle \ope{A}_{t}^{\mor{V}}(t_1) \rangle_{ \psi_{t}^{\mor{V}} }^{t_1}
\bigl( = \langle \mor{A}_\gamma(t) \rangle_{\Psi_\gamma}^{t} \bigr),
\\	\label{7.18}
\psi_{t}^{\ope{V}}(t_1) &:= \ope{V}(t_1,t)\psi(t) \in \Hil,
\\	\label{7.19}
\ope{A}_{t}^{\ope{V}}(t_1) &:=
\ope{V}(t_1,t) \circ \ope{A}(t) \circ \ope{V}^{-1}(t_1,t) \colon  \Hil\to\Hil
	\end{align}
where $\ope{V}(t_1,t) \colon  \Hil\to\Hil$ is the linear and unitary,
i.e.\ $\ope{V}^\dag(t_1,t) = (\ope{V}(t,t_1))^{-1}$, operator corresponding
to the map
\( \mor{V}(t_1,t)\colon \bHil_{\gamma(t)}\to \bHil_{\gamma(t_1)} \)
via (cf.~(\ref{4.7}))
	\begin{equation}	\label{7.20}
\mor{V}_\gamma(t_1,t)
= l_{\gamma(t_1)}^{-1}\circ \ope{V}(t_1,t)\circ l_{\gamma(t)}.
	\end{equation}

	 The description of the quantum evolution in the Hilbert space $\Hil$
via $\psi_{t}^{\ope{V}}(t_1)$ and $\ope{A}_{t}^{\ope{V}}(t_1)$
is the $\mor{V}$\ndash\emph{picture} of motion in $\Hil$. Besides,
due to~(\ref{4.3a}), (\ref{6.0}), and~(\ref{7.15})--(\ref{7.20}), the
following relations are valid:
	\begin{align}	\label{7.21}
\Psi_{\gamma,t}^{\ope{V}}(t_1) &:=
  l_{\gamma(t_1)}^{-1} \left( \psi_{t}^{\ope{V}}(t_1) \right) =
\Psi_{\gamma,t}^{\mor{V}}(t_1),
\\	\label{7.22}
\mor{A}_{\gamma,t}^{\ope{V}}(t_1)   &:=
l_{\gamma(t_1)}^{-1} \circ \ope{A}_{t}^{\ope{V}}(t_1) \circ l_{\gamma(t_1)} =
\mor{A}_{\gamma,t}^{\mor{V}}(t_1).
	\end{align}
According to~(\ref{7.20})--(\ref{7.22}), the sets of
equalities~(\ref{7.14})--(\ref{7.16}) and~(\ref{7.17})--(\ref{7.19})
are equivalent; they are, respectively, the Hilbert bundle and the
(usual) Hilbert space descriptions of the $\mor{V}$-picture of motion.

\subsubsection{Equations of motion}
	\label{VII.3.2}

	The equations of motion in the $\mor{V}$-picture cannot be obtained
directly by differentiating~(\ref{7.15}) and~(\ref{7.16}) with
respect to $t$ because derivatives like
$\partial \mor{V}_\gamma(t_1,t)/\partial t$
are not (`well') defined due to
$\mor{V}_\gamma(t_1,t)\colon \bHil_{\gamma(t)} \to  \bHil_{\gamma(t_1)} $.
They can be derived by differentiating the corresponding
to~(\ref{7.15}) and~(\ref{7.16}) matrix equations, but below we shall
describe another method which explicitly reveals the connections between the
conventional and the bundle descriptions of  quantum evolution.
	The easiest way for deriving the equations of motion in the
$\mor{V}$\ndash picture, is to transform the conventional Schr\"odinger
equations (by means of~\eref{7.18}) into the $\mor{V}$\ndash picture and
then to transform the obtained equations into their bundle versions. With
respect to the observables, a procedure similar to the one of
Subsect.~\ref{VII.2.1} should be followed.

	Differentiating~(\ref{7.18}) with respect to $t$, substituting
into the so-obtained result the Schr\"odinger equation~(\ref{2.5}),
and introducing the modified Hamiltonian
	\begin{align}	\label{7.23}
\widetilde{\Ham}(t) &:= \Ham(t) - \,_\ope{V}\!\Ham(t_1,t),
\\	\label{7.23a}
_\ope{V}\! \Ham(t_1,t) &:=
\ih \frac{\partial\ope{V}^{-1}(t_1,t)}{\partial t} \circ \ope{V}(t_1,t) =
-\ih \ope{V}^{-1}(t_1,t) \circ \frac{\partial\ope{V}(t_1,t)}{\partial t},
	\end{align}
we find the \emph{equation of motion} for the state vectors in the
$\mor{V}$-picture as
	\begin{equation}	\label{7.24}
\ih \frac{\partial\psi_{t}^{\ope{V}}(t_1)}{\partial t} =
\widetilde{\Ham}_{t}^{\ope{V}}(t_1) \psi_{t}^{\ope{V}}(t_1).
	\end{equation}
Here
	\begin{equation}	\label{7.25}
\widetilde{\Ham}_{t}^{\ope{V}}(t_1) =
\ope{V}(t_1,t) \circ \widetilde{\Ham}(t) \circ \ope{V}^{-1}(t_1,t) =
\Ham_{t}^{\ope{V}}(t_1) - \,_\ope{V}\!\Ham_{t}^{\ope{V}}(t_1) ,
	\end{equation}
where
	\begin{equation}	\label{7.26}
	\begin{split}
\Ham_{t}^{\ope{V}}(t_1) &:=
	\ope{V}(t_1,t) \circ {\Ham}(t) \circ \ope{V}^{-1}(t_1,t),
\\
\,_\ope{V}\!\Ham_{t}^{\ope{V}}(t_1) &:=
\ope{V}(t_1,t) \circ _\ope{V}\!\Ham(t_1,t) \circ \ope{V}^{-1}(t_1,t) =
-\ih \frac{\partial \ope{V}(t_1,t)}{\partial t} \circ \ope{V}^{-1}(t_1,t) ,
	\end{split}
	\end{equation}
is the $\mor{V}$-form of~(\ref{7.23}).

	The \emph{equation of motion for the observables in the
$\mor{V}$-picture} in $\Hil$ is obtained in an analogous way.
Differentiating~(\ref{7.19})  with respect to $t$ and applying~(\ref{7.26}),
we find
	\begin{equation}	\label{7.27}
\ih \frac{\partial\ope{A}_{t}^{\ope{V}}(t_1)}{\partial t} =
\left[
\ope{A}_{t}^{\ope{V}}(t_1), _\ope{V}\!\Ham_{t}^{\ope{V}}(t_1)
\right]_{\_}
+
\ih\left( \frac{\partial\ope{A}(t)}{\partial t} \right)
_{\!t}^{\!\ope{V}} (t_1) .
	\end{equation}

	The bundle equations of motion in the $\mor{V}$-picture are
corollaries of the already obtained ones in $\Hil$. In fact,
differentiating the first equalities from~(\ref{7.21}) and~(\ref{7.22})
with respect to $t$ and then using~(\ref{7.24}), (\ref{7.27}),
(\ref{7.21}), and~(\ref{7.22}), we, respectively, get:
	\begin{align}	\label{7.28}
\ih \frac{\partial \Psi_{\gamma,t}^{\mor{V}}(t_1) }{\partial t}  &=
\widetilde{\bHam}_{\gamma,t}^{\mor{V}}(t_1) \Psi_{\gamma,t}^{\mor{V}} (t_1),
\\	\label{7.29}
\ih \frac{\partial \mor{A}_{\gamma,t}^{\mor{V}}(t_1) }{\partial t}     &=
\left[
\mor{A}_{\gamma,t}^{\mor{V}}(t_1), _\mor{V}\!\bHam_{\gamma,t}^{\mor{V}}(t_1)
\right]_{\_}
+
\ih
\left(
\frac{\partial \ope{A}(t)} {\partial t}
\right)_{\!\gamma,t}^{\!\mor{V}}(t_1).        \ \ \ \ \
	\end{align}
Here
	\begin{equation}	\label{7.28-29}
	\begin{split}
\widetilde{\bHam}_{\gamma,t}^{\mor{V}}(t_1) &=
l_{\gamma(t_1)}^{-1} \circ \widetilde{\Ham}_{t}^{\ope{V}}(t_1) \circ
l_{\gamma(t_1)} =
\mor{V}_\gamma(t_1,t) \circ \widetilde{\bHam}_\gamma(t) \circ
\mor{V}_{\gamma}^{-1}(t_1,t) ,
\\
_\mor{V}\!\bHam_{\gamma,t}^{\mor{V}}(t_1) &=
l_{\gamma(t_1)}^{-1} \circ  _\ope{V}\!\Ham_{t}^{\ope{V}}(t_1) \circ
l_{\gamma(t_1)} =
\mor{V}_\gamma(t_1,t) \circ _\ope{V}\!\bHam_\gamma(t_1,t) \circ
\mor{V}_{\gamma}^{-1}(t_1,t),
	\end{split}
	\end{equation}
where
\(
\widetilde{\bHam}_\gamma(t) :=
l_{\gamma(t)}^{-1} \circ \widetilde{\Ham}(t)  \circ l_{\gamma(t)}
\)
and
\(
_\ope{V}\!\bHam_\gamma(t_1,t) =
-\ih
\mor{V}^{-1}(t_1,t) \circ l_{\gamma(t_1)} \circ
\frac{\partial \ope{V}(t_1,t)}{\partial t} \circ l_{\gamma(t)},
\)
are, respectively, the modified and `additional' Hamiltonians in the
$\mor{V}$-picture (cf.~(\ref{7.23}), (\ref{7.16}), and~(\ref{7.22})).

\subsubsection{Evolution operator and transport}
	\label{VII.3.3}

	In the  $\mor{V}$-picture the evolution operator $\ope{U}^{\ope{V}}$
in $\Hil$ and evolution transport $\mor{U}^{\mor{V}}$ in $\HilB$ are define,
respectively, by (cf.~(\ref{2.1}) and~(\ref{4.4}))
	\begin{align}	\label{7.30}
\psi_{t}^{\ope{V}}(t_1) &=
		\ope{U}^\ope{V}(t,t_1,t_0)\psi_{t_0}^{\ope{V}}(t_1),
\\	\label{7.31}
\Psi_{\gamma,t}^{\mor{V}}(t_1 ) &=
\mor{U}_\gamma^\mor{V}(t,t_1,t_0)\Psi_{\gamma,t_0}^{\mor{V}}(t_1).
	\end{align}
Due to~(\ref{7.24}) and~(\ref{7.28}), they satisfy the
following initial-value problems:
	\begin{align}	\label{7.32}
\ih \frac{\partial \ope{U}^\ope{V}(t,t_1,t_0) }{\partial t} &=
\widetilde{\Ham}_{t}^{\ope{V}}(t_1)\circ\ope{U}^\ope{V}(t,t_1,t_0),&
\ope{U}^\ope{V}(t_0,t_1,t_0) &= \id_\Hil,
\\				\label{7.33}
\ih \frac{\partial \mor{U}_\gamma^\mor{V}(t,t_1,t_0) }{\partial t} &=
\widetilde{\bHam}_{\gamma,t}^{\mor{V}}(t_1)\circ
				\mor{U}_\gamma^\mor{V}(t,t_1,t_0),
&
\mor{U}_\gamma^\mor{V}(t_0,t_1,t_0) &= \id_{\bHil_{\gamma(t_1)}}.
	\end{align}

	The relations between the evolution operator or evolution transport
in the Schr\"odinger picture and $\mor{V}$\ndash picture can be found as
follows.  From one hand, combining~(\ref{7.30}) , (\ref{7.18}),
and~(\ref{2.1}) and, from another hand, using~(\ref{7.31}), (\ref{7.15}),
and~(\ref{4.4}), we respectively obtain:
	\begin{align}	\label{7.34}
\ope{U}^\ope{V}(t,t_1,t_0) &= \ope{V}(t_1,t)\circ \ope{U}(t,t_0)\circ
				\ope{V}^{-1}(t_1,t_0)\colon \Hil\to\Hil,
\\			\label{7.35}
\mor{U}_\gamma^\mor{V}(t,t_1,t_0) &=
\mor{V}_\gamma(t_1,t)\circ \mor{U}_\gamma(t,t_0)\circ
				\mor{V}_\gamma^{-1}(t_1,t_0)
\colon \bHil_{\gamma(t_1)}\to \bHil_{\gamma(t_1)}.
	\end{align}

	Notice, in the Heisenberg picture, we have
	\begin{equation}	\label{7.35a}
\ope{U}^\mathrm{H}(t,t_0,t_0) =  \id_\Hil ,
\qquad
\mor{U}_\gamma^\mathrm{H}(t,t_0,t_0) =  \id_{\bHil_{\gamma(t_0)}}.
	\end{equation}

	Substituting in~(\ref{7.35}) the equalities~(\ref{7.20})
and~(\ref{4.7}) and taking into account~(\ref{7.34}), we find the
connection between the evolution operator and transport in the
$\mor{V}$-picture as
	\begin{equation}	\label{7.36}
\mor{U}_\gamma^\mor{V}(t,t_1,t_0) =
l_{\gamma(t_1)}^{-1}\circ \ope{U}^\ope{V}(t,t_1,t_0)  \circ l_{\gamma(t_1)}.
	\end{equation}

\subsubsection{Interaction interpretation}
	\label{VII.3.4}

	The derived here equations of motion have a direct practical
applications in connection with the approximate treatment of the problem of
quantum evolution of state vectors and observables (cf.~\cite[ch.~VIII,
\S~14]{Messiah-QM}). Indeed, by~(\ref{7.23}) is fulfilled
$\Ham(t) = _\ope{V}\!\Ham(t,t_1) + \widetilde{\Ham}(t) $.
We can consider
	\begin{equation}	\label{7.37}
\Ham^{(0)}(t) := _\ope{V}\!\Ham(t_1,t) =
\ih \frac{\partial \ope{V}^{-1}(t_1,t)}{\partial t} \circ \ope{V}(t_1,t)
	\end{equation}
as a given approximate (unperturbed) Hamiltonian of the quantum system with
evolution operator
$\ope{U}^{(0)}(t_1,t) = \ope{V}(t_1,t)$.
(In this case $\bHam^{(0)}(t)$ is independent of $t_1$ and
$\ope{V}^{-1}(t_1,t) = \ope{V}(t,t_1).$)
	Then $\widetilde{\Ham}(t)$ may be regarded, in some `good'
cases, as a `small' correction to $\bHam^{(0)}(t)$. In other words, we can
say that $\bHam^{(0)}(t)$ is the Hamiltonian of the `free' system, while
$\Ham(t)$ is its Hamiltonian when a given interaction with Hamiltonian
$\widetilde{\Ham}(t)$ is introduced.

	In this interpretation the $\mor{V}$-picture is the well known
\emph{interaction picture}.
In it one supposes to be given the basic (zeroth order) Hamiltonian
$\Ham^{(0)}(t) := \mbox{$_\ope{V}\!\Ham(t,t_1)$}$
and the interaction Hamiltonian
$\Ham^{(I)}(t) = \widetilde{\Ham}(t)$.
On their base can be computed all other quantities of the system described by
them. In particular, all of the above results hold true for
\(
\ope{V}(t_1,t) = \ope{U}^{(0)}(t_1,t) =
\Texp\Bigl(
\int\limits_{t}^{t_1} \Ham^{(0)}(\tau) d\tau / \ih \Bigr)
\).
	Besides, in this case the total evolution operator
\(
\ope{U}(t,t_0) =
\Texp\Bigl(
\int\limits_{t_0}^{t}\Ham(\tau) d\tau / \ih \Bigr)
\)
splits into
	\begin{equation}	\label{7.38}
\ope{U}(t,t_0) =
\ope{U}^{(0)}(t,t_0) \circ \ope{U}^{(I)}(t,t_0)
	\end{equation}
with
\(
\ope{U}^{(I)}(t,t_0) :=
\Texp\Bigl(
\int\limits_{t_0}^{t}
\left(\Ham^{(I)}\right)_{\tau}^{\ope{U}^{(0)}(t_0)}
d\tau / \ih \Bigr),
\)
where
$\left(\Ham^{(I)}\right)_{\tau}^{\ope{U}^{(0)}(t_0)}$
is an operator given by~(\ref{7.25}) for
$\widetilde{\Ham}=\Ham^{(I)}$, $t_1=t_0$,
and $\ope{V}(t_0,t)=\ope{U}^{(0)}(t_0,t)$.
Now the equations of motion~(\ref{7.24}) and~(\ref{7.27}) take,
respectively, the form:
	\begin{align}	\label{7.39}
\ih \frac{\partial \psi^{(I)}(t)}{\partial t} &=
\left(\Ham^{(I)}\right)_{ \!\ope{U}^{(0)} }^{\! t}(t_0) \psi^{(I)}(t),
\\	\label{7.40}
\ih \frac{\partial \ope{A}^{(I)}(t)}{\partial t} &=
\Bigl[
\ope{A}^{(I)}(t),
\left(\Ham^{(I)}\right)_{ \!\ope{U}^{(0)} }^{\! t}(t_0)
\Bigr]_{\_}
+ \ih
\left(
\frac{\partial \ope{A}}{\partial t}
\right)^{\! \ope{U}^{(0)} }_{\! t} {\!} (t_0)
	\end{align}
where
$\psi^{(I)}(t):=\psi^{\ope{U}^{(0)} }_{t} (t_0)$
and
$\ope{A}^{(I)}(t):=\ope{A}^{\ope{U}^{(0)} }_{t} (t_0)$.
	Up to notation, the last two equations coincide respectively with
equations~(55) and~(56) of~\cite[ch.VIII, \S~15]{Messiah-QM}.

	The bundle form of the interaction interpretation of the
$\mor{V}$-picture of motion will not be presented here as an almost evident
one.

\subsubsection{Some inferences}
	\label{VII.3.5}

	Partially the conclusions of Subsect.~\ref{VII.2.3} are valid
\emph{mutatis mutandis} in the general  $\mor{V}$\ndash picture of motion. In
short, their essence is the following.

	In the bundle $\mor{V}$-picture the system's state is represented by
a, generally, time\ndash dependent bundle state vector~\eref{7.15} from a
fixed fibre over the reference path $\gamma$. The dynamical variables are
described via, generally, time\ndash depending maps acting on this single
fibre. Due to ~\eref{7.13b8}, the Schr\"odinger and the $\mor{V}$\ndash
picture  are identical from the view\ndash point of predicting physical
results.

	If $\mor{V}$ happens to be a (Hermitian linear) transport along paths
in $\Hil$, then the whole concluding part of Subsect.~\ref{VII.2.3},
beginning with the paragraph containing equation~\eref{7.13b1}, is valid
in the case of the  $\mor{V}$\ndash picture provided $\mor{V}$ is taken for
${^\circ\!\mor{U}}$, $\Psi^{\mor{V}}$ for $\Psi^{\mathrm{H}}$, and by
$\mor{D}$ is understood the generated by $\mor{V}$ derivation along paths.
(The particular choice $\mor{V}={^\circ\!\mor{U}}$ reduces the
$\mor{V}$\ndash picture to the Heisenberg one.) But if $\mor{V}$ is not a
transport along paths, the conclusions from the last part of
Subsect.~\ref{VII.2.3} cannot be applied.

\section {Integrals of motion}
\label{VIII}
\setcounter{equation} {0}

	The integrals of motion, called also constants of motion, are the
quantum\ndash mechanical analogues of the preserved quantities in classical
physics~\cite[chapter~V, \S\S~19, 20; chapter~VIII, \S~12]{Messiah-1}. They
provide invariant characteristics of a quantum system which do not change in
time. An important example of this kind is the energy of a system with
explicitly time\ndash independent Hamiltonian. In more special cases, such
quantities are the angular momentum, parity, etc.\ The aim of this section is
the development of the general formalism of integrals of motion in the bundle
version of quantum mechanics.

\subsection{Hilbert space description}
	\label{VIII.1}

	Usually~\cite[ch.~VIII, \S~12]{Messiah-QM},
\cite[\S~28]{Dirac-PQM} an
\emph{explicitly not depending on time}
dynamical variable is called an integral (or a constant) of motion
if the corresponding to it observable is time-independent in the
Heisenberg picture of motion. Due to~(\ref{7.8}) this means
	\begin{equation}	\label{8.1}
0 = \ih \frac{\partial \ope{A}_{t}^{\mathrm{H}}(t_0)}{\partial t} =
\left[ \ope{A}_{t}^{\mathrm{H}}(t_0) ,
			\Ham_{t}^{\mathrm{H}}(t_0) \right]_{\_} .
	\end{equation}
Hence, if
\(
{\partial \ope{A}(t)}/{\partial t} = 0,
\)
then $\ope{A}$ is an integral of motion if and only if it commutes with the
Hamiltonian. By virtue of~(\ref{7.7}), (\ref{7.19}), and~(\ref{7.27}), this
result is true in any picture of motion.

	If~(\ref{8.1}) holds, then
$\partial\ope{A}(t)/\partial t = 0 $ and~(\ref{7.7}) imply
	\begin{equation}	\label{8.2}
\ope{A}(t)=\ope{A}(t_0)=\ope{A}_{t}^{\mathrm{H}}(t_0)=
					\ope{A}_{t_0}^{\mathrm{H}}(t_0).
	\end{equation}
 From~(\ref{7.7}) and~(\ref{8.2}) one easily obtains
that~(\ref{8.1}) (under the assumption
\(
{\partial \ope{A}(t)}/{\partial t} = 0
\))
is equivalent to the commutativity of the observable and the evolution
operator:
	\begin{equation}	\label{8.3}
[ \ope{A}(t_0) , \ope{U}(t_0,t) ]_{\_} = 0
	\end{equation}
which, in connection with further generalizations, is better to be
written as
	\begin{equation}
	\tag{\ref{8.3}$^\prime$}	\label{8.3a}
 \ope{A}(t_0) \circ\ope{U}(t_0,t) = \ope{U}(t_0,t)\circ\ope{A}(t).
	\end{equation}

	It is almost evident that the mean values of the integrals of
motion are constant:
	\begin{equation}	\label{8.4}
\langle\ope{A}(t)\rangle_{\psi}^{t} =
\langle\ope{A}_{t}^{\mathrm{H}}(t_0)\rangle_{\psi}^{t_0} =
\langle\ope{A}_{t_0}^{\mathrm{H}}(t_0)\rangle_{\psi}^{t_0} =
\langle\ope{A}(t_0)\rangle_{\psi}^{t_0} .
	\end{equation}
In particular, if $\psi^{\mathrm{H}}(t)=\psi(t_0)$ is an eigenvector of
$\ope{A}_{t}^{\mathrm{H}}(t_0)$ with eigenvalue $a$, i.e.\
\(
\ope{A}_{t}^{\mathrm{H}}(t_0)\psi^{\mathrm{H}}(t) = a \psi^{\mathrm{H}}(t),
\)
then $a=\const$ as
$\langle \ope{A}_{t}^{\mathrm{H}}(t_0) \rangle_{\psi^{\mathrm{H}}}^{t_0} = a$.
Besides, in the Schr\"odinger picture we have
$\ope{A}(t_0)\psi(t)=a\psi(t).$

	Evidently, the identity map $\id_\Hil$, which plays the r\^ole of
the unit operator in $\Hil$, is an integral on motion. For it every state
vector is an eigenvector with $1\in\mathbb{R}$ as eigenvalue.

	Now we shall generalize the above material in the case when
\(	{\partial \ope{A}(t)}/{\partial t}	\)
may be different from zero.

	\begin{Defn}	\label{Defn8.1}
	A dynamical variable, which  may be explicitly time\ndash dependent,
is an integral (or a constant) of motion if the mean values of the
corresponding to it observable are time-independent.
	\end{Defn}

According to~(\ref{6.1''}),
(\ref{7.14}) and~(\ref{7.17}) this definition does not depend on the used
concrete picture of motion. Hence, without a lost of generality, we consider
at first the Schr\"odinger picture in $\Hil$.

	So, by definition,
$\ope{A}(t)\colon \Hil\to\Hil$ is an integral of motion if
	\begin{equation}	\label{8.5}
\langle\ope{A}(t)\rangle_{\psi}^{t} = \langle\ope{A}(t_0)\rangle_{\psi}^{t_0}
	\end{equation}
for some given instant of time $t_0$.

	Due to~(\ref{2.9}), (\ref{2.1}), (\ref{2.4}), and~(\ref{7.7}) the last
equation is equivalent to
	\begin{equation}	\label{8.6}
\ope{A}(t)=
\ope{U}(t,t_0)\circ\ope{A}(t_0)\circ\ope{U}(t_0,t) =
					\ope{A}_{t_0}^{\mathrm{H}}(t)
	\end{equation}
or to
	\begin{equation}	\label{8.7}
\ope{U}(t_0,t)\circ\ope{A}(t)=\ope{A}(t_0)\circ\ope{U}(t_0,t).
	\end{equation}
Thus~(\ref{8.3a}) remains true in the general case, when it generalizes the
commutativity of an observable and the evolution operator; in fact, in this
case we can say, by definition, that $\ope{A}$ and $\ope{U}$ commute
iff~(\ref{8.7}) holds.

	Differentiating~(\ref{8.6}) with respect to $t$ and
using~(\ref{2.7}), we see that $\ope{A}$ is an integral of motion iff
	\begin{equation}	\label{8.8}
\ih \frac{\partial \ope{A}(t)}{\partial t} +[\ope{A}(t),\Ham(t)]_{\_} = 0 .
	\end{equation}
For ${\partial \ope{A}(t)}/{\partial t}=0$ this equation reduces
to~(\ref{8.1}). Indeed, according to equations~(\ref{7.7}) and~(\ref{7.8}),
in the Heisenberg picture~(\ref{8.8}) is equivalent to
	\begin{equation}	\label{8.9}
0 =
\ih \left( \frac{\partial \ope{A}(t)}{\partial t}
					\right)_{\! t}^{\mathrm{H}}(t_0) +
 \left[ \ope{A}_{t}^{\mathrm{H}}(t_0) ,
			\Ham_{t}^{\mathrm{H}}(t_0) \right]_{\_} =
\ih \frac{\partial \ope{A}_{t}^{\mathrm{H}}(t_0)}{\partial t}
	\end{equation}
which proves our assertion. Besides, from~(\ref{8.9}) follows
	\begin{equation}	\label{8.10}
\ope{A}_{t}^{\mathrm{H}}(t_0) =
		\ope{A}_{t_0}^{\mathrm{H}}(t_0) = \ope{A}(t_0)
	\end{equation}
but now $\ope{A}(t_0)$ is generally different from $\ope{A}(t)$.
In this way we have proved that an
\emph{%
observable is an integral of motion iff in the Heisenberg picture it
coincides with its initial value in the Schr\"odinger picture.%
}

	Analogously to the explicitly time-independent case considered above,
now one can easily prove that, if some state vector is an eigenvector for
$\ope{A}$ with an eigenvalue $a$, then $\ope{A}$ is an integral of motion iff
$a$ is time\ndash independent, i.e.\ $a=\const$.%
\footnote{%
In fact, in this case we have $\ope{A}(t)\psi(t)=a(t)\psi(t)$ for $\psi(t)$
satisfying $\ih\frac{d\psi(t)}{d t}=\Ham(t)\psi(t)$. The integrability
condition for this system of equations (with respect to $\psi(t)$) is
\(
\ih\frac{\partial\ope{A}(t)}{\partial t} +[\ope{A}(t),\Ham(t)]_{\_} =
\ih\frac{d a(t)}{d t} \id_\Hil
\)
from where the above result follows.%
}

\subsection{Hilbert bundle description}
	\label{VIII.2}

	The next definition is a bundle version of definition~\ref{Defn8.1}.
\renewcommand{\theVarDefn}{\ref{Defn8.1}$^{\prime}$} 	
       \begin{VarDefn}	\label{Defn8.1'}
	A dynamical variable is called an
integral of motion if the corresponding to it observable lifting has
time-independent mean values.
	\end{VarDefn}

	So, if $\mor{A}$ is the lifting of paths corresponding to a dynamical
variable $\dyn{A}$ (see Sect.~\ref{VI}), then $\dyn{A}$ (or $\mor{A}$) is an
integral of motion iff
	\begin{equation}	\label{8.11}
\langle \mor{A} \rangle_{\Psi}^{t,\gamma} :=
\langle \mor{A}_\gamma(t)\rangle_{\Psi_\gamma}^{t} =
\langle \mor{A}_\gamma(t_0)\rangle_{\Psi_\gamma}^{t_0} =:
\langle \mor{A} \rangle_{\Psi}^{t_0,\gamma}
	\end{equation}
which, due to~(\ref{6.1''}) is equivalent (and equal) to~(\ref{8.5}).
Consequently, definitions~\ref{Defn8.1} and ~\ref{Defn8.1'} are equivalent:
$\dyn{A}$ is an integral of motion in the Hilbert space description iff it is
such in the Hilbert bundle one. Therefore we can simply say that a dynamical
variable is integral of motion if its mean values are time\ndash independent.

	From~(\ref{8.6}), (\ref{4.4}), (\ref{6.1'}), (\ref{4.11}),
(\ref{4.14}), (\ref{7.12}), and~\eref{7.13b1}, we see that
equation~(\ref{8.11}) is 	equivalent to
	\begin{equation}	\label{8.12}
\mor{A}_\gamma(t)
 = \mor{U}_\gamma(t,t_0)\circ \mor{A}_\gamma(t_0) \circ \mor{U}_\gamma(t_0,t)
 = {}^\circ\! U_\gamma(t,t_0)(\mor{A}(t_0))
 = \mor{A}_{\gamma,t_0}^{\mathrm{H}}(t)
	\end{equation}
where ${}^\circ\! U$is the associated to $\mor{U}$ transport in
$\morf_\base\HilB$.

	A feature of the Hilbert bundle description is that in it, for the
difference of the Hilbert space one, we cannot directly differentiate with
respect to $t$ maps like
\(
\mor{A}_\gamma(t)\colon \bHil_{\gamma(t)} \to \bHil_{\gamma(t)}
\)
and
\(
\mor{U}_\gamma(t,t_0)\colon \bHil_{\gamma(t_0)} \to \bHil_{\gamma(t)} .
\)
So, to obtain the differential form of~(\ref{8.12})
(or~(\ref{8.11})), we differentiate with respect to $t$ the matrix form
of~(\ref{8.12}) in a given field of bases (see Sect.~\ref{V}).
Thus, using~(\ref{5.6}), we find
	\begin{equation}	\label{8.13}
\ih \frac{\partial\mmor{A}_\gamma(t)}{\partial  t} +
\left[
{\mmor{A}_\gamma(t)} , {\mbHam_{\gamma}(t)}
\right]_{\_}
 = 0.
	\end{equation}
Because of~(\ref{5.10}) and~(\ref{6.4}), this equation is the local
matrix form of the invariant equation%
\footnote{%
Recall, the induced derivation $\widetilde{\mor{D}}$ along paths was defined
via~\eref{3.32}\Ndash\eref{3.34}.%
}
	\begin{equation}	\label{8.14}
\bigl( \widetilde{\mor{D}}_{t}^{\gamma} (\mor{A}) \bigr) (\Psi) = 0
	\end{equation}
for every state lifting $\Psi$.

	Consequently a
\emph{%
dynamical variable is an integral of motion iff the induced
derivative along paths  of the corresponding to it observable lifting has a
vanishing action on the state liftings.%
}

	If in some basis
\( \mmor{A}_\gamma(t)=\const=\mmor{A}_\gamma(t_0) \),
then, with the help of~(\ref{8.8}), we get
\(
[
\mmor{A}_\gamma(t) , \mbHam_{\gamma}(t)
]_{\_}
= 0,
\)
i.e.\ the matrix of $\mor{A}_\gamma$
and the matrix-bundle Hamiltonian commute.
It is important to note, from here does not follow the commutativity of
the maps $\mor{A}_\gamma(t)$, representing an observable by~(\ref{6.0}),
and the bundle Hamiltonian~(\ref{6.2'}) because the matrix of the latter
is connected with the matrix\ndash bundle Hamiltonian through~(\ref{6.2''}).

	If the bundle state vector $\Psi_\gamma(t)$ is an eigenvector for
$\mor{A}_\gamma(t)$, that is
\(
\mor{A}_\gamma(t)\Psi_\gamma(t) = a(t) \Psi_\gamma(t),\
a(t)\in\mathbb{R}
\),
then $\langle \mor{A}_\gamma(t)\rangle_{\Psi_\gamma}^{t} = a(t)$.
Hence from~(\ref{8.6}) follows that $\mor{A}_\gamma$  is an integral of
motion iff $a(t)=\const=a(t_0)$.

	Rewriting equation~\eref{8.13} in the form of Lax pair
equation~\cite{Lax}
	\begin{equation}	\label{Lax-equation}
\frac{\partial}{\partial t} \mmor{A}_\gamma(t) =
- \iih [ \mmor{A}_\gamma(t) ,\mHam_\gamma(t) ]_- =
[ \mmor{A}_\gamma(t) , \boldsymbol{\Gamma}_\gamma(t) ]_- ,
	\end{equation}
where~\eref{5.10} was taken into account, we see that
\emph{%
$\dyn{A}$ is an
integral of motion iff in some (and hence in any) field of bases the matrices
$\mmor{A}_\gamma(t)$ and $\boldsymbol{\Gamma}_\gamma(t)$ form a Lax pair.%
}

	It is known~\cite[sect.~2]{Rosquist&Goliath} that the Lax pair
equation~\eref{Lax-equation} is invariant under  transformations of a form
	\begin{equation}	\label{Lax-invariance}
\mmor{A}_\gamma(t) \mapsto \mmor{W}\mmor{A}_\gamma(t)\mmor{W}^{-1}, \qquad
\boldsymbol{\Gamma}_\gamma(t) \mapsto
	\mmor{W}\boldsymbol{\Gamma}_\gamma(t)\mmor{W}^{-1} -
	\frac{\partial\mmor{W}}{\partial t}\mmor{W}^{-1}
	\end{equation}
where $\mmor{W}$ is a nondegenerate matrix, possibly depending on $\gamma$
and $t$ in our case. Hence $\mmor{A}_\gamma(t)$ transforms as a tensor while
$\boldsymbol{\Gamma}_\gamma(t)$ transforms as the matrix of the coefficients
of a linear connection.  These observations fully agree with our results of
Sect.~\ref{V}, expressed by equations~\eref{5.02b} and~\eref{5.11} with
$(\boldsymbol{\Omega}^\top(t;\gamma))^{-1}=\mmor{W}$,
and give independent arguments for treating (up to a constant) the
matrix-bundle Hamiltonian as a gauge (connection) field.

	Another invariant bundle necessary and sufficient condition for a
dynamical variable to be an integral of motion can be found as follows.
In the Heisenberg picture~(\ref{8.11}) transforms to
	\begin{equation}	\label{8.15}
\langle \mor{A}_{\gamma,t}^{\mathrm{H}}(t_0) \rangle_{\Psi_\gamma}^{t_0} =
\langle \mor{A}_{\gamma,t_0}^{\mathrm{H}}(t_0) \rangle_{\Psi_\gamma}^{t_0}
	\end{equation}
which is equivalent to (cf.~(\ref{8.10}))
	\begin{equation}	\label{8.16}
\mor{A}_{\gamma,t}^{\mathrm{H}}(t_0) = \mor{A}_{\gamma,t_0}^{\mathrm{H}}(t_0)
\ \left( = \mor{A}_\gamma(t_0) \right) .
	\end{equation}
So, due to~(\ref{7.10}), an observable lifting $\mor{A}$ is an integral of
motion if and only if (cf.~(\ref{8.9}))
	\begin{equation}	\label{8.17}
0 = \ih \frac{\partial \mor{A}_{\gamma,t}^{\mathrm{H}}(t_0)}{\partial t} =
\left[
\mor{A}_{\gamma,t}^{\mathrm{H}}(t_0) , \bHam_{\gamma,t}^{\mathrm{H}}(t_0)
\right]_{\_}
+
\ih
\left(
\frac{\partial \ope{A}}{\partial t}
\right)_{\! \gamma,t}^{\!\mathrm{H}}(t_0).
	\end{equation}

	This equation, according to~\eref{7.13b7}, is equivalent to
(cf.~\eref{8.14})
	\begin{equation}	\label{8.17a}
^\circ{\!}\mor{D}_{t}^{\gamma}(\mor{A}) = 0.
	\end{equation}
Since $\gamma$ and $t$ are arbitrary, we can rewrite the last equation as
	\begin{equation}	\label{8.19new}
^\circ{\!}\mor{D}(A) = 0  .
	\end{equation}
We can paraphrase this result by saying that a dynamical variable is an
integral of motion iff the corresponding to it observable lifting is linearly
transported (along paths) by means of the transport ${^\circ{\!}\mor{U}}$
associated to the evolution transport $\mor{U}$. The last result is
explicitly expressed by~\eref{8.12}.

	Therefore a
\emph{%
dynamical variable is an integral of motion iff the corresponding to it
observable lifting of paths in $\morf_\base\HilB$ has a vanishing derivative
with respect to the derivation along paths in
$\morf_\base{\HilB}$ induces by the evolution transport.%
}
According to definition~\ref{3-Defn2.4} (see also~\eref{3-2.20new}
and~\eref{3-2.20}) and equations~\eref{8.12} and~\eref{8.17a}, the same
result can be expressed by saying that a
\emph{%
dynamical variable is an integral of motion iff the corresponding to it
observable lifting is a lifting of paths generated by the evolution
transport.%
}
Paraphrasing (see~\eref{8.12}), we can also assert that a dynamical variable
is an integral of motion iff the corresponding to it observable lifting of
paths is ${}^\circ\! U$\ndash transported along the paths in the base $\base$
of the bundle $\morf_\base\HilB$.

	Ending, we notice that the descriptions of the integrals of
motion in the Hilbert space $\Hil$ and in the Hilbert bundle $\HilB$
are completely equivalent because of~(\ref{6.0}) and~(\ref{4.3}).

\section{Mixed states}
\label{XI}
\setcounter{equation}{0}

	The most general description of the state of a quantum system in
the framework of standard quantum mechanics is provide by the so-called
density (or statistical) operator (or matrix). It allows a uniform
description of pure and mixed states. At a physical degree of rigor, this
formalism is given, for instance, in~\cite[chapter~VIII,
sect.~IV]{Messiah-QM} and~\cite[\S~33]{Dirac-PQM}; for mathematically
rigorous exposition of the problem, see~\cite[chapter~IV,
sect.~8]{Prugovecki-QMinHS}. The formalism of density operator has also a
bundle analogue which is described in the present section.

\subsection{Hilbert space description (review)}
\label{XI.1}

	Below we briefly recall the notions of a mixed state and density
operator in the conventional Hilbert space description of quantum
mechanics~\cite{Messiah-QM} (see also~\cite{Uhlmann-91a, Anandan-90a}).

	Consider a quantum system which at a moment
$t$ with a probability $p_i$ can be found in a state with a state vector
$\psi_i(t)\in\Hil$. Here $i$ belongs to some set $I$ of indices and the
statistical weights $p_i$ are assumed explicitly time-independent:
	\begin{equation}	\label{11.1}
0\le p_i \le 1,\quad \sum_{i\in I}p_i = 1,\quad
\frac{\partial p_i}{\partial t} = 0.
	\end{equation}
The state of such a system is described by the \emph{density operator}
	\begin{equation}	\label{11.2}
\rho(t) := \sum_{i\in I}\psi_i(t)
\frac{p_i}{\langle\psi_i(t)|\psi_i(t)\rangle}
\psi_i^{\dagger}(t)
\colon\Hil\to\Hil.
	\end{equation}
Here with a dagger as a superscript are denoted the dual Hermitian conjugate
vectors and spaces with respect to the inner product
$\langle\cdot|\cdot\rangle$, i.e.\  if $\psi\in\Hil$, then
$\psi^\dagger\in\Hil^\dagger$ is a map
$\psi^\dagger\colon\Hil\to\mathbb{C}$ such that
 $\psi^\dagger\colon\chi\mapsto\langle\psi|\chi\rangle$ for
 $\chi\in\Hil$, and a product like
 $\psi\chi^\dagger$,  $\psi,\chi\in\Hil$ is defined as an operator
 $\psi\chi^\dagger\colon\Hil\to\Hil$ via
 $(\psi\chi^\dagger)(\varphi):=(\chi^\dagger(\varphi))\psi
			      =\langle\chi|\varphi\rangle\psi$ for
 $\varphi\in\Hil$.%
\footnote{%
In Dirac's notation  $\psi,\ \psi^\dagger,$ and $\psi\chi^\dagger$ will look
like $|\psi\rangle,\ \langle\psi|$ and $|\psi\rangle\langle\chi|$
respectively~\cite{Dirac-PQM,Messiah-QM}. Notice that
 $(\psi\chi^\dagger)^\dagger=\chi\psi^\dagger$ corresponds to
 $( |\psi\rangle\langle\chi| )^\dagger = |\chi\rangle\langle\psi|$
in Dirac's notation.%
}
The density operator~\eref{11.2} is Hermitian, positive definite,
trace-class, and of unit trace~\cite{Messiah-QM,Prugovecki-QMinHS}. So, we
have
	\begin{equation}	\label{11.3}
\rho^\dagger(t) = \rho(t),\quad
\langle\psi(t)|\rho(t)\psi(t)\rangle \ge 0,\quad
\Tr\rho(t) = 1,
	\end{equation}
where $\Tr$ denotes the trace of an operator. Conversely, every such operator
is a density operator (in the absence of superselection
rules)~\cite[chapter~IV, subsect.~8.6]{Prugovecki-QMinHS}.

	By definition the \emph{mean (expectation) value} of an observable
$\ope{A}$ is
	\begin{equation}	\label{11.4}
\langle\ope{A}\rangle_{\rho}^{t} :=
\langle\ope{A}(t)\rangle_{\rho}^{t} :=
	\Tr( \rho(t)\circ\ope{A}(t) )
	\end{equation}
for a system whose state is described by a density operator $\rho(t)$. It is
interpreted as physically observable value of a dynamical variable $\dyn{A}$
represented by $\ope{A}$.

	The time evolution of the density operator is described by
postulating the Schr\"odinger equation for it, called also
\emph{von Neumann's or Liouville equation}:
	\begin{equation}	\label{11.5}
\ih \frac{d\rho(t)}{dt} =
[ \Ham(t),\rho(t) ]_{\_}  := \Ham(t)\circ\rho(t) - \rho(t)\circ\Ham(t)
	\end{equation}
where $\Ham(t)$ is system's Hamiltonian. If $\rho(t_0)$ is known for some
instant of time $t_0$, then
	\begin{equation}	\label{11.6}
\rho(t) = \ope{U}(t,t_0)\circ\rho(t_0)\circ\ope{U}^{-1}(t,t_0),
	\end{equation}
where $\ope{U}(t,t_0)$ is the system's evolution operator (see
Sect.~\ref{II}). In fact,~\eref{11.6} is the general solution of~(\ref{11.5})
with respect to $\rho(t)$.%
\footnote{%
Equation~(\ref{11.5}) is equivalent to the assumption that every vector
$\psi_i(t)$ in~\eref{11.2} evolves according to the Schr\"odinger
equation~(\ref{2.5}). Respectively, equation~(\ref{11.6}) is
equivalent to ~\eref{2.1} for the vectors $\psi_i(t).$%
}

	If the sum in~\eref{11.2} contains only one term or terms which are
proportional up to a phase factor to one of them, the system is said to be in
a \emph{pure state} (described by (one of) the corresponding state vector(s)
entering in~\eref{11.2}); otherwise the system's state is called
\emph{mixed}.  The equality $\rho^2(t)=\rho(t)$ is a criterion for a state to
be pure~\cite{Messiah-QM,Prugovecki-QMinHS}.

\subsection{Hilbert bundle description}
\label{XI.2}

	The fibre bundle approach to quantum mechanics developed until now
concerns only pure states of the quantum systems. In this subsection we are
going to show that it admits a natural modification which covers the bundle
description of systems' mixed states. The idea for including mixed states in
the new approach is quite simple: applying the `principle of invariance of
the observed values' (see the comments after postulates~\ref{Post6.1} in
Subsect.~\ref{VI.2}) one should `mend' the definition of a `bundle mean
value' in a way analogous to the transition from~\eref{2.9} to~\eref{11.4}
and, besides this, on the base of~\eref{11.2} postulate~\ref{Post4.2}
should appropriately be changed. A realization of such kind of procedure is
proposed in the present subsection.

\subsubsection{Heuristic considerations}
	\label{XI.2.1}

	The easiest way to introduce the bundle analogue of the density
operator $\rho(t)$ is via~(\ref{11.4}). Expressing $\ope{A}$ from~\eref{6.0}
and substituting the result into~\eref{11.4}, we find
	\begin{equation}	\label{11.7}
\langle \ope{A}(t) \rangle_{\rho}^{t} =
\langle \mor{A}_\gamma(t) \rangle_{\Rho_\gamma}^{t}
	\end{equation}
where
	\begin{equation}	\label{11.8}
\langle \mor{A}_\gamma(t) \rangle_{\Rho_\gamma}^{t} =
\Tr( \Rho_\gamma(t)\circ\mor{A}_\gamma(t) )
	\end{equation}
is the (bundle) mean value of the morphism $\mor{A}$, corresponding to
$\ope{A}$, in a state characterized by the
\emph{density lifting of paths}
\(
\Rho\colon \gamma\mapsto\Rho_\gamma \colon t \mapsto \Rho_\gamma(t)
\)
defined via (cf.~\eref{6.0})
	\begin{equation}	\label{11.9}
\Rho_\gamma(t) := l_{\gamma(t)}^{-1}\circ\rho(t)\circ l_{\gamma(t)}^{}
\colon\bHil_{\gamma(t)}\to\bHil_{\gamma(t)}.
	\end{equation}

	Meanwhile, equation~(\ref{11.7}) expresses the natural requirement
that the expectation value of a dynamical variable $\dyn{A}$ must
be independent of the (mathematical) way we calculate it.

	The density lifting has also a representation
like~(\ref{11.2}). In fact, substituting~\eref{4.3} and its Hermitian
conjugate, \ie
	\begin{equation}	\label{11.10}
\psi^\dagger(t) = \Psi_\gamma^\ddagger(t)\circ l_{\gamma(t)},
	\end{equation}
where
 $\Psi_x^\ddagger\colon\bHil_x\to\mathbb{C}$ is defined by
\(
\Psi_x^\ddagger\colon\Phi_x\mapsto \langle \Psi_x | \Phi_x \rangle_x
\)
for $\Psi_x,\Phi_x\in\bHil_x$, $x\in \base$,%
\footnote{%
Equation~\eref{11.10} is a consequence of the unitarity of $l_x$
(see~\eref{4.10b}).
}
for the vectors $\psi_i(t)$ (appearing in~\eref{11.2}) into~\eref{11.2},
we get
	\begin{equation}	\label{11.11}
\rho(t) = l_{\gamma(t)}^{}\circ\Rho_\gamma(t)\circ l_{\gamma(t)}^{-1}
	\end{equation}
which is equivalent to~\eref{11.9} with
	\begin{equation}	\label{11.12}
\Rho_\gamma(t) =
\sum_{i\in I} \Psi_{i,\gamma}(t)
\frac{p_i}{\langle\Psi_{i,\gamma}(t) | \Psi_{i,\gamma}(t)\rangle}
\Psi_{i,\gamma}^\ddagger(t).
	\end{equation}
Here a product like $\Phi_x\Psi_x^\ddagger$ is considered as an operator
$\Phi_x\Psi_x^\ddagger\colon\bHil_x\to\bHil_x$ such that
 \(
(\Phi_x\Psi_x^\ddagger)\mathrm{X}_x :=
\langle\Psi_x | \mathrm{X}_x\rangle_x \Phi_x,\
\mathrm{X}_x\in\bHil_x.
 \)

	The above results show that the transition from Hilbert space to
Hilbert bundle description of mixed states is achieved by replacing the
vectors in $\Hil$ (resp.\ the operators acting on $\Hil$) with
liftings of paths according to the general rules of sections~\ref{new-I}
and~\ref{VI}. As we shall see, this observation has a general validity.

\subsubsection{Rigorous considerations}
	\label{XI.2.2}

	Above we have made a number of implicit assumptions which should be
clearly formulated as explicit assertions. As mentioned earlier, for this end
some modifications of the basic postulates of the bundle approach to quantum
mechanics of pure states is required. Postulate~\ref{Post4.1} introduces and
describes the general mathematical arena on which the objects of bundle
version of quantum mechanics play their roles. It is so formulated that a
modification for the inclusion of mixed states in the scheme is not required.
But the situation with postulate~\ref{Post4.2} is completely different: it
concerns the concrete way by means of which the pure states are represented
on the scene provided by postulate~\ref{Post4.1}. Therefore this assertion
needs a radical change for the inclusion of mixed states in its range of
validity. In a `middle' position is postulate~\ref{Post6.1}: in it the only
referring to description of pure states is via equation~\eref{6.1'} defining
the mean values. Hence it needs only a `cosmetic' repairing.

	Taking into account the material of Subsect.~\ref{XI.2.1}, the new
forms of postulates~\ref{Post4.2} and~\ref{Post6.1} and
definition~\ref{Defn4.2}, which describe mixed and pure states equally well,
are the following.

	\begin{Post}	\label{Post11.1}
	Let $J\subseteq\mathbb{R}$ be real interval representing the period
of time in which a quantum system is investigated, $\HilB$ be its Hilbert
bundle, and $\gamma\colon J\to\base$ be a $C^1$ path in the base $\base$. In
$\HilB$ the state of the system at a moment $t\in J$ is described by a map
$\Rho$ assigning to the pair $(\gamma,t)$ a mapping
$\Rho_\gamma(t)\colon\bHil_{\gamma(t)}\to\bHil_{\gamma(t)}$
such that
	\begin{equation}	\label{11.00}
\Rho_\gamma(t) := l_{\gamma(t)}^{-1}\circ\rho(t)\circ l_{\gamma(t)}^{}
	\end{equation}
where $\rho(t)\colon\Hil\to\Hil$ is the conventional density operator
describing the system's state at a moment $t$ in the ordinary quantum
mechanics.
	\end{Post}

	\begin{Post}	\label{Post11.2}
	Let $\HilB$ be the Hilbert bundle of a quantum system,
$\gamma\colon J\to\base$, and $t\in J$. In the bundle description of
quantum mechanics, every dynamical variable $\dyn{A}$ characterizing the
system is represented by a map $\mor{A}$ assigning to the pair $(\gamma,t)$
a map $\mor{A}_\gamma(t)\colon\pi^{-1}(\gamma(t))\to\pi^{-1}(\gamma(t))$ such
that
	\begin{equation}	\label{11.01}
\mor{A}_\gamma(t) = l_{\gamma(t)}^{-1} \circ\ope{A}(t)\circ l_{\gamma(t)}
	\end{equation}
where $\ope{A}(t)\colon\Hil\to\Hil$ is the linear Hermitian operator (in the
system's Hilbert space $\Hil$) representing $\dyn{A}$ in the conventional
quantum mechanics. If at a moment $t\in J$ the system is in a state
characterized by a mapping $\Rho_\gamma(t)$, provided by
postulate~\ref{Post11.1}, the observed value of $\dyn{A}$ (or of $\mor{A}$)
with respect to $\gamma$ at a moment $t$ is equal to the mean value of
$\mor{A}_\gamma(t)$ in $\Rho_\gamma(t)$ which, by definition, is
	\begin{equation}	\label{11.02}
\langle \mor{A} \rangle_{\Rho}^{t,\gamma} =
\langle \mor{A}_\gamma(t) \rangle_{\Rho_\gamma}^{t} =
\Tr( \Rho_\gamma(t)\circ\mor{A}_\gamma(t) )
	\end{equation}
	\end{Post}

	\begin{Defn}	\label{Defn11.1}
	The Hilbert bundle (resp.\ Hilbert space) description of the (quantum
mechanics of) a quantum system is the description of its state and dynamical
variables via the mappings $\Rho$ and $\mor{A}$ (resp.\ operators $\rho$ and
$\ope{A}$) in the Hilbert bundle $\HilB$ (resp.\ Hilbert space $\Hil$).
	\end{Defn}

	Notes~\ref{Note4.1}--\ref{Note4.4} remain completely valid now, with
respect to the case of mixed states.

	The mapping $\Rho$, introduced by postulate~\ref{Post11.1}, can
naturally be considered as a lifting of paths in the bundle
$\morf_\base\HilB$ of restricted morphisms over $\base$ in the Hilbert bundle
of states $\HilB$. Actually, defining $\Rho$ as
$\Rho\colon\gamma\mapsto\Rho_\gamma$ with $\Rho_\gamma\colon
t\mapsto\Rho(t)$, we get
	\begin{equation}	\label{11.03}
\Rho\in\PLift(\morf_\base\HilB).
	\end{equation}
Evidently, the map $\Rho$ can be regarded also as a, generally,
multiple\ndash valued section along paths
$\Rho\colon\gamma\mapsto\lindex[\Rho]{\gamma}{}$ with
\(
\lindex[\Rho]{\gamma}{}\colon x\mapsto
	\{\Rho_\gamma(t) | \gamma(t)=x,\ t\in J\}
\)
for $x\in\gamma(J)$. This situation is practically the same as the one with
the mapping (state lifting/section) $\Psi$ considered in
Subsect.~\ref{new-I.3} due to which we are not going to repeat here the
corresponding discussion.

	\begin{Defn}	\label{Defn11.2}
	The unique lifting of paths $\Rho$ or (multiple-valued) section along
paths corresponding to the density operator $\rho$ from conventional quantum
mechanics will be called density lifting (of paths) or density section (along
paths)
	\end{Defn}

	For brevity, we call, by abuse of the language, a particular value of
$\Rho$, say $\Rho_\gamma(t)$, a bundle density operator (at a moment $t$
along $\gamma$).

	Analogously to~\eref{4.2new}, we have the bijective correspondences
	\begin{multline}	\label{11.04}
\text{\scshape density operator} \boldsymbol{\iff}
\text{\scshape density lifting of paths}
\\
\boldsymbol{\iff}
\text{\scshape density section along paths} .
	\end{multline}

	What concerns the observables, the transition from the description of
pure to mixed states is formally expressed through the replacement
of~\eref{6.1'} with~\eref{11.02} as a definition of the mean value. Since the
explicit definition of the mean value was not used in Sect.~\eref{VI} after
postulate~\ref{Post6.1}, all of this section after postulate~\ref{Post6.1} is
\emph{in extenso} completely valid when mixed states are investigated.
Consequently, all of the results contained in the mentioned piece of text
can and will be applied freely in what follows in the present section. In
particular, we notice the the validity of~\eref{6.1new-1} and the full
preservation of definition~\ref{Defn6.1} in the case of mixed states.

	Substituting~\eref{11.00} into~\eref{11.02} and using the invariance of
the trace of composition of maps under their cyclic permutations, we
get~\eref{11.7}, \ie
	\begin{equation}	\label{11.05}
\langle \mor{A} \rangle^{t,\gamma}_{\Rho} = \langle \ope{A} \rangle_\rho .
	\end{equation}
This simple equality is a concrete expression of the `principle of invariance of
the mean values' (Subsect.~\ref{VI.2}) which, in the present context, means
the coincidence of the mean values of a dynamical variable calculated in the
Hilbert bundle and Hilbert space descriptions of quantum mechanics. Expressed
in other words, we can assert the complete coincidence of the predictions of
the both, Hilbert bundle and Hilbert space, versions of quantum mechanics in
the case when mixed states are involved.

	If we insert~\eref{11.2} into~\eref{11.00} and take into
account~\eref{4.3a}, we obtain the representation~\eref{11.12} for the
density lifting $\Rho$. It could be interpreted as follows: if a quantum
system with probability $p_i$ has a state lifting $\Psi_i$, then this system
can be described via a density lifting given by~\eref{11.12}.

	From equation~\eref{11.12} or
from~\eref{11.3}, \eref{11.9} and~\eref{4.10b} follows that the
bundle density operator $\Rho_\gamma(t)$  in $\bHil_{\gamma(t)}$ is Hermitian,
positive definite, trace-class, and of trace one, i.e.\
	\begin{equation}	\label{11.13}
\Rho_\gamma^\ddagger(t) = \Rho_\gamma(t),\quad
	\langle
	\Psi_\gamma(t) | \Rho_\gamma(t)\Psi_\gamma(t)
	\rangle_{\gamma(t)} \ge 0,\quad
\Tr\left( \Rho_\gamma(t) \right) = 1.
	\end{equation}
In this sense, the bundle density operators are completely analogous to the
ordinary density operator but they act on the corresponding fibres over the
reference path $\gamma$, not in the Hilbert space $\Hil$.

\subsection{Time evolution and equations of motion}
	\label{XI.new}

	The time evolution of the density lifting of paths, i.e.\  the
relation between $\Rho_\gamma(t)$ and $\Rho_\gamma(t_0)$ for every
$t,t_0\in J$, can be found as follows. Substituting~\eref{11.11} for $t=t_0$
into~\eref{11.6}, then inserting the result into~\eref{11.9}, and,
at last, applying~\eref{4.7}, we get
	\begin{equation}	\label{11.14}
\Rho_\gamma(t) =
\mor{U}_\gamma(t,t_0)\circ\Rho_\gamma(t_0)\circ\mor{U}_{\gamma}^{-1}(t,t_0),
\qquad t,t_0\in J,
	\end{equation}
where $\mor{U}_\gamma(t,t_0)$ is the evolution transport from
$\bHil_{\gamma(t_0)}$ to $\bHil_{\gamma(t)}$.
Therefore the system's evolution transport $\mor{U}$ governs the time\ndash
evolution (along $\gamma$) of the mixed states via~\eref{11.14}.

	A simple verification proves that the linear map
 $\Rho_\gamma(t_0)\mapsto\Rho_\gamma(t)$, defined by~\eref{11.14},
satisfies~\eref{3.1} and~\eref{3.2}. So, freely speaking, we may say that
this is a `transport-like' map by means of which $\Rho_\gamma$  is
`transported' along $\gamma$. This is something like `representation' of the
evolution transport in the space of liftings of $\gamma$. The rigorous
study of this problem reviles that the pointed map is a
linear transport along $\gamma$ in the bundle $\morf_\base\HilB$ of
point\ndash restricted morphisms of $\HilB$.
Indeed, comparing~\eref{11.14} with~\eref{3.35} and using~\eref{4.7a}, we
see that~\eref{11.14} can equivalently be rewritten as (cf.~\eref{7.13b1})
	\begin{equation}	\label{11.06}
\Rho_\gamma(t) =   \lindex[\mor{U}]{}{\circ\!}_\gamma(t) (\Rho_\gamma(t_0))
 \in(\pi_{0}^{\base})^{-1}(\gamma(t))
	\end{equation}
where $\lindex[\mor{U}]{}{\circ\!}$ is the transport along paths in
$\morf_\base\HilB$ associated to $\mor{U}$. Consequently, the density
lifting~\eref{11.03} is $\lindex[\mor{U}]{}{\circ\!}$\ndash transported along
the paths in $\base$. Thus, with respect to the density liftings, the
transport $\lindex[\mor{U}]{}{\circ\!}$ plays the same role as the evolution
transport $\mor{U}$ with respect to the state liftings (in the case of pure
states).

	\begin{Defn}	\label{Defn11.3}
	The linear transport $\lindex[\mor{U}]{}{\circ\!}$ along paths,
associated to the evolution transport $\mor{U}$, will be called associated
evolution transport.
	\end{Defn}

	Earlier we met the associated evolution transport in
Subsect~\ref{VII.2.3} during the exploration of the bundle Heisenberg picture
of motion: we proved that it maps the Schr\"odinger observables to the
Heisenberg ones via~\eref{7.13b2}. It was established in
Subsect.~\ref{VII.2.3} that $\lindex[\mor{U}]{}{\circ\!}$ (by means of the
generated by if lifting $\lindex[\overline{\mor{U}}]{}{\circ\!}$ of paths)
satisfies the invariant equation~\eref{7.24new-1} and is a solution of the
initial\ndash value problem~\eref{7.13b8} in which
$\lindex[\mor{D}]{}{\circ\!}$ is the generate by
$\lindex[\mor{U}]{}{\circ\!}$ derivation along paths. The associated
evolution transport also appeared in Subsect~\ref{VIII.2} where we proved
that it governs the time evolution of the integrals of motion
via~\eref{8.12}.

	In conformity with definition~\ref{3-Defn2.4}, the above means that
the density lifting $\Rho$ is a lifting generated by the associate evolution
transport $\lindex[\mor{U}]{}{\circ\!}$ and, by virtue of~\eref{3-2.20},
satisfies the equation
	\begin{equation}	\label{11.07}
\lindex[\mor{D}]{}{\circ\!} (\Rho) = 0
	\end{equation}
which replace the bundle Schr\"odinger equation~\eref{5.24new} for the states
liftings in the general case of mixed states.%
\footnote{%
Equation~\eref{11.07} is equivalent to the assumption that the liftings
$\Psi_i$ in~\eref{11.12} evolve according to the bundle Schr\"odinger
equation~\eref{5.24new}.%
}
We call~\eref{11.07} \emph{bundle Schr\"odinger equation for the density
lifting} $\Rho$. This equation is covariant in a sense that it is completely
independent of any coordinates or observers (reference paths).

	Now the time evolution of a quantum system described by a density
lifting $\Rho$ can be formulated entirely in bundle terms. Let the system is
characterized by the derivation $\lindex[\mor{D}]{}{\circ\!}$ along paths in
$\morf_\base\HilB$.%
\footnote{%
Equivalently, the system can be characterized via the bundle Hamiltonian
$\mor{H}$, the evolution transport $\mor{U}$ or the generate by it (resp.\
associated to it) derivation $\mor{D}$ (resp.\ transport
$\lindex[\mor{U}]{}{\circ\!}$) along paths. Anyone of these quantities
uniquely determines (and is determined by) the derivation
$\lindex[\mor{D}]{}{\circ\!}$ generated by the associated evolution transport
$\lindex[\mor{U}]{}{\circ\!}$.%
}
Suppose the system's density lifting $\Rho_\gamma^0$ is known along some path
$\gamma\colon J\to\base$ at the point $t_0\in J$. Then the density lifting
$\Rho$ is the unique solution of~\eref{11.07} under the initial condition
	\begin{equation}	\label{11.08}
\Rho_\gamma(t_0) = \Rho_\gamma^0 .
	\end{equation}
Due to~\eref{11.06}, the explicit form of $\Rho$ is given via
	\begin{equation}	\label{11.09}
\Rho_\gamma(t)
	= \lindex[\mor{U}]{}{\circ\!}_\gamma(t,t_0) (\Rho_{\gamma}^{0})
	\end{equation}
where the associated evolution transport $\lindex[\mor{U}]{}{\circ\!}$ can be
found as, e.g., the unique solution of the initial\ndash value
problem~\eref{7.13b8}. Notice, this situation is similar to the one
described in the beginning of Subsect.~\ref{V.2} when we studied bundle
equations of motion for pure states.

	An equivalent to~\eref{11.07} differential equation, corresponding to
the evolution law~\eref{11.14}, can be derived in the following way.
Differentiating the matrix form of~\eref{11.14} with respect to $t$ and
applying equation~\eref{5.3}, we obtain the \emph{matrix-bundle
Schr\"odinger equation for the density lifting} as
	\begin{equation}	\label{11.15}
\ih\frac{d\mmor{\Rho}_\gamma(t)}{dt} =
[ \mbHam_\gamma(t) , \mmor{\Rho}_\gamma(t) ]_{\_}
	\end{equation}
with $\mbHam_\gamma(t)$ being the matrix-bundle Hamiltonian (given
by~\eref{5.2}). This equation is the matrix-bundle analogue of~\eref{11.5},
to which it is equivalent as it can be proved via the substitution of the
matrix form of~\eref{11.11} into the one of~\eref{11.5} (see
also~\eref{6.2''}). Consequently, the results of Sect.~\ref{V} show
that~\eref{11.14} gives the general solution ~\eref{11.15} with respect to
$\Rho_\gamma(t)$.

	The matrix equation~\eref{11.15} can be written in an invariant
form too. To this end we shall use the following result which is a simple
corollary of~\eref{11.15}, \eref{5.8}, and~\eref{5.10}: If $\Psi$  is a
lifting of paths in $\HilB$ and one of the equations
 $\mor{D}_{t}^{\gamma}( \Psi ) = 0$,
 $ \mor{D}_{t}^{\gamma} [\Rho (\Psi)] = 0$,
or~\eref{11.15} is valid, then the remaining two of them are equivalent.%
\footnote{%
The action  of $\Rho\in\PLift(\morf_\base\HilB)$ on $\Psi\in\PLift\HilB$ is
defined by~\eref{3.31}.%
}
From here follows that~\eref{11.15} is equivalent to the system
	\begin{subequations}	\label{11.16}
\begin{align}	\label{11.16a}
(\mor{D}_{t}^{\gamma}\circ\Rho) \Psi &= 0, \\
		\label{11.16b}
\mor{D}_{t}^{\gamma}(\Psi) &= 0.
\end{align}
	\end{subequations}

	If we denote by  $\tilde{\Rho}_\gamma(t)$ the restriction of
 ${\Rho}_\gamma(t)\colon\bHil_{\gamma(t)}\to\bHil_{\gamma(t)}$
on the set of (state) liftings of $\gamma$ which are (linearly)
transported along $\gamma$ by means of the evolution transport, \ie the ones
satisfying~\eref{11.16b}, then~\eref{11.16} is equivalent to
	\begin{equation}	\label{11.17}
\tilde{\mor{D}}( \tilde{\Rho} ) = 0
	\end{equation}
where $\tilde{\mor{D}}$ is the differentiation along paths of bundle
$\morf_\base\HilB$ of point\ndash restricted morphisms (see
equation~\eref{6.3}). The above discussion shows the equivalence
of~\eref{11.17} and~\eref{11.15}, a fact which is also an evident corollary
of~\eref{6.5} and~\eref{5.10}. The equation~\eref{11.17} is a version of the
bundle\emph{ Schr\"odinger equation~\eref{11.07} for the density liftings}.

\subsection{Representations in the different pictures of motion}
\label{XI.3}

	Let us turn now our attention to the description of mixed states in
the different pictures of motion (see Sect.~\ref{VII}).

	In the Schr\"odinger picture of the Hilbert bundle (resp.\ space)
description of quantum mechanics, which, in fact, was investigated until now
in this section, the motion of a quantum system is described by pairs like
$(\Rho_\gamma(t),\mor{A}_\gamma(t))$
(resp.\ $(\rho_\gamma(t),\ope{A}_\gamma(t))$)
of generally time-depending mappings representing the density lifting
(resp.\ operator) and some dynamical variable $\dyn{A}$.

	The transition to the \emph{Heisenberg picture} is achieved via the
general formulae~\eref{7.13b2} for the observable liftings and~\eref{7.7}
for the operators in $\Hil$. In particular, for the density lifting $\Rho$
and operator $\rho$ they, respectively, give:
	\begin{align}
		\label{11.18}
\Rho_{\gamma,t}^{\mathrm{H}}(t_0) &:=
\mor{U}_{\gamma}^{-1}(t,t_0)\circ\Rho_\gamma(t)\circ\mor{U}_\gamma(t,t_0)
={^\circ{\!}\mor{U}}_\gamma(t_0,t)(\Rho_\gamma(t))
\\	\notag
&=
\Rho_\gamma(t_0)
\colon\bHil_{\gamma(t_0)}\to\bHil_{\gamma(t_0)},
\\		\label{11.19}
\rho_{t}^{\mathrm{H}}(t_0) &:=
\ope{U}^{-1}(t,t_0)\circ\rho(t)\circ\ope{U}(t,t_0) =
\rho_\gamma(t_0)
\colon\Hil\to\Hil
	\end{align}
where~\eref{11.14} and~\eref{11.6} were used. Consequently in the
Heisenberg picture the Hilbert bundle and Hilbert space descriptions
are by means of pairs like
\(
\left(
\Rho_{\gamma,t}^\mathrm{H}(t_0),\mor{A}_{\gamma,t}^\mathrm{H}(t_0)
\right)
=
\left(
\Rho_{\gamma}(t_0),\mor{A}_{\gamma,t}^\mathrm{H}(t_0)
\right)
\)
and
\(
\left( \rho_{t}^\mathrm{H}(t_0),\ope{A}_{t}^\mathrm{H}(t_0) \right) =
\left( \rho(t_0),\ope{A}_{t}^\mathrm{H}(t_0) \right),
\)
respectively, in which the time dependence is entirely shifted from the
density liftings and operators to the observables. So, in this picture the
density liftings and operators are replaced by constant in time mappings,
while the observables' evolution is governed by the Heisenberg equation of
motion for them (see~\eref{7.10}, or~\eref{7.8}, or~\eref{7.13b9}).

	Of course, in the Heisenberg picture the mean values remain
unchanged:
	\begin{equation}	\label{11.20}
\langle\mor{A}_{\gamma,t}^{\mathrm{H}}(t_0)\rangle
_{ \Rho_{\gamma,t}^{\mathrm{H}} }^{t_0} =
\langle\ope{A}_{t}^{\mathrm{H}}(t_0)\rangle
_{ \rho_{t}^{\mathrm{H}} }^{t_0} =
\langle\mor{A}_{\gamma}^{}(t)\rangle_{ \Rho_{\gamma}^{} }^{t} =
\langle\ope{A}(t)\rangle_{ \rho }^{t},
	\end{equation}
where
	\begin{equation}	\label{11.21}
\langle\mor{A}_{\gamma,t}^{\mathrm{H}}(t_0)\rangle
_{ \Rho_{\gamma,t}^{\mathrm{H}} }^{t_0} :=
\Tr\left( \Rho_\gamma(t_0)\circ\mor{A}_{\gamma,t}^{\mathrm{H}}(t_0) \right),
\quad
\langle\ope{A}_{t}^{\mathrm{H}}(t_0)\rangle
_{ \rho_{t}^{\mathrm{H}} }^{t_0} :=
\Tr\left( \rho(t_0)\circ\ope{A}_{t}^{\mathrm{H}}(t_0) \right).
	\end{equation}
The chain equation~\eref{11.20} is a corollary of the invariance of the
trace of a product (composition) of operators with respect to a cyclic
permutation of the multipliers.

	The shift from the Schr\"odinger to `general' picture can be
achieved by means of the general equations~\eref{7.16} and~\eref{7.19}.
Hence, in the $\mor{V}$-picture of motion the density lifting $\Rho$ and
operator $\rho$ are replaced, respectively, by
	\begin{align}
	\label{11.22}
\Rho_{\gamma,t}^\mor{V}(t_1) &=
\mor{V}_\gamma(t_1,t) \circ \Rho_\gamma(t) \circ
\mor{V}_\gamma^{-1}(t_1,t)\colon
\bHil_{\gamma(t_1)} \to \bHil_{\gamma(t_1)}, \\
	\label{11.23}
\rho_{t}^\ope{V}(t_1) &=
\ope{V}_\gamma(t_1,t) \circ \rho_\gamma(t) \circ \ope{V}^{-1}(t_1,t)
\colon \Hil \to \Hil.
	\end{align}

	Since in the $\mor{V}$-picture the Hilbert bundle (resp.\ space)
description is via pairs like
\(
\left( \Rho_{\gamma,t}^\mor{V}(t_1) , \mor{A}_{\gamma,t}^\mor{V}(t_1) \right)
\)
\(\left(
\text{resp.\
\( \left( \rho_{t}^\ope{V}(t_1) , \ope{A}_{t}^\ope{V}(t_1) \right) \)%
}
\right),
\)
the mean values of the observables remain unchanged, as in the Heisenberg
picture:
	\begin{equation}	\label{11.24}
\langle \mor{A}_{\gamma,t}^\mor{V}(t_1) \rangle
_{\Rho_{\gamma,t}^\mor{V}}^{t_1} =
\langle \ope{A}_{t}^\ope{V}(t_1) \rangle _{\rho_{t}^\ope{V}}^{t_1} =
\langle \mor{A}_{\gamma}(t) \rangle
_{\Rho_{\gamma}}^{t} =
\langle \ope{A}(t) \rangle _{\rho}^{t},
	\end{equation}
where
	\begin{equation}	\label{11.25}
	\begin{split}
\langle \mor{A}_{\gamma,t}^\mor{V}(t_1) \rangle
_{\Rho_{\gamma,t}^\mor{V}}^{t_1} &:=
\Tr\left(
\Rho_{\gamma,t}^{\mor{V}}(t_1) \circ \mor{A}_{\gamma,t}^\mor{V}(t_1)
\right),
\\
\langle \ope{A}_{t}^\ope{V}(t_1) \rangle _{\rho_{t}^\ope{V}}^{t_1} &:=
\Tr\left(
\rho_{t}^\ope{V}(t_1) \circ \ope{A}_{t}^\ope{V}(t_1)
\right).
	\end{split}
	\end{equation}

	In the $\mor{V}$-picture the density liftings,
operators, and observables generally change in time. For all of them this
evolution is governed by the equations~\eref{7.29} and~\eref{7.27}, but for
the density liftings and operators they can be written in a more concrete
form. For this purpose, we have to calculate the last terms in the r.h.s.\
of~\eref{7.29} and~\eref{7.27}.

	Using~\eref{7.16} and~\eref{11.5}, we obtain
\[
\left( \frac{\partial\rho(t)}{\partial t} \right)_{\!t}^{\!\ope{V}} (t_1) =
\ope{V}(t_1,t)\circ
\frac{\partial\rho(t)}{\partial t}\circ\ope{V}^{-1}(t_1,t) =
\iih
\left[ \Ham_{t}^{\ope{V}}(t_1) , \rho_{t}^{\ope{V}}(t_1) \right]_{\_}.
\]
Analogously, applying~\eref{7.22} and the just get equation, we find
\[
\left( \frac{\partial\rho(t)}{\partial t} \right)
_{\!\gamma,t}^{\!\mor{V}} (t_1) =
l_{\gamma(t_1)}^{-1} \circ
\left( \frac{\partial\rho(t)}{\partial t} \right)_{\!t}^{\!\ope{V}} (t_1)
\circ l_{\gamma(t_1)} =
\iih\left[
\bHam_{\gamma,t}^{\mor{V}}(t_1) , \Rho_{\gamma,t}^{\mor{V}}(t_1)
\right]_{\_}.
\]
At last, substituting the above two equations into~\eref{7.29}
and~\eref{7.27}, we, respectively, get
	\begin{align}
		\label{11.26}
\ih
\frac{\partial\Rho_{\gamma,t}^{\mor{V}}(t_1)}{\partial t} &=
\left[
\widetilde{\bHam}_{\gamma,t}^{\mor{V}}(t_1) ,
\Rho_{\gamma,t}^{\mor{V}}(t_1)
\right]_{\_},	\\
		\label{11.27}
\ih
\frac{\partial\rho_{t}^{\ope{V}}(t_1)}{\partial t} &=
\left[
\widetilde{\Ham}_{t}^{\ope{V}}(t_1) ,
\rho_{t}^{\ope{V}}(t_1)
\right]_{\_}
	\end{align}
where~\eref{7.25} and~\eref{7.28-29} were taken into account. These are the
equations of motion for the density lifting and operator in the $V$-picture.

	If the evolution transport and operator are known (in the
$V$-picture), then combining, from one hand,~\eref{11.22}, \eref{11.14}
and~\eref{7.35} and, from other hand, \eref{11.23}, \eref{11.6},
and~\eref{7.34}, we get the general solution of~\eref{11.26}
and~\eref{11.27}, respectively, in the form
	\begin{align}
		\label{11.28}
\Rho_{\gamma,t}^{\mor{V}}(t_1) &=
\mor{U}_\gamma^\mor{V}(t,t_1,t_0)\circ
\Rho_{\gamma,t_0}(t_1)\circ
\left(\mor{U}_\gamma^\mor{V}(t,t_1,t_0)\right)^{-1},	\\
		\label{11.29}
\rho_{t}^{\ope{V}}(t_1) &=
\ope{U}^\ope{V}(t,t_1,t_0)\circ
\rho_{t_0}(t_1)\circ
\left(\ope{U}^\ope{V}(t,t_1,t_0)\right)^{-1}.
	\end{align}

	As one can expect, in the case of Heisenberg picture, due
to~\eref{7.35a}, these formulae reproduce~\eref{11.18} and~\eref{11.19}
respectively.

\section{Curvature of the evolution transport}
\label{IX}
\setcounter{equation} {0}

	The curvature of a linear transport along paths could be considered as
a measure of its dependence on the transportation
path~\cite{bp-LTP-Cur+Tor,bp-LTP-Cur+Tor-prop}; in particular, the
transportation of a vector by means of a curvature\ndash free transport
depends only on the initial and final points of the transportation (in the
base) but not on the concrete path along which it is performed. These results
are important for us in connection with the interpretation of the reference
path $\gamma\colon J\to\base$ as a trajectory (or world line) of some
observer. In this sense the curvature of the evolution transport is a measure
of its observer\ndash dependence and, consequently, it reflects the
observer\ndash dependence of the quantum evolution as a whole. In particular,
the evolution of a system described via curvature\ndash free evolution
transport has an absolute character in a sense that it is independent of the
reference path (observer) with respect to which it is investigated. By these
reasons, below we shall find a necessary and sufficient condition for the
evolution transport to be with zero curvature.

	At first, we define the curvature of the evolution transport
(cf.~\cite[sect.~3]{bp-LTP-Cur+Tor}).

	Let $\mathsf{D}\colon\gamma\mapsto\mathsf{D}^\gamma$,
$\gamma\colon J\to\base$ be the section-derivation along paths induced by the
derivation $\mor{D}$ along paths generated by the evolution transport
$\mor{U}$, \ie (see~\eref{3-2.17}, \eref{3-2.17*} and the comments after
them)
	\begin{equation}	\label{9.01}
\mathsf{D}^\gamma \colon
	\Sec^1\HilB|_{\gamma(J)} \to \Sec^0\HilB|_{\gamma(J)}
	\end{equation}
and if $\sigma$ is a $C^1$ section (defined in a (neighborhood of)
$\gamma(J)$) and $x\in\gamma(J)$, we have
	\begin{equation}	\label{9.02}
(\mathsf{D}^\gamma \sigma) (x) := \mor{D}_{s}^{\gamma} \Hat{\sigma},
\qquad
\gamma(s)=x,\ s\in J,\  \Hat{\sigma}:=\sigma\circ\gamma
	\end{equation}
for injective path $\gamma$ and
\(
(\mathsf{D}^\gamma \sigma) (x)
 := \{ \mor{D}_s^\gamma \Hat{\sigma} |
	\gamma(s)=x,\ s\in J \}
\)
if $\gamma$ has self\ndash intersection points.

	Let $J$ and $J'$ be real intervals and the mapping
$\eta\colon J\times J'\to\base$ be such that the paths
 $\eta(\cdot,t)\colon s\mapsto\eta(s,t)$
and
 $\eta(s,\cdot)\colon t\mapsto\eta(s,t)$
for $(s,t)\in J\times J'$ be injective, \ie without self\ndash intersections,
and
 $\eta(s,J')\cap\eta(s',J') = \eta(J,t)\cap\eta(J,t') = \varnothing$
for $(s,t),(s',t')\in J\times J'$,
 $s\not=s'$ and $t\not=t'$; so the paths $\eta(\cdot,t)$, $t\in J'$ or
$\eta(s,\cdot)$, $s\in J$ do not intersect each other.

	The curvature (operator) $R$ of the evolution transport $\mor{U}$ is
a mapping $R\colon\eta\mapsto R^\eta$ with
 $R^\eta\colon(s,t)\mapsto R^\eta(s,t)$, $(s,t)\in J\times J'$
where the mapping
	\begin{equation}	\label{9.03}
R^\eta(s,t) \colon \Sec^2\HilB|_{\eta(J,J')} \to \pi^{-1}(\eta(s,t))
	\end{equation}
is defined as follows. For every $\sigma\in\Sec^2\HilB|_{\eta(J,J')}$, we
define the sections $\sigma_1,\sigma_2\in\Sec^1\HilB|_{\eta(J,J')}$ by
	\begin{gather*}
\sigma_1(\eta(s,t)) :=
 \bigl( \mathsf{D}^{\eta(s,\cdot)} (\sigma|_{\eta(s,J')}) \bigr) (\eta(s,t))
 = \mor{D}_{t}^{\eta(s,\cdot)} (\Hat{\sigma})
	\in \pi^{-1}(\eta(s,t)),
\\
\sigma_2(\eta(s,t)) :=
 \bigl( \mathsf{D}^{\eta(\cdot,t)} (\sigma|_{\eta(J,t)}) \bigr) (\eta(s,t))
 = \mor{D}_{s}^{\eta(\cdot,t)} (\Hat{\sigma})
	\in \pi^{-1}(\eta(s,t)),
	\end{gather*}
where $\Hat{\sigma}\in\PLift\HilB|_{\eta(J,J')}$ and
$\Hat{\sigma}\colon\gamma\mapsto\Hat{\sigma}_\gamma:=\sigma\circ\gamma$
for every path $\gamma$ in $\eta(J,J')$.

	The action of the mapping~\eref{9.03} on $\sigma$ is
	\begin{multline}	\label{9.04}
(R^\eta(s,t))\sigma
:= \bigl( \mathsf{D}^{\eta(\cdot,t)}(\sigma_1|_{\eta(J,t)}) \bigr) (\eta(s,t))
- \bigl( \mathsf{D}^{\eta(s,\cdot)}(\sigma_2|_{\eta(s,J')}) \bigr) (\eta(s,t))
\\
= \mor{D}_{s}^{\eta(\cdot,t)} (\Hat{\sigma}_1)
	- \mor{D}_{t}^{\eta(s,\cdot)} (\Hat{\sigma}_2)
	\end{multline}
where $\Hat{\sigma}_1,\Hat{\sigma}_2\in\PLift\HilB|_{\eta(J,J')}$
and
\(
\Hat{\sigma}_\alpha\colon\gamma\mapsto(\Hat{\sigma}_\alpha)_\gamma
 := \sigma_\alpha\circ\gamma,
\)
 $\alpha=1,2$, for every path $\gamma$ in $\eta(J,J')$.

	Symbolically, by abuse of the notation, one may write
	\begin{equation}	\label{9.1}
R^\eta(s,t) :=
\mathsf{D}^{\eta(\cdot,t)} \!\circ \mathsf{D}^{\eta(s,\cdot)} -
\mathsf{D}^{\eta(s,\cdot)} \!\circ \mathsf{D}^{\eta(\cdot,t)}
	\end{equation}
but, as a consequence of~\eref{9.01}, compositions like
$\mathsf{D}^{\eta(\cdot,t)} \!\circ \mathsf{D}^{\eta(s,\cdot)}$
are not quite correctly defined since the action of this expression on
$\sigma$ must be defined as
 $\mathsf{D}^{\eta(\cdot,t)}(\sigma_1|_{\eta(J,t)})$.

	If $\{e_a(x)\}$, $x\in\eta(J,J')$, is a basis in
$\proj^{-1}(x)=\bHil_x$, the local expansion~\eref{5.8} can be apply for
explicit calculation of the derivatives
 $\mor{D}_t^{\eta(s,\cdot)} \Hat{\sigma}$,
 $\mor{D}_s^{\eta(\cdot,t)} \Hat{\sigma}$,
 $\mor{D}_s^{\eta(\cdot,t)} \Hat{\sigma}_1$,
 $\mor{D}_t^{\eta(s,\cdot)} \Hat{\sigma}_2$.
The substitution of these expansions into~\eref{9.04} results in the
assertion that the mapping~\eref{9.03} is linear, that is
	\begin{equation}	\label{9.05}
(R^\eta(s,t))(\sigma)
 = \Sprindex[(R^\eta(s,t))]{b}{a} \, \sigma^b(\eta(s,t)) e_a(\eta(s,t)),
	\end{equation}
and its local components in $\{e_a\}$ are
	\begin{multline}
\left( R^\eta(s,t) \right)_{{\ }b}^a  =
\frac{\partial}{\partial s}
\left[ \Gamma_{{\ }b}^a(t;\eta(s,\cdot)) \right] -
\frac{\partial}{\partial t}
\left[ \Gamma_{{\ }b}^a(s;\eta(\cdot,t)) \right]
\\	\label{9.2}
+
\Gamma_{{\ }c}^a(s;\eta(\cdot,t)) \Gamma_{{\ }b}^c(t;\eta(s,\cdot)) -
\Gamma_{{\ }c}^a(t;\eta(s,\cdot)) \Gamma_{{\ }b}^c(s;\eta(\cdot,t)).
	\end{multline}
Here $\Gamma_{\ldots}^{\ldots}(\cdots)$ are the corresponding coefficients of
the evolution transport in $\{e_i(x)\}$ defined by~\eref{5.9}.
Using the fundamental relation~(\ref{5.10}), we can
explicitly calculate the curvature. The easiest way to do this is to
choose the bases $\{e_a(x)\}$ and $\{f_a(t)\}$ such that
$\boldsymbol{l}_x(t)=[\delta_b^a]=\openone$ (see remark~\ref{rem5.1}).
Then
$\boldsymbol{E}(t)=0$ and
\(
\mbHam_{\gamma}(t) =
\bHamM(t)=\HamM(t),
\)
 $\gamma=\eta(\cdot,t),\eta(s,\cdot)$.
Hence, due to~\eref{5.10}, now equation~(\ref{9.2}) reduces to
	\begin{equation}	\label{9.3}
\boldsymbol{R}^\eta(s,t)=\frac{1}{(-\ih)^2}
[ \HamM(s) , \HamM(t) ]_{\_}
	\end{equation}
where we have assumed, as usual, that the conventional Hamiltonian $\Ham(t)$
is independent of the observer's trajectory (reference path) $\gamma$.%
\footnote{%
If $\Ham$ depends explicitly on the reference path $\gamma$,
$\Ham=\Ham(t;\gamma)$, in the r.h.s.\ of~\eref{9.3} the term
\(
\bigl[
\frac{\pd}{\pd s} \HamM(t;\eta(s,\cdot))
-
\frac{\pd}{\pd t} \HamM(s;\eta(\cdot,t))
\bigr]
/(-\ih)^2
\)
will appear.%
}
(The last equality is valid only in the special bases in which it is
derived!)

	From~\eref{9.3} we infer that the
\emph{%
evolution transport is curvature free if and only if the values of the
Hamiltonian at different moments commute,%
}
viz.\
	\begin{equation}	\label{9.4}
{R}^\eta = 0 \iff [ {\Ham}(s) , {\Ham}(t) ]_{\_} = 0.
	\end{equation}
In particular, this is true for explicitly time-independent Hamiltonians, \ie
for ones with $\partial\Ham(t)/\partial t = 0$ . According to~(\ref{6.085}),
the bundle form of~\eref{9.4} is
	\begin{equation}	\label{9.5}
{R}^\eta = 0 \iff
\left[
{\Breve{\bHam}}_{\gamma,s}(r) , {\Breve{\bHam}}_{\gamma,t}(r)
\right]_{\_} = 0.
	\end{equation}
for some (and hence any) path $\gamma\colon J\to\base$. Here $r,s,t\in J$ and
${\Breve{\bHam}}_{\gamma,t}(r)$ is (the transported by means of
$l_{s\to t}^{\gamma}$ Hamiltonian which is) calculated via~(\ref{6.082}).

	Consider now a curvature free evolution transport on
$W=\eta(J,J')$, \ie  $R^\eta(s,t)\equiv0$ for every $(s,t)\in J\times J'$.
From~\cite[proposition~3.3]{bp-LTP-Cur+Tor} we know that in this case
there exists a field of base $\{e_i\}$ over $W$ in which the transport's
coefficients vanish along any path $\gamma$ in $W$. In it, due to~\eref{5.3}
and~\eref{5.10}, we have
	\begin{equation}	\label{9.6}
\widetilde{\mmor{U}_\gamma}(t,t_0) = \openone,	\quad
\widetilde{\boldsymbol{\Gamma}}_\gamma(t) = \boldsymbol{0},	\quad
\widetilde{\mbHam_\gamma}(t) = \boldsymbol{0}
\qquad\text{for every $\gamma$}.
	\end{equation}
Notice, equations~\eref{7.3} are valid for \emph{arbitrary} evolution
transports along any \emph{fixed} path in appropriate bases along it,
while equations~\eref{9.6} hold only for \emph{curvature free} evolution
transports in a suitably chosen field of bases in a \emph{whole} set $W$. The
connection of the special bases in which~\eref{9.6} is true with the
Heisenberg picture is the same as discussed in Subsect~\ref{VII.2} for the
bases in which~\eref{7.3} are satisfied.

	We want to emphasize on the fact that according to~\eref{5.2} the
local vanishment of the matrix\ndash bundle Hamiltonian does not imply the
same property for the Hamiltonian as an operator or lifting.

	According to~\cite[theorem~6.1]{bp-LTP-Cur+Tor-prop} a linear
transport along paths is path\ndash independent iff it is flat, \ie with
vanishing curvature, provided the corresponding paths lie entirely in the
region of flatness of the transport. Applying this result to the evolution
transport $\mor{U}$, we can assert that only the curvature free evolution
transports are path independent. Physically, due to~\eref{4.4}
and/or~\eref{11.14}, this means that the evolution of quantum systems with
curvature free evolution transports is independent of the reference path
$\gamma$ with respect to which it is described. Since we interpret  $\gamma$
as observer's trajectory (world line),%
\footnote{%
Physically we interpret the paths
 $\eta(s,\cdot)\colon s\mapsto\eta(s,t)$
and
 $\eta(\cdot,t)\colon s\mapsto\eta(s,t)$
as trajectories (world lines) of observers with (proper) times $t$ and $s$
respectively.%
}
the last result has a meaning of observer\ndash independence of the
time\ndash evolution of such systems. As we proved above, for a sufficient
general class of quantum systems the evolution transport is flat (=curvature
free) iff equation~\eref{9.4} holds, \ie if the (ordinary Hilbert space)
Hamiltonian of a system is such that its different time values commute. An
evident sub\ndash class of quantum systems with flat evolution transports is
the one of systems described via explicitly time\ndash independent
Hamiltonians. This sub\ndash class is also broad enough and covers a great
number of physically important cases~\cite{Messiah-QM,Dirac-PQM,Fock-FQM}.

\section{On observer's r\^ole and theory's interpretation}
\label{XII}

	The concept of an `observer' is more physical than mathematical one
as it is not very well mathematically rigorously defined; sometimes its
meaning is more intuitive than strict one. Generally an `\emph{observer}' is
a physical system whose state is assumed to be `completely' known and which
has a double r\^ole with respect to the other systems(s) which is (are) under
consideration. From one hand, it provides certain reference point, generally
a set of objects and their properties, with respect to which are determined
(all of) the quantities characterizing the investigated system(s) in some
problem. From the other hand, it is supposed the `observer' can perform
certain procedures, called `measurements' or `observations', by means of
which `he' finds (determines) the parameters, properties, quantities, etc.
describing the studied system(s). This second r\^ole of the observers is out
of the subject of the present investigation and will not be discussed here
(for some its aspects see, e.g.,~\cite{Messiah-QM,Prugovecki-QMinHS}).

	In this work the observers are supposed to be local and point-like,
i.e.\   they are material points that can perform measurements at the points at
which they are situated. They are moving (along paths) in some differentiable
manifold $\base$.%
\footnote{%
Mathematically the developed here theory is sensible also if $\base$ is
considered as a `more general' object than a manifold, but, at present, there
are not indications that such a theory can be physically important.%
}
Namely their trajectories (world lines in special or general relativity
interpretation (see below)) are the reference objects with respect to which
we study the behaviour of the quantum systems. The `observational' properties
of an observer are assumed fixed and such that: (i) Allow the observer to
determine the initial values of the quantities characterizing the state of
the studied system(s) at some instant of time $t_0$; (ii) Give certain
correspondence rules according to which to any dynamical variable $\dyn{A}$,
connected to the investigated system(s), is assigned some observable which is
a Hermitian operator (resp.\ lifting of paths), say $\ope{A}$ (resp.\
$\mor{A}$), in the Hilbert space (resp.\ Hilbert bundle) description that has
a complete set of (maybe orthonormal) eigenvectors (resp.\ eigen\ndash
liftings of paths).

	All quantities in the present work are referred to an observer moving
along some path $\gamma\colon J\to\base$ parameterized with $t\in J$, where
$J\subseteq\mathbb{R}$  is a real interval. Our intention is to interpret
$t$ as a `time'. The possibility for such an interpretation is connected with
the specific choice of the manifold $\base$.

If we assume $\base$ to be the classical, coordinate, 3-dimensional
Euclidean space $\mathbb{E}^3$ of the classical and quantum mechanics, which
we have done more implicitly than explicitly throughout this investigation,
there exists a \emph{global time}
$t_\mathrm{g}\in(-\infty,+\infty)=\mathbb{R}$
(in the Newtonian sense). In this case the trajectories, such as $\gamma$,
of \emph{all} observers is natural and convenient to be parameterized
with this global time which, in fact, is done in the conventional, Hilbert
space, quantum mechanics. (One can also parameterize each observer's
trajectory with its own local time which is in bijective, usually $C^1$,
correspondence with $t_\mathrm{g}$, but this is an inessential
generalization as $t_\mathrm{g}$ itself is defined with some arbitrariness
(usually a $C^1$ map $\mathbb{R}\to\mathbb{R}$).)
Consequently the
assumptions $\base=\mathbb{E}^3$, $J=\mathbb{R}$ and $t=t_\mathrm{g}$ are
necessary and sufficient for the full equivalence of the Hilbert space and
Hilbert bundle descriptions of quantum mechanics. These assumptions lead to
one inessential mathematical complication: since the observer's trajectory
$\gamma$ can has self-intersections, the sections along $\gamma$ and
morphisms along $\gamma$ can be multiple-valued at the points of
self-intersection of $\gamma$, if any. But this does not have some serious
consequences, especially if one works with liftings of paths which are always
single\ndash valued.

	Analogous to $\base=\mathbb{E}^3$ is the case when the Hilbert bundle
is identified with the system's configuration space. The only difference is
that $\gamma\colon J\to\base$ has to be interpreted as the trajectory of the
system in this space rather then the one of some observer. Practically the
same is the case when as $\base$ is taken the system's phase space but, since
this situation has some peculiarities, we shall comment on it in
Sect.~\ref{discussion}.

	A similar, but slightly different, is the situation when $\base$ is
taken to be the four dimensional Minkowski space\ndash time $M^4$ of special
relativity.%
\footnote{%
A like construction, under the name `Schr\"odinger bundle', is introduced in
the paragraph containing equation~(4) of~\cite{Aldaya&Azcarraga-81}. This is
a Hilbert bundle over $M^4$ having as a (standard) fibre
$\Hil\times\Hil^\ast$ instead of the conventional Hilbert space $\Hil$ in our
case.%
}
Nevertheless that now every observer has its own local preferred
time, called proper or eigen-time, there is a global (e.g.\ coordinate) time
$t_\mathrm{g}$ which is in bijective correspondence with these local times.
In this case the observers are moving along paths, such as $\gamma$, which
are their \emph{world lines (paths or curves)} and usually are parameterized
with the global time $t_\mathrm{g}$. This is an important moment because the
world lines of the real objects observed until now can not have
self-intersections.  Therefore, since the observers are supposed to be such,
their world lines are without self-intersections. This implies the absence of
the complication mentioned for $\base=\mathbb{E}^3$, viz.\ now
the sections along $\gamma$ and morphisms along $\gamma$ are single-valued
and, in fact, are respectively sections and morphisms of the restriction
$\left.\HilB\right|_{\gamma(J)}$ of the Hilbert bundle $\HilB$ on the set
$\gamma(J)$. Otherwise the case $\base=M^4$ is identical with the one with
$\base=\mathbb{E}^3$. Therefore, we can say that it represents the Hilbert
bundle description of the nonrelativistic quantum mechanics over the
special relativity space-time.  This point is worth-mentioning as at it meet
the relativistic and non-relativistic concepts whose unification leads to the
relativistic quantum theory which will be considered elsewhere.

	Another important possibility is $\base=V_4$ where  $V_4$ is the four
dimensional pseudo-Riemannian space-time of general relativity.%
\footnote{%
Almost the same is the case when $\base$ represents the space-time model of
other gravitational theories, like Einstein-Cartan and the metric-affine
ones, in which $\base$ is a curved (non-flat) manifold with respect to some
linear connection.%
}
The crucial point here is the generic non-existence of some global time, so the
\emph{world line} of any particular observer, say $\gamma$, with necessity
has to be parameterize with its (specifically local) \emph{proper time} $t$.
A consequence of this is that in the (conventional) Hilbert space description
the parameter (`time') $t$ also has to be considered as a local (proper) time
for the observer which describes the quantum system under consideration.
Hence the global sections and morphisms defined via~\eref{4.3b}
and~\eref{6.1} have no physical meaning now. As in special relativity case,
$\base=M^4$, now $\gamma$ cannot has self\ndash intersections; so the
sections and morphisms along $\gamma$ are single\ndash valued. What concerns
other aspects of the case $\base=V_4$, it is identical with the one for
$\base=\mathbb{E}^3$.  Consequently, it represents the non-relativistic
quantum mechanics over the general relativity space\ndash time. Here we see,
again, a meeting of (general) relativistic and nonrelativistic concepts whose
unification will be investigated elsewhere.

	Now we want to call attention to a particular `degenerate' case which
falls out of our general interpretation of $\gamma:J\to M$ as an observer's
world line (trajectory). Namely, it is possible to put $\base=J$, $J$ being a
real interval. For instance, as we pointed at the end of Subsect.~\ref{V.2},
the considerations of~\cite{Asorey&Carinena&Paramio} correspond to the choice
$\base=\mathbb{R}_+=\{t: t\in\mathbb{R},\ t\ge0\}$. Thus, if $\base=J$,
then $\gamma:J\to J$ and, as we tend to interpret $t\in J$ as time, it is
natural now to assume that $\gamma$ is bijective smooth ($C^1$) map.%
\footnote{%
The case considered in reference~\cite{Asorey&Carinena&Paramio} corresponds
to $\gamma=\id_{\mathbb{R}_+}$.%
}
If so, $\gamma(t)\in J$ is also a `presentation' of the time, but in
other (re)parameterization. In this case $\gamma$ is without
self\ndash intersections and, consequently, the sections and morphisms along
$\gamma$ are simply (single-valued) sections and morphisms of the bundle
$(\bHil,\proj,J)$ over the one\ndash dimensional base $J$. Therefore, in this
situation, we can say that the evolution of a quantum system is described via
linear transportation (of the state) sections of $(\bHil,\proj,J)$ along the
time, respectively the evolution of the observables is represented by linear
transportation of (the observable) liftings/morphisms of $(\bHil,\proj,J)$
along the time. Let us note that now the connection with the observers is not
completely lost as the time does not exist by `itself' in the theory; it is
connected with (measured by) a concrete observer regardless of the fact that
it can be global of local (see above).

	Another degenerate case is when $\base$ consist of a single point,
$\base=\{x\}$. Then $\bHil=\proj^{-1}(x)=:\bHil_{x}$ and
$\gamma\colon J\to\{x\}$. Therefore $\gamma(t)\equiv x$ for every $t\in J$
and, if $J$ is not compressed into a single real number, $\gamma$
self-intersects at $x$ infinitely many times. Also we have
 $l_{\gamma(t)}=l_{\gamma(s)}=l_{x}=:l$, $s,t\in J$ with
$l\colon\bHil\to\Hil$ being an isomorphism. So, in this way (see
note~\ref{Note4.4}), we obtain an isomorphic image in $\bHil$ of the quantum
mechanics in $\Hil$.  Evidently, the conventional quantum mechanics is
recovered by the choices $\Hil=\bHil$ and $l=\id_{\bHil}$. Of course, now we
can not interpreted $\gamma$ as observer's trajectory or world line but the
interpretation of $t$ as a `time' can be preserved.

	If the quantum system under consideration has a classical analogue,
then the manifold $\base$ can be identified with the system's configuration
or phase space. In this case the path $\gamma\colon J\to\base$ can be taken
to be the trajectory of the system's classical analogue in the corresponding
space. Thus we obtain an interesting situation: the (bundle) quantum evolution
is described with respect to (is referred to) the corresponding classical
evolution of the same system.

	One can also take $\base$ to be the configuration or phase space of
some observer. Then $\gamma$ can naturally be defined as the observer's
trajectory in the corresponding space.

	At this point we want to say a few words on the possibility to
identify the Hilbert bundle's base $\base$ with the phase space of certain
system and to make some comments on~\cite{Reuter}, where this case is taken
as a base for a bundle approach to quantum mechanics. Our generic opinion is
that the phase space is not a `suitable' candidate for a bundle's base, the
reason being the Heisenberg uncertainty principle by virtue of which the
points of the phase space have no physical
meaning~\cite[chapter~IV]{Messiah-1}. This reason does not apply if as a base
is taken the phase space of some observer as, by definition, the observers
are treated as classical objects (systems). Therefore one can set the base
$\base$ of the Hilbert bundle $\HilB$ to be the phase space of some observer.
Then the reference path $\gamma\colon J\to\base$ can be interpreted as the
observer's phase\ndash space trajectory which, generally, can have self\ndash
intersections. The further treatment of this case is the same as of
$\base=\mathbb{E}^3$.
	Regardless of the above\ndash said, one can always identify $\base$
with the system's phase space, if it exists, as actually $\base$ is a free
parameter in the present work.

	An interesting bundle approach to quantum mechanics
is contained in~\cite{Reuter}. In it the evolution of a quantum system is
described in a Hilbert bundle \emph{over the system's phase space} with the
ordinary system's Hilbert space as a (typical) fibre which is, sometimes,
identified with the fibre over an arbitrary fixed phase-space point. The
evolution itself is presented as a parallel transport in the bundle space
generated via non\ndash dynamical linear (and symplectic) connection which is
closely related to the symplectic structure of the phase space.
	An important feature of~\cite{Reuter} is that in it the bundle
structure is derived from the physical content of the paper. In this
sense~\cite{Reuter} can be considered as a good motivation for the general
constructions in the present investigation.

	Before comparing the mathematical results of~\cite{Reuter} with the
ones of this work, we have to say that the \emph{loc.\ cit.}\ contains some
incorrect `bundle' expressions which, however, happily do not influence most
of the conclusions made on their base.
       In~\cite{bp-Reuter-comments} we point to and show possible ways for
improving of a number of mathematically non-rigorous or wrong expressions,
assertions, and definitions in~\cite{Reuter}. We emphasize that all this
concerns only the `bundle' part of the mathematical structure of \emph{loc.\
cit.}\ and does not deal with its physical contents.
	The general moral of the critical remarks
of~\cite{bp-Reuter-comments} is: most of the final results and conclusions
of~\cite{Reuter} are valid provided the pointed in~\cite{bp-Reuter-comments}
(and other minor) corrections are made. Below we shall suppose that this
is carefully done. On this base we will compare~\cite{Reuter} with the
present work.

	The main common point between~\cite{Reuter} and this investigation is
the consistent application of the fibre bundle theory to (nonrelativistic)
quantum mechanics. But the implementation of this intention is quite
different: in~\cite{Reuter} we see a description of quantum mechanics in a
new, but `frozen', geometrical background based on a non-dynamical linear
connection deduced from the symplectical structure of the system's phase
space, while the present work uses a `dynamical geometry' (linear transport
along paths, which may turn to be a parallel one generated by a linear
connection) whose properties depend on the system's Hamiltonian, \ie on the
physical system under consideration itself.

	The fact that in~\cite{Reuter} the system's phase space is taken as a
base of the used Hilbert bundle is not essential since nothing can prevent us
from making the same choice as, actually, the base is not fixed here.
In~\cite{Reuter}  is partially considered the dynamics of multispinor fields.
This is an interesting problem, but, since it is not primary related to
conventional quantum mechanics, we think it is out of the scope of our work.
The methods of its solution are outlined in~\cite{Reuter} and can easily be
incorporated within our bundle quantum mechanics.

	The fields of (metaplectic) spinors used in~\cite{Reuter} are simply
sections of the Hilbert bundle, while the ``world-line spinors'' in
\emph{loc.\ cit.}\ are sections along paths in our terminology. The family of
operators $\mathcal{O}_{\phi^a}$ or
$\mathcal{O}_{f}(\phi)$~\cite[equations~(4.8) and~(4.9)]{Reuter} acting on
$\bHil_\phi$ are actually bundle morphisms.

	A central r\^ole in both works plays the `principle of invariance of
the mean values' (see the paragraph after postulate~\ref{Post6.1}): the mean
values (mathematical expectations) of the liftings of paths or the morphisms
along paths corresponding to the observables (dynamical variables) are
independent of the way they are calculate. We have used this assumption many
times (see, e.g., Sections~\ref{VI}, \ref{VII}, and~\ref{XI}, in particular,
equations~\eref{6.1''}, \eref{7.5}, \eref{7.11}, \eref{7.17},
and~\eref{11.17}). In~\cite{Reuter}  `the invariance of the mean values' is
mentioned several times and it is used practically in the form of the
`background-quantum split symmetry' principle, explained
in~\cite[sect.~4]{Reuter} (see, e.g.,~\cite[equation~(4.18)]{Reuter} and the
comments after it).  Its particular realizations are written
as~\cite[equation~(4.17) and second equation~(4.42)]{Reuter} which are
equivalent to it in the corresponding context. A consequence of the
mean-value invariance is the `Abelian' character of the compatible with it
connections, expressed by~\cite[equation~(4.14)]{Reuter}, which is a special
case of our result~\cite[equation~(4.4)]{bp-BQM-preliminary}.
In~\cite{Reuter} the mean values are independent of the point at which they
are determined. In our bundle quantum mechanics this is not generally the
case as different points correspond to different time values (see,
e.g.,~\eref{6.1''}). This difference clearly reflects the dynamical character
of our approach and the `frozen' geometrical one of~\cite{Reuter}.

	In both works the quantum evolution is described via appropriate
transport along paths:
In~\cite[see, e.g., equations~(3.54) and~(4.53)]{Reuter} this is an `Abelian'
parallel transport along curves, whose holonomy group is
$U(1)$~\cite[equation~(4.38)]{Reuter}, while in our investigation is employed
a transport along paths uniquely determined by the Hamiltonian
(see Sect.~\ref{IV}) which, generally, need not to be a parallel translation.

	Now we turn our attention on the bundle equations of motion: in the
current work we have a \emph{single} bundle Schr\"odinger
equation~\eref{5.13} (see also its matrix version~\eref{5.1}), while
in~\cite[equation~(5.54)]{Reuter} there is an \emph{infinite} number of such
equations, one Schr\"odinger equation in each fibre $\bHil_\phi$ for the
system's state vector $|\psi(t)\rangle_\phi$ at every point $\phi\in\base$.%
\footnote{%
Note that the appearing in~\cite[equations~(4.54)--(4.56)]{Reuter} operator
$\mathcal{O}_H$ is an analogue of our matrix-bundle Hamiltonian
(see Sect.~\ref{V}).%
}
Analogous is the situation with the statistical operator (compare our
equation~\eref{11.17} or~\eref{11.15} with~\cite[equation~(4.56)]{Reuter}).
This drastical difference is due to the \emph{different objects} used to
describe systems states: for the purpose we have used \emph{liftings
of paths or sections along paths} (see Sect.~\ref{new-I}), while
in~\cite{Reuter} are utilized \emph{(global) sections} of the bundle defined
via~\eref{4.3b} (cf.~\cite[equation~(4.41)]{Reuter}). Hence, what actually is
done in~\cite{Reuter} is the construction of an isomorphic images of the
quantum mechanics from the fibre $\Hil$ in every fibre $\bHil_\phi$,
$\phi\in\base$ (see the comments after~\eref{4.3b}).

	To summarize the comments on part of the mathematical
structures in~\cite{Reuter}: It contains a fibre
bundle description of quantum mechanics. The state vectors are replaced by
(global) sections of a Hilbert bundle with the system's phase space as a base
and their (bundle) evolution is governed through Abelian parallel transport
arising from the symplectical structure of the phase space. Locally, in any
fibre of the bundle, the evolution is presented by a Schr\"odinger
equation, specific for each fibre of the bundle. The work contains a number
of incorrect mathematical constructions which, however, can be corrected so
that the final conclusions remain valid. Some ideas of the paper are near to
the ones of this investigation but their implementation and development is
quite different in both works.

\section{Summary}
\label{summary}
\setcounter{equation} {0}

	In this work we have proposed and developed  a new invariant
fibre bundle formulation of nonrelativistic quantum mechanics.
With respect to the  physically predictable results, it is fully
equivalent to the usual formulation of the theory in the case when the
underlying manifold is the (coordinate) three dimensional Euclidean space of
classical/quantum mechanics. In the new description a pure state of a
quantum system is describe by a state lifting of paths or a state section
along paths of a Hilbert bundle. The time evolution of the state
liftings/sections obeys the bundle Schr\"odinger equation~(\ref{5.13}). A
mixed state of a quantum system is described via the density lifting of paths
(in the bundle of point\ndash restricted morphisms of system's Hilbert
bundle) or the density morphism along paths (in system's Hilbert bundle)
satisfying the Schr\"odinger (type) equation~\eref{11.17}. In the proposed
bundle approach to any dynamical variable corresponds a unique observable
which is a lifting of paths (in the bundle of point\ndash restricted
morphisms of system's Hilbert bundle) or morphism along paths (in the Hilbert
bundle of the investigated system).  The observed value of a dynamical
variable is equal (by definition) to the mean value (the mathematical
expectation) of the corresponding lifting/morphism and it is calculated by
means of the bundle state lifting/section or density lifting/morphism
corresponding to the system's state at the moment.

	The correspondence between the conventional Hilbert space description
and the new Hilbert bundle description of non\ndash relativistic quantum
mechanics is given in table~\ref{table:space/bundle} on
page~\pageref{table:space/bundle}.

	A feature of the bundle form of quantum mechanics is the inherent
connection between physics and geometry: the system's Hamiltonian
(via equation~\eref{5.10}) completely determines the concrete properties of
the system's Hilbert bundle. In this sense, here the Hamiltonian plays the
same r\^ole as the energy\ndash momentum tensor in general relativity.
Another view\nobreakdash-point (also based on~\eref{5.10}) is to look on the
Hamiltonian as a gauge field in the sense of Yang-Mills theories. In any
case, we see in the bundle quantum mechanics a realization of the
intriguing idea, going back to Albert Einstein and Bernhard Riemann,
that the physical properties of the systems are responsible
for the geometry of the spaces used for their description.
\newpage%
\thispagestyle{plain}%
\addtolength{\oddsidemargin}{-6em}
\renewcommand{\arraystretch}{1.2}
%
\begin{table}
\vspace{-20.0ex}
\footnotesize
\centering
\caption{\normalsize
Comparison between Hilbert space and Hilbert bundle descriptions.
\vspace{1.2ex} }
\label{table:space/bundle}
\begin{tabular}{||l|l|l||}
\hline
\hline
\multicolumn{1}{||c||}
{
\begin{tabular}{c}
\small 
\\[-1.75ex] \bfseries Hilbert space description \\[-1.75ex] \\
\end{tabular}
} &
\multicolumn{1}{||c||}
{
\begin{tabular}{c}
\small 
\\[-1.75ex] \bfseries Hilbert bundle description \\[-1.75ex] \\
\end{tabular}
} &
\multicolumn{1}{||c||}
{
\begin{tabular}{c}
\small 
\\[-1.75ex] \bfseries Remark(s) \\[-1.75ex] \\
\end{tabular}
}
\\  
 \hline \hline
Hilbert space $\Hil$ &
Hilbert bundle $\HilB$ &
$l_x:\bHil_x\to\Hil$
\\ \cline{1-3}
Vector $\varphi\in\Hil$ &
Section $\overline{\Phi}\in\Sec{\HilB}$ &
$\overline{\Phi}\colon x \mapsto l_x^{-1}(\varphi)$
\\ \cline{1-3}
Operator $\ope{A}:\Hil\to\Hil$ &
Morphism $\overline{\mor{A}}\in\Morf_\base{\HilB}$ &
$\overline{\mor{A}}_x=l_x^{-1}\circ\ope{A}\circ l_x$
\\ \cline{1-3}
State vector $\psi\in\Hil $ &
State lifting $\Psi$ of paths in $\HilB$ &
$\Psi_\gamma (t)=l_{\gamma(t)}^{-1}(\psi(t))$
\\ \cline{1-3}
Observable $\ope{A}\colon\Hil\to\Hil$ &
Observable lifging $\mor{A}$ of paths &
$\mor{A}_\gamma(t)=l_{\gamma(t)}^{-1}\circ\ope{A}(t)\circ l_{\gamma(t)}^{}$
\\ \cline{1-3}
Hermitian scalar product $\langle\phi | \psi\rangle$
&
Hermitian bundle scalar product
  $\langle\Phi_x | \Psi_x\rangle_x$
&
$\langle\cdot | \cdot\rangle_x = \langle l_x\cdot | l_x \cdot\rangle$
\\ \cline{1-3}
\multicolumn{1}{||l|}
{
\begin{tabular}{@{}l@{}}
Hermitian conjugate operator $\ope{A}^\dag$    \\
   corresponding to an operator $\ope{A}$:         \\
     $\langle\ope{A}^\dag\phi | \psi\rangle =
   \langle \phi | \ope{A}\psi\rangle$
\end{tabular}
}
&
\multicolumn{1}{l|}
{
\begin{tabular}{@{}l@{}}
1) Hermitian conjugate map
   $\mor{A}^\ddag_{x\to y}\colon\bHil_x\to\bHil_y$    \\
\ \ \ \       to a bundle map $\mor{A}_{y\to x}:\bHil_y\to \bHil_x$: \\
\ \ \ \   $\langle \mor{A}_{x\to y}^\ddag\Phi_x | \Psi_y\rangle_y =
 \langle\Phi_x | \mor{A}_{y\to x} \Psi_y\rangle_x$ \\[1 ex]
2)  Hermitian conjugate morphism $\mor{A}^\ddag$   \\
\ \ \ \    to a morphism $\mor{A}$ along paths: \\
\ \ \ \   $\langle \mor{A}_x^\ddag\Phi_x | \Psi_x\rangle_x =
  \langle\Phi_x | \mor{A}_x \Psi_x\rangle_x$
\end{tabular}
}
&
\multicolumn{1}{l||}
{
\begin{tabular}{@{}l@{}}
$\mor{A}_{x\to y}^\ddag = $ \\
$ = l_y^{-1}\circ(l_x\circ \mor{A}_{y\to x}\circ l_y^{-1})^\dag \circ l_x$
\\
\\[1 ex]
\\
$\mor{A}_x^\ddag = l_x^{-1}\circ(l_x\circ \mor{A}_x\circ l_x^{-1})^\dag \circ l_x$
\end{tabular}
}
\\ \cline{1-3}
Unitary operator: $\ope{A}^\dag=\ope{A}^{-1}$ &
\multicolumn{1}{|l|}
{
\begin{tabular}{@{}l@{}}
1) Unitary  bundle map:  \\
\ \ \ \   $\mor{A}_{x\to y}^\ddag=\mor{A}_{y\to x}^{-1}:=
    (\mor{A}_{y\to x})^{-1}$
\\[1ex]
2) Unitary bundle morphism: $\mor{A}^\ddag=\mor{A}^{-1}$
\end{tabular}
}
&
 \multicolumn{1}{l||}
{
\begin{tabular}{@{}l@{}}
$\mor{A}_{x\to y}^{-1}$ is the left   \\
inverse of $\mor{A}_{x\to y}$  \\[1ex]
$\mor{A}_x^\ddag=\mor{A}_x^{-1}$
\end{tabular}
}
\\ \cline{1-3}
Hermitian operator: $\ope{A}^\dag=\ope{A}$ &
Hermitian morphism: $\mor{A}^\ddag=\mor{A}$ &
$\mor{A}_x^\dag=\mor{A}_x$;
$\mor{A}^\ddag=\mor{A}\iff\ope{A}^\dag=\ope{A}$
\\ \cline{1-3}
Basis $\{f_a(t)\}$ in $\Hil$ &
Basis $\{e_a\}$ in $\Sec\HilB$ &
\begin{tabular}{@{}l@{}}
A good choice is \\ $e_a(x)=l_x^{-1}(f_a(t))$
\end{tabular}
\\ \cline{1-3}
\multicolumn{2}{||l|}
{
\begin{tabular}{@{}l@{}}
Matrix corresponding
to a linear map or a vector in a given basis (bases):
\\
the same notation but the kernel letter is in \textbf{boldface}
\end{tabular}
}
&
\multicolumn{1}{l||}
{
\begin{tabular}{@{}l@{}}
For example:
\(
\boldsymbol{\ope{A}}(t),\ \mmor{A},\
\boldsymbol{\psi}(t),\)
\\
\(
\boldsymbol{\Psi}_\gamma(t),\
\boldsymbol{l}_x,\
\mope{U}(t,s),\  \mmor{U}_\gamma(t,s)
\)
\end{tabular}
}
\\ \cline{1-3}
\begin{tabular}{@{}l@{}}
Mean value of operator $\ope{A}(t)$: \\
  $\langle \ope{A}(t)\rangle_\psi^t =
  \frac{\langle \psi(t) | \ope{A}(t)\psi(t)\rangle}
  {\langle \psi(t) | \psi(t)\rangle}$
\end{tabular}
&
\begin{tabular}{@{}l@{}}
Mean value of observable lifting $\mor{A}(t)$: \\
\(
\langle \mor{A}(t)\rangle_{\Psi_\gamma}^t =
\frac{\langle \Psi_\gamma(t) | \mor{A}(t)\Psi_\gamma(t)\rangle_{\gamma(t)}}
{\langle \Psi_\gamma(t) | \Psi_\gamma(t)\rangle_{\gamma(t)} }
\)
\end{tabular}
&
$\langle \mor{A}(t)\rangle_{\Psi_\gamma}^t =
   \langle \ope{A}(t)\rangle_\psi^t$
\\ \cline{1-3}
Evolution operator $\ope{U}$ &
Evolution transport $\mor{U}$ along paths &
 \multicolumn{1}{l||}
{
\begin{tabular}{@{}l@{}}
$\mor{U}_\gamma(t,s)=l_{\gamma(t)}^{-1}\circ\ope{U}(t,s)\circ l_{\gamma(s)}$;
\\
see~\eref{2.1} and~\eref{4.4}
\end{tabular}
}
\\ \cline{1-3}
1) Hamiltonian ${\Ham}$ &
1) Bundle Hamiltonian $\bHam$ &
$\bHam_\gamma(t)=l_{\gamma(t)}^{-1}\circ\Ham(t)\circ l_{\gamma(t)}$ \\
2) Matrix Hamiltonian $\mHam$ &
2) Matrix-bundle Hamiltonian $\mbHam_\gamma$ &
See~(\ref{5.08}), (\ref{5.2}), and~(\ref{5.2'}) \\[1ex]
 &
\begin{tabular}{@{}l@{}}
3) Matrix $\boldsymbol{\Gamma}$ of the coefficients  \\
\ \ \ \  of the evolution transport
\end{tabular}
&
$\boldsymbol{\Gamma}_\gamma(t)=-\mbHam_{\gamma}(t)/\ih$ \\
 &
4) Derivation $D$ along paths &
 $D_{t}^{\gamma}e_i = \Sprindex[\Gamma]{i}{j}(t;\gamma) e_j(t;\gamma)$
\\ \cline{1-3}
\multicolumn{1}{||l|}
{
\begin{tabular}{@{}l@{}}
1) Schr\"odinger equation: \\
\ \ \ \  \( \ih \frac{d \psi(t)}{d t} = \Ham(t)\psi(t) \) \\[1ex]
2) Matrix Schr\"odinger equation: \\
\ \ \ \   \( \ih \frac{d \boldsymbol{\psi}(t)}{d t} =
  \mHam(t)\boldsymbol{\psi}(t) \)
\end{tabular}
}
&
\multicolumn{1}{l|}
{
\begin{tabular}{@{}l@{}}
1) Bundle Schr\"odinger equation: \\
\ \ \ \  $\mor{D}_t^\gamma \Psi_\gamma = 0$ or $D\Psi=0$  \\[1ex]
2) Matrix-bundle Schr\"odinger equation: \\
\ \ \ \    \( \ih \frac{d \boldsymbol{\Psi}_\gamma(t)}{d t} =
 \mbHam_{\gamma}(t)\boldsymbol{\Psi}_\gamma(t) \)
\end{tabular}
}
&
\multicolumn{1}{l||}{Equivalent equations}
\\ \cline{1-3}
Density operator $\rho$ &
Density lifting $\Rho$ of paths &
 $\Rho_\gamma(t)=l_{\gamma(t)}^{-1}\circ\rho(t)\circ l_{\gamma(t)}$
\\ \cline{1-3}
\begin{tabular}{@{}l@{}}
Mean value of operator $\ope{A}(t)$: \\
\(
\langle\ope{A}\rangle_{\rho}^{t} :=
\langle\ope{A}(t)\rangle_{\rho}^{t} :=
	\Tr( \rho(t)\circ\ope{A}(t) )
\)
\end{tabular}
&
\begin{tabular}{@{}l@{}}
Mean value of observable lifting $\mor{A}(t)$: \\
\(
\langle \mor{A} \rangle_{\Rho}^{t,\gamma} =
\langle \mor{A}_\gamma(t) \rangle_{\Rho_\gamma}^{t} =
\Tr( \Rho_\gamma(t)\circ\mor{A}_\gamma(t) )
\)
\end{tabular}
&
\(
\langle\ope{A}\rangle_{\rho}^{t} =
\langle \mor{A} \rangle_{\Rho}^{t,\gamma}
\)
\\ \cline{1-3}
\begin{tabular}{@{}l@{}}
Density operator evolution: \\
$ \ih\frac{d\rho(t)}{d t} = [\Ham(t),\rho(t)]_{\_} $
\end{tabular}
&
\begin{tabular}{@{}l@{}}
Density lifting evolution: \\
$ \widetilde{\mor{D}}(\widetilde{\Rho}) = 0 $
or
$ ^\circ\!{\mor{D}}(\Rho) = 0 $
\end{tabular}
&
Equivalent equations
\\ \cline{1-3}
{
\begin{tabular}{@{}l@{}}
Schr\"odinger picture of motion: \\
$\psi(t),\ \ope{A}(t)$; \,  $\rho(t),\ \ope{A}(t)$
\end{tabular}
}
&
\multicolumn{1}{l|}
{
\begin{tabular}{@{}l@{}}
Bundle Schr\"odinger picture of motion: \\
$\Psi_\gamma(t),\ \mor{A}_\gamma(t)$; \,
$\Rho_\gamma(t),\ \mor{A}_\gamma(t)$
\end{tabular}
}
&
\begin{tabular}{@{}l@{}}
See:~(\ref{2.5}), (\ref{5.13}); \\
  \eref{11.5}, \eref{11.15}, \eref{11.17}
\end{tabular}
\\ \cline{1-3}
\multicolumn{1}{||l|}
{
\begin{tabular}{@{}l@{}}
Heisenberg picture of motion:  \\
$\psi_t^{\mathrm{H}}(t_0),\ \ope{A}_t^{\mathrm{H}}(t_0)$; \,
$\rho_t^{\mathrm{H}}(t_0),\ \ope{A}_t^{\mathrm{H}}(t_0)$
\end{tabular}
}
&
\multicolumn{1}{l|}
{
\begin{tabular}{@{}l@{}}
Bundle Heisenberg picture of motion: \\
$\Psi_{\gamma,t}^{\mathrm{H}}(t_0),\ \mor{A}_{\gamma,t}^{\mathrm{H}}(t_0)$; \,
$\Rho_{\gamma,t}^{\mathrm{H}}(t_0),\ \mor{A}_{\gamma,t}^{\mathrm{H}}(t_0)$
\end{tabular}
}
&
\begin{tabular}{@{}l@{}}
See:~(\ref{7.11a}), (\ref{7.8}), (\ref{7.13}), (\ref{7.13a}); \\
\eref{11.18},~\eref{11.19}
\end{tabular}
\\ \cline{1-3}
\multicolumn{1}{||l|}
{
\begin{tabular}{@{}l@{}}
`General' picture of motion:  \\
$\psi_t^\ope{V}(t_0),\ \ope{A}_t^\ope{V}(t_0)$; \,
$\rho_t^\ope{V}(t_0),\ \ope{A}_t^\ope{V}(t_0)$
\end{tabular}
}
&
\multicolumn{1}{l|}
{
\begin{tabular}{@{}l@{}}
Bundle `general' picture of motion:  \\
$\Psi_{\gamma,t}^\mor{V}(t_0),\ \mor{A}_{\gamma,t}^\mor{V}(t_0)$; \,
$\Rho_{\gamma,t}^\mor{V}(t_0),\ \mor{A}_{\gamma,t}^\mor{V}(t_0)$
\end{tabular}
}
&
\begin{tabular}{@{}l@{}}
See:~(\ref{7.24}), (\ref{7.27}), (\ref{7.20}), (\ref{7.28}), \\
(\ref{7.29});\,~\eref{11.26}--\eref{11.29}
\end{tabular}
\\ \cline{1-3}
\begin{tabular}{@{}l@{}}
Integral of motion $\ope{A}$:\\
$\ih \frac{\partial \ope{A}(t)}{\partial t} +[\ope{A}(t),\Ham(t)]_{\_} = 0$
\end{tabular}
&
\begin{tabular}{@{}l@{}}
Integral of motion $\mor{A}$: \\
$( \widetilde{\mor{D}}_{t}^{\gamma} (\mor{A}))\Psi=0$
or
$^\circ\!{\mor{D}} (\mor{A})=0$
\end{tabular}
&
\begin{tabular}{@{}l@{}}
Equivalent concepts. See also:\\
\eref{8.9},~\eref{8.17};\,~\eref{8.10},~\eref{8.19new}
\end{tabular}
\\ \hline\hline
\end{tabular}
\end{table}
\newpage%
\addtolength{\oddsidemargin}{+6em}
\noindent%

\section{Discussion}
\label{discussion}
\setcounter{equation} {0}

	Since the set of all sections of a vector bundle is a module over the
ring of all ($C^0$) functions on its base with values in the field with
respect to which it has a vector
structure~\cite[chapter~3, propositions~1.6]{Husemoller},
the set $\Sec{\HilB}$ is a module over the ring of functions
$f\colon\base\to\mathbb{C}$.
Besides, this module is equipped with a scalar product. In
fact, if $\Phi,\Psi\in\Sec{\HilB}$, and $f,g\colon\base\to\mathbb{C}$, the
vector structure of $\Sec{\HilB}$ is given by
	\begin{equation}	\label{concl.1}
(f\Phi+g\Psi)\colon x\mapsto f(x)\Phi(x) + g(x)\Psi(x) \in\bHil_x,
\qquad x\in\base
	\end{equation}
and its inner product is defined via
 $(\Phi,\Psi)\mapsto\langle\Phi | \Psi\rangle\colon \base \to \mathbb{C}$
where
	\begin{equation}	\label{concl.2}
\langle\Phi | \Psi\rangle \colon x\mapsto \langle\Phi | \Psi\rangle (x) :=
\langle\Phi(x) | \Psi(x)\rangle_x \in\mathbb{C}, \qquad x\in\base.
	\end{equation}
Such a structure, module with an inner product, can naturally be called a
\emph{Hilbert module}.

	Moreover, any morphism $\mor{A}\in\Morf{\HilB}$ can be
considered as an operator $\mor{A}\colon\Sec{\HilB}\to\Sec{\HilB}$ of the
sections over $\HilB$ whose action is defined by
	\begin{equation}	\label{concl.3}
(\mor{A}\Phi)\colon x\mapsto\mor{A}_x(\Phi(x)),\qquad
\Phi\in\Sec{\HilB},\quad x\in\base
	\end{equation}
and vice versa, to any operator $\mor{B}\colon\Sec{\HilB}\to\Sec{\HilB}$
there corresponds a unique morphism of $\HilB$ whose restriction $\mor{B}_x$
on $\bHil_x$ is given via
	\begin{equation}	\label{concl.4}
\mor{B}_x(\Phi(x)):=(\mor{B}\Phi)(x),\qquad
\Phi\in\Sec{\HilB},\quad x\in\base.
	\end{equation}

	These mathematical results allow us, if needed, to reformulate
(equivalently) the Hilbert bundle description of quantum mechanics in terms
of vectors and operators, but now in the Hilbert module of sections of the
Hilbert bundle over the space-time $\base$.

	Since any pure state of a quantum system can be described via
a suitable density operator~\cite[chapter~VIII, \S~24]{Messiah-QM}, the
nonrelativistic quantum mechanics is possible to be formulated entirely in
terms of liftings of paths in the bundle of restricted morphisms of the
Hilbert bundle of the considered system.

	If unbounded states are investigated, the system's Hilbert space has
to be replaced with a more general space. In our formalism, this will result
in the identification of the fibre $\Hil$ with that last space. If this is
the case, some problems with the infinite norms of some vectors may arise,
but this is a technical task which does not change the main ideas.

	In the present investigation we have not fixed the base $\base$ of the
Hilbert bundle $\HilB$. We did not used even some concrete
assumptions about $\base$, except the self-understanding ones, e.g.\ such as
that it is an non-empty topological space. In Subsect.~\ref{new-I.3},
we assume as a working hypothesis, $\base$ to be the (coordinate)
3\ndash dimensional Euclidean space $\mathbb{E}^3$ of quantum (or classical)
mechanics. This is required for the physical interpretation of the developed
here theory. This interpretation holds true also for any differentiable
manifold $\base$ with $\dim \base \ge 3$. This is important in connection
with further generalizations. For instance, such are the cases when $\base$
is chosen as the 4\ndash dimensional Minkowski space\ndash time $M^4$ of
special relativity, or the pseudo\ndash Riemannian space\ndash time $V_4$ of
general relativity, or the Riemann\ndash Cartan space\ndash time $U_4$ of the
$U_4$\ndash gravitational theory.  All this points to the great
arbitrariness in the choice of the geometrical structure of~$\base$.
Generally it has to be determined by a theory different from quantum
mechanics, such as the classical mechanics  or the special or general
relativity. As a consequence of this, there is a room for some kind of
unified theories, for instance for a unification of quantum mechanics and
general relativity.  These problems will be investigated elsewhere.

	The developed in the present investigation theory is global in a sense
that in it we interpret $\base$ as being the whole space(\ndash time) where the
studied objects `live'. If by some reasons one wants or is forced to consider
a system resided into a limited region of $\base$, then all the theory can be
localized by replacing $\base$ with this region or simply via
\emph{mutatis mutandis} restricting the already obtained results on it. At
the same time, our theory is local in a sense that such are the used in it
observers which are assumed to can make measurements only at the points they
reside. The theory can slightly be generalized by admitting that the
observers can perform observations (measurements) at points different from
their own residence. This puts the problem of defining the mean values of the
observables at points different from the one at which the observer is
situated in such a way as they to be independent of the possibly introduce for
this purpose additional constrictions
(cf.~\cite[sect.~3]{bp-BQM-preliminary}).  This problem will be solved
elsewhere.

	The observers we have been dealing until now in this work can be
called scalar and point\ndash like as it is supposed that they have no internal
structure and the only their characteristics in the theory are their positions
(generally) in the space\ndash time $\base$. Notice that as a set the manifold
$\base$ coincides with the variety of all possible positions of all possible
observers. This observation suggests that in the general case $\base$ has to
be replaced with a set $\widetilde\base$ consisting of all values of
all parameters describing completely the states of every possible observer.
Naturally, the set $\widetilde\base$ has to be endowed with some topological
or/and smooth geometrical structure. For instance, consider an `anisotropic'
point\ndash like observers characterized by their position $x$ in $\base$ and some
vector $\mor{V}_x$ at it, i.e.\  by pairs like $(x,\mor{V}_x)$, $x\in\base$,
$\mor{V}_x\in \mor{L}_x$ with $\mor{L}_x$ being some vector space. In this
case $\widetilde\base$ is naturally identified with the (total) bundle space
$\mor{L}$ of the arising over $\base$ vector bundle
$(\mor{L},\pi_\mor{L},\base)$ with
$\mor{L}:=\bigcup_{x\in\base}\bigcup_{\mor{V}_x\in\mor{L}_x}(x,\mor{V}_x)$
and
$\pi_\mor{L}(x,\mor{V}_x):=x\in\base$,
i.e.\  we can put $\widetilde\base=\mor{L}$. In particular, if $\mor{V}_x$
is the observer's velocity at $x$, we have $\widetilde\base=T(\base)$
with $(T(\base),\pi_{T(\base)},\base)$
being the bundle tangent to $\base$. Another sensible example is when
$\mor{V}_x$ is interpreted as observer's spin etc.
	In connection with some recent investigations (see,
e.g.,~\cite{Ashtekar&Schilling,Reuter}) it is worth to be studied the special
case when $\widetilde\base$ is taken to be the (classical) phase space of the
observer.
Returning to the general
situation, we see that our theory can easily be modified to cover such
generalizations. For this purposed we have simply to replace the Hilbert
bundle $\HilB$ over $\base$ with the Hilbert bundle
$(\bHil,\proj,\widetilde\base)$ over $\widetilde\base$.
This, together with other evident corresponding changes, such as
$(x\in\base)\mapsto (x\in\widetilde\base)$ and
 $(\gamma\colon J\to\base)\mapsto(\gamma\colon J\to\widetilde\base)$,
allows us to apply the developed in the present investigation Hilbert bundle
description to far more general situations than the one we were speaking
about until now. In principle, the afore\ndash described procedure is
applicable to non\ndash local observers too, but this is out of the subject of the
present work.

	We want also to mention that since any fibre of the Hilbert bundle
$\HilB$ is an isomorphic image of the Hilbert space $\Hil$, the conventional
probabilistic interpretation of the nonrelativistic quantum
mechanics~\cite{Messiah-QM,Fock-FQM} remains
\emph{mutatis mutandis} completely
valid in the Hilbert bundle description too. For this purpose one has to
replace state vectors and operators acting on $\Hil$ with the corresponding
state liftings of paths and liftings of paths in $\HilB$ and then to
follow the general rules outlined in this work (see the paragraph after
remark~\ref{Rem4.4}) and, for instance, in~\cite{Messiah-QM}.

	In connection with further applications of the bundle approach to the
quantum field theory, we notice the following. Since in this theory
the matter fields are represented by operators acting on (wave) functions
from some space, the matter fields in their bundle modification should be
described via morphisms along paths of a suitable fibre bundle whose
sections (along paths) will represent the wave functions. We can also,
equivalently, say that in this way the matter fields would be sections of the
fibre bundle of bundle morphisms of the mentioned suitable bundle. An
important point here is that the matter fields are primary related to the
bundle arising over the space\ndash time (or other space which includes it)
and not to the space-time itself to which other structures are directly
related, such as connections and the principle bundle over it.


	The bundle approach to nonrelativistic quantum mechanics, developed in
the present investigation, seems also applicable to classical mechanics,
statistical mechanics, relativistic quantum mechanics, and field theory. We
hope that such a novel treatment of these theories will reveal new
perspectives for different generalizations, in particular for their
unification with the theory of gravitation.


\addcontentsline{toc}{section}{References}

\bibliography{bozhopub,bozhoref}
\bibliographystyle{unsrt}

\addcontentsline{toc}{subsubsection}{\vspace{1ex}This article ends at page}

\addtocontents{toc}{\vspace{3ex}\noindent\protect\enlargethispage{0ex}
\textbf{Distribution of the material:}%
\protect\\[2.3ex]\protect\hspace*{-3.4em}%
\protect\begin{tabular}{lllll}\protect\hline\protect\small
Part 	& Sections	& Title
		& quant-ph/\protect\# & Ref. \\
\protect\hline
0	&  		& Preliminary considerations
		& 9803083 	& \protect\cite{bp-BQM-preliminary}	\\
I	&1--5	 	&Introduction. The evolution transport
		& 9803084 & \protect\cite{bp-BQM-introduction+transport}	\\
II	&6\protect\&7 	&Equations of motion and observables
		& 9804062 & \protect\cite{bp-BQM-equations+observables}	\\
III	&8\protect\&9	&Pictures and integrals of motion
		& 9806046 & \protect\cite{bp-BQM-pictures+integrals} 	\\
IV	&10\protect\&11 &Mixed states and evolution transport's curvature
		& 9901039 & \protect\cite{bp-BQM-mixed_states+curvature}\\
V	&12--14 	&Theory's interpretation, summary and discussion
		& 9902068 & \protect\cite{bp-BQM-interpretation+discussion}\\
\protect\hline
\protect\end{tabular}
}

\end{document}